\documentclass[aps,letterpaper,twocolumn,nofootinbib]{revtex4}

\pdfoutput=1

\usepackage{amsthm}         
\usepackage{graphicx} 	    
\usepackage{amsmath} 	   
\usepackage{amssymb}       
\usepackage{amsfonts}
\usepackage{setspace}	    
\usepackage{capt-of}
\usepackage{algorithm}
\usepackage{algorithmic}

\begin{document}

\title{Simulation of Asymptotically AdS$_5$ Spacetimes with a Generalized Harmonic\\ Evolution Scheme}

\author{Hans Bantilan}
\author{Frans Pretorius}
\author{Steven S.\ Gubser}
\address{Joseph Henry Laboratories, Princeton University, Princeton, NJ 08544}

\begin{abstract}

Motivated by the gauge/gravity duality, we introduce a numerical scheme based 
on generalized harmonic evolution to solve the Einstein field equations on  
asymptotically anti-de Sitter (AdS) spacetimes. We work in global AdS$_5$, which 
can be described by the $(t,r,\chi,\theta,\phi)$ spherical coordinates adapted 
to the $\mathbb{R} \times S^3$ boundary. We focus on solutions that preserve an 
$SO(3)$ symmetry that acts to rotate the 2-spheres parametrized by $\theta,\phi$. 
In the boundary conformal field theory (CFT), the way in which this symmetry manifests 
itself hinges on the way we choose to embed Minkowski space in $\mathbb{R} \times S^3$.
We present results from an ongoing study of prompt black hole formation via 
scalar field collapse, and explore the subsequent quasi-normal ringdown. 
Beginning with initial data characterized by highly distorted apparent horizon geometries,
the metrics quickly evolve, via quasi-normal ringdown, to equilibrium static black hole
solutions at late times. The lowest angular number quasi-normal modes are consistent with 
the linear modes previously found in perturbative studies, whereas the higher angular 
modes are a combination of linear modes and of harmonics arising from non-linear mode-coupling. 
We extract the stress energy tensor of the dual CFT on the boundary, and find that 
despite being highly inhomogeneous initially, it nevertheless evolves from the outset 
in a manner that is consistent with a thermalized $\mathcal{N}=4$ SYM fluid. 
As a first step towards closer contact with relativistic heavy ion collision physics, we map 
this solution to a Minkowski piece of the $\mathbb{R} \times S^3$ boundary, and obtain 
a corresponding fluid flow in Minkowski space. 
\end{abstract}

\maketitle
\begin{picture}(0,0)(0,0){\put(450,300){PUPT-2403}}\end{picture}
\vspace{-1.1cm}

\section{Introduction}

It has been hoped since the beginnings of the AdS/CFT correspondence, first 
formulated in~\cite{Maldacena:1997re,Gubser:1998bc,Witten:1998qj}, that 
gauge-gravity dualities would eventually be used to relate gravitational 
calculations to experimentally  testable predictions in gauge theory. 
Over the past several years, there has been a flurry of activity in applying 
AdS/CFT to problems in high energy and condensed matter physics, which  
began with the discovery of the strongly-coupled quark-gluon plasma formed in 
heavy ion collisions. Recent proposals have suggested application to other systems, 
including those that exhibit superconductivity, the quantum hall effect, and 
superfluidity. For some review articles 
see~\cite{Gubser:2009md,Herzog:2009xv,Hartnoll:2009sz,McGreevy:2009xe,
Sachdev:2010ch,Gubser:2010nc,Horowitz:2010gk,Gubser:2011qv}. 
Though conformal field theories cannot model the exact properties of all of these 
physical phenomena, it is hoped that they nevertheless
capture aspects of the essential physics. In the strong coupling limit of the 
boundary CFT, the duality maps CFT states to bulk gravity configurations,  
and there are many cases where the latter theory is more tractable than the 
former. In many if not most of the model problems studied to date, the gravity 
description has involved black holes; aside from the interesting philosophical 
questions this poses, the implication is that one needs to study bulk solutions 
within the strong-field regime of general relativity. In situations where exact 
solutions are not known, or perturbative expansions about known solutions are 
inadequate to capture the non-linear dynamics, numerical solution of the 
Einstein field equations for an asymptotically AdS spacetime is required.

Numerical relativity has seen significant progress in recent
years modeling dynamical, strong-field geometries, though the majority
of applications have been to compact object collisions in asymptotically flat 
spacetimes. For some review articles see~\cite{Pretorius:2007nq,Duez:2009yz,
Campanelli:2010ac,Centrella:2010mx}. The asymptotic structure of AdS is drastically 
different from that of flat space; in particular the boundary of AdS is 
time-like and in causal contact with bulk geometric structures, on time scales 
relevant to the boundary physics. This poses unique challenges for numerical evolution. 
The majority of existing literature on numerical solution in AdS has focused on 
black hole formation in spherically-symmetric 
3-dimensional~\cite{Pretorius:2000yu,Husain:2000vm,Hwang:2011mn}, 
4-dimensional~\cite{Bizon:2011gg}, 5-dimensional~\cite{Garfinkle:2011hm} 
and general $D$-dimensional~\cite{Husain:2002nk,Birukou:2002ge} spacetimes,
i.e. $1+1$ dimensional simulations where the 
boundary is a single point, significantly simplifying its treatment. A 
notable exception is a study of colliding shockwaves in 5-dimensional AdS 
(AdS$_5$), with application to heavy ion collisions~\cite{Chesler:2010bi} (and
see~\cite{Wu:2011yd,Chesler:2011ds} for follow-up studies).
Planar symmetry was imposed in two spatial dimensions, reducing the problem to 
a $2+1$ dimensional simulation. Their approach was based on a null 
(characteristic) evolution scheme which is well-adapted to describing such 
colliding plane waves. This method simplifies the treatment of the AdS$_5$ 
boundary, though is difficult to generalize to situations with 
less symmetry. A more recent related study of boundary hydrodynamics
via numerical solution of the full field equations in the bulk was presented
in~\cite{Heller:2011ju}; though the evolution was effectively $1+1$ dimensional, 
the space-plus-time decomposition, as used here, in principle allows for a
straightforward extension to situations with less symmetry. 

Inspired by the growing success of gauge/gravity dualities, we are initiating 
a new program to solve the Einstein field equations in asymptotically AdS$_5$ 
spacetimes, based on the generalized harmonic (GH) evolution scheme presented 
in~\cite{Pretorius:2004jg,Pretorius:2005gq}. These methods were introduced in 
the context of asymptotically flat spacetimes, and it turns out to be a rather 
non-trivial exercise to adapt them to AdS$_5$. The main purpose of this paper 
is to describe, in detail, the steps taken towards achieving a stable Cauchy evolution 
code in spacetimes that are asymptotically AdS$_5$. In this first study we impose an 
$SO(3)$ symmetry for simplicity; since the method is based on Cauchy evolution, it should 
be straightforward to relax this symmetry restriction for future studies, at the expense 
of computational complexity. As a sample application, we study the quasi-normal ringdown 
of initial data describing highly distorted black holes formed from scalar field collapse, 
and extract the stress energy tensor of the dual conformal field theory defined on the 
$\mathbb{R} \times S^3$ boundary.  The $SO(3)$ symmetry we impose here is precisely the one 
identified in analytical studies of colliding shock waves in AdS$_5$ \cite{Gubser:2008pc,Gubser:2009sx}.

An outline of the rest of the paper is as follows. We begin in 
Sec.~\ref{section:gravity_in_aads_spacetimes} with relevant facts about 
asymptotically AdS$_5$ spacetimes. In Sec.~\ref{section:initial_data}, we 
describe a method for constructing AdS$_5$ initial data based on a conformal 
decomposition, restricted to time-symmetric initial conditions. For this study 
we use a scalar field to source deviations from pure AdS$_5$, and find that we 
can solve for arbitrarily strong initial data: the initial slice can 
be made to contain an apparent horizon, and this horizon can be made arbitrarily 
large by adjusting (for example) the amplitude of the initial scalar field profile. 
In Sec.~\ref{section:evolution}, we describe the GH formalism that we use to 
evolve the initial data forward in time. Crucial to the stability of this scheme 
is the asymptotic nature of the so-called {\em source functions} that in GH 
evolution are traditionally associated with coordinate degrees of freedom. Here, 
in contrast to asymptotically flat spacetimes, we find that their asymptotic 
form can {\em not} be freely specified if the metric deviation is to be 
non-singular in the approach to the AdS$_5$ boundary. This and other 
subtleties related to evolution are clarified and addressed in the latter 
parts of the section. We discretize the field equations using finite difference 
methods, and the details of the numerical methods that we use to solve the initial 
data and evolution equations are discussed in Sec.~\ref{section:numerical_scheme}. 

Results from a study of highly deformed black holes, their subsequent evolution and 
ringdown, and the stress tensor of the corresponding states in the dual 
boundary CFT, are presented in Sec.~\ref{results}. What is novel about this 
study is that the initial horizon geometry cannot be considered a small 
perturbation of the final static horizon, and hence we are probing 
an initial non-linear phase of the evolution of the bulk spacetime. 
Shortly after the initial time, the metric can be described by 
a combination of quasi-normal modes and what appear to be gauge modes.
We find frequencies that are consistent with the linear modes 
found in perturbative studies of static black holes, and in modes at higher angular number, 
we find evidence of non-linear mode-coupling. On the boundary, the dual CFT stress tensor 
behaves like that of a thermalized $\mathcal{N}=4$ super-Yang-Mills (SYM) fluid.
We find that the equation of state $\epsilon=3P$ of this fluid is consistent with conformal 
invariance (here, $\epsilon$ and $P$ are the rest frame density and pressure of the fluid, 
respectively), and that its transport coefficients match those previously 
calculated for an $\mathcal{N}=4$ SYM fluid via holographic methods. 
Modulo a brief transient that is numerical in nature, this matching 
appears to hold from $t=0$ onwards. This is not {\em a priori} inconsistent 
with earlier results from quasi-normal modes~\cite{Kovtun:2006pf,Janik:2006gp,Friess:2006kw}
and numerical analyses~\cite{Chesler:2010bi,Heller:2011ju,Wu:2011yd,Chesler:2011ds},
where a certain ``thermalization time'' is observed before the boundary
physics begins to correspond to that of a fluid. In contrast to those studies, 
we begin with a large, distorted black hole, i.e. an inhomogeneous thermal 
state, and the physics studied here is more of equilibration rather than of 
thermalization.

In the final section, we transform solutions computed in global AdS onto 
a Minkowski piece of the boundary, and examine the temperature of the corresponding fluid flows. Under 
this transformation, the spatial profile of temperature at the initial time resembles a 
Lorentz-flattened pancake centered at the origin of Minkowski space. By interpreting the 
direction along which the data is flattened as the beam-line direction, our initial data 
can be thought of as approximating a head-on heavy ion collision at its moment of impact.

We relegate to Appendix~\ref{app:effect_of_scalar_backreaction_on_metric_falloff} 
some technical details on the effect of matter backreaction on the asymptotic 
metric fall-off. In Appendix~\ref{app:boundary_operator}, 
we discuss how our asymptotic metric boundary conditions relate to the 
boundary conditions derived in~\cite{Ishibashi:2004wx} for linearized 
gravitational perturbations on an AdS background that lead to well-defined 
dynamics, and in Appendix~\ref{app:tables_of_scalar_qnm_frequencies} we 
tabulate the scalar field quasi-normal mode frequencies. Throughout, we use 
geometric units where Newton's constant $G$ and the speed of light $c$ are set to $1$.

\section{Gravity in Asymptotically AdS$_5$ Spacetimes}\label{section:gravity_in_aads_spacetimes}

\subsection{The AdS$_5$ Spacetime}

The Lagrangian density of gravity with cosmological constant $\Lambda$,
coupled to matter with a Lagrangian density $\mathcal{L}_m$, is given 
by
\begin{equation}\label{eqn:gravitylagrangian}
\mathcal{L} = \frac{1}{16 \pi} \left( R - 2 \Lambda \right) + \mathcal{L}_m,
\end{equation}
where $R$ is the Ricci scalar.
The corresponding equations of motion then take the local form
\begin{equation}\label{eqn:efe}
R_{\mu \nu} - \frac{1}{2} R g_{\mu \nu} + \Lambda g_{\mu \nu} = 8\pi T_{\mu\nu}.
\end{equation}
Here, $R_{\mu \nu}$ is the Ricci tensor, $g_{\mu \nu}$ is the metric tensor and 
$T_{\mu\nu}$ is the stress energy tensor of the matter. The metric of AdS$_5$ is 
the maximally symmetric vacuum solution of~(\ref{eqn:efe}) in $D=5$ dimensions, 
with a negative cosmological constant $\Lambda \equiv \Lambda_5 < 0$. In terms 
of the global coordinates $x^{\mu} = (t,r,\chi,\theta,\phi)$ that cover the 
whole spacetime, this solution is
\begin{eqnarray}\label{eqn:originalmetric}
ds^2 &\equiv& \hat{g}_{\mu\nu} dx^\mu dx^\nu \\
&=& - f(r) dt^2 + \frac{1}{f(r)} dr^2 +r^2 d{\Omega_3}^2\nonumber
\end{eqnarray}
where we have defined $f(r)=1+r^2/L^2$ for convenience. Here, 
$d{\Omega_3}^2 = d\chi^2 + \sin^2\chi \left( d\theta^2 + \sin^2\theta d\phi^2 \right)$ 
is the metric of the 3-sphere parametrized by angles $\chi,\theta,\phi$, 
and $L$ is the AdS radius of curvature, related to the cosmological constant 
by $\Lambda_5 = -(D-1)(D-2)/(2 L^2) = -6/L^2$. We are free to set the AdS 
length scale $L$, which is the scale with respect to which all other lengths 
are measured. In the code we set $L=1$, though we will continue to explicitly 
display $L$ in all of the following, unless otherwise indicated.

Due to the significance of the AdS$_5$ boundary in the context of AdS/CFT, it 
is useful to introduce a ``compactified'' radial coordinate $\rho$ so that the 
boundary is at a finite $\rho$. We choose 
\begin{equation}\label{eqn:r_def}
r=\frac{\rho}{1-\rho/\ell},
\end{equation}
where $\ell$ is an arbitrary compactification scale, independent of the AdS 
length scale $L$, such that the AdS$_5$ boundary is reached when $\rho=\ell$. 
In our code and in all of the following, we set $\ell=1$, though note that this
scale is implicitly present since $\rho$ has dimensions of length. Transforming 
to the $\rho$ coordinate, the metric~(\ref{eqn:originalmetric}) takes the form:
\begin{equation}\label{eqn:compactifiedmetric}
ds^2 =\frac{1}{(1-\rho)^2} \left( -\hat{f}(\rho) dt^2 + 
\frac{1}{\hat{f}(\rho)} d\rho^2 +\rho^2 d{\Omega_3}^2 \right) 
\end{equation}
where $\hat{f}(\rho)=(1-\rho)^2+\rho^2/L^2$. 

AdS$_5$ can also be described as the universal cover of the hyperboloid $X^A$ in 
$\mathbb{R}^{4,2}$ defined by the locus
\begin{equation}\label{eqn:embedding1}
-(X^{-1})^2-(X^0)^2+\sum_{i=1}^{i=4} (X^i)^2 = -L^2.
\end{equation}
This space has symmetry group $SO(4,2)$ whose transformations preserve the 
quadratic form (\ref{eqn:embedding1}). The metric of AdS$_5$ is then the 
metric $\hat{g}_{\mu \nu}$ induced on the above hyperboloid from the flat 
metric $\hat{G}_{A B}$ of an $\mathbb{R}^{4,2}$ ambient space. Here the metric 
$\hat{G}_{A B}$ is simply given by $\hat{G} = diag(-1,-1,1,1,1,1)$. The 
hyperboloid can be more efficiently described in terms of embedding 
coordinates $x^\mu$ defined by a set of embedding functions $X^A(x^\mu)$, so 
that the induced metric $\hat{g}_{\mu \nu}$ on the hyperboloid is
\begin{equation}\label{eqn:embedding2}
\hat{g}_{\mu\nu} = \left( \frac{\partial X^A}{\partial x^\mu} \right) 
\left( \frac{\partial X^B}{\partial x^\nu} \right) \hat{G}_{A B}.
\end{equation}

The global coordinates $x^{\mu} = (t,r,\chi,\theta,\phi)$ can then be thought of 
as corresponding to a choice of embedding functions 
\begin{eqnarray}\label{eqn:embedding3}
X^{-1} &=& \sqrt{r^2 + L^2} \cos(t/L) \nonumber \\
X^{0} &=& \sqrt{r^2 + L^2} \sin(t/L) \nonumber \\
X^{1} &=& r \sin\chi \sin\theta \sin\phi \nonumber \\
X^{2} &=& r \sin\chi \sin\theta \cos\phi \nonumber \\
X^{3} &=& r \sin\chi \cos\theta \nonumber \\
X^{4} &=& r \cos\chi.
\end{eqnarray}
Notice that $t=0$ and $t=2\pi L$ are identified on the hyperboloid, so that 
there are closed time-like curves in this space. AdS$_5$ is defined as the 
hyperboloid's universal cover precisely to remove these closed time-like 
curves. This universal cover is obtained by unwrapping the $S^1$ parametrized 
by $t$ on the hyperboloid, which would then run from $-\infty$ to $\infty$ in 
AdS$_5$. 

\subsection{Boundary of AdS$_5$}

\begin{figure}[h]
        \centering
        \includegraphics[width=4.0in,clip=true]{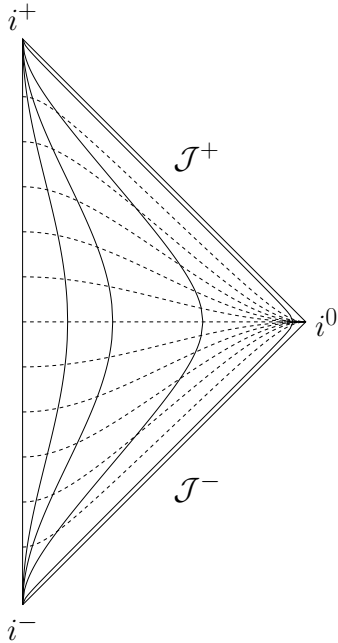}
\parbox{3.3in}{\vspace{-2cm} \caption{The conformal diagram of Minkowski space. The boundary 
consists of the point at spatial infinity $i^0$, and the null surfaces at future 
null infinity $\mathcal{J}^+$ and past null infinity $\mathcal{J}^-$. In this 
compactification, future time-like infinity $i^+$ and past time-like infinity 
$i^-$ are represented by points. Dashed lines are constant $t$ surfaces, and solid 
lines are constant $r$ surfaces.
        }\label{fig:minkowski}}
\end{figure}

The boundary of AdS$_5$ differs dramatically from that of asymptotically flat 
spacetimes. To see this, let us first remember how the analogous story unfolds 
in the familiar setting of Minkowski space, whose metric in polar coordinates 
$\eta_{\mu \nu} dx^\mu dx^\nu = -dt^2 + dr^2 + r^2 d\Omega^2$ can be rewritten 
through a series of coordinate transformations $u_\pm = t \pm r$, 
$\tilde{u}_\pm = \arctan u_\pm = (T \pm R)/2$ to read
\begin{equation}
ds^2 = \frac{1}{4 \cos^2u_+ \cos^2u_-} \left( -dT^2 + dR^2 + 
\sin^2 R d\Omega^2 \right).
\end{equation}
Through this conformal compactification, the infinite region 
$t\in (-\infty,\infty)$, $r \in [0,\infty)$ is mapped to the interior of a 
compact region defined by $|T \pm R| < \pi$. Given that $T$ is a time-like 
coordinate and $R>0$ is a space-like coordinate, this region is the triangle 
$T \in [-\pi,\pi]$, $R \in [0,\pi]$ in the $T,R$ plane---see 
Fig.~\ref{fig:minkowski}. Consequently, the boundary consists of the two null 
surfaces of future/past null infinity, meeting at spatial infinity. Spatial 
infinity in Minkowski space is thus not in causal contact with its interior. 
Notice that the geometry of Minkowski space is conformal to a patch 
$T \in [-\pi,\pi]$ of the Einstein static universe.

\begin{figure}[h]
        \centering
        \includegraphics[width=4.0in]{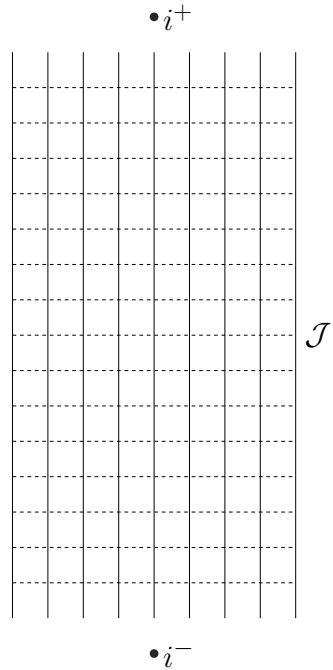}
\parbox{3.3in}{\vspace{-2cm} \caption{The conformal diagram of anti-de Sitter space. The 
boundary consists of the time-like surface $\mathcal{J}$; past and future 
time-like infinity are represented by the points $i^-$ and $i^+$, respectively. 
Conventions used here are the same as those in Fig.~\ref{fig:minkowski}.
        }\label{fig:pureads}}
\end{figure}

The conformal compactification of AdS$_5$ is achieved\footnote{In practice, 
we compactify using (\ref{eqn:r_def}), but for the current didactic discussion 
we choose the compactification that most closely resembles its Minkowski space 
analog.} by the spatial coordinate transformation $r/L = \tan R$, which brings its 
metric (\ref{eqn:originalmetric}) into the form
\begin{equation}
ds^2 = \frac{L^2}{\cos^2 R} \left( -dt^2 + dR^2 + \sin^2 R {d\Omega_3}^2 \right)
\end{equation}
where the infinite region $r \in [0,\infty]$ is mapped to the interior of a 
compact region $R \in [0,\pi/2]$---see Fig.~\ref{fig:pureads}. This halved range 
of $R$ implies that anti-de Sitter space is conformal to one-half of the 
Einstein static universe. More crucially, we have not rescaled the time 
coordinate $t$ in this compactification. The important consequence is a 
spatial infinity that runs along the $t$ direction: it is time-like and thus 
causally connected to the interior. 

A proper treatment of the AdS boundary is crucial to a solution of the Cauchy 
problem in an asymptotically AdS spacetime. Without some specification of 
boundary conditions at time-like infinity, only a small wedge to the causal
future of an initial space-like foliation can be solved for. This is in contrast 
with asymptotically Minkowski spacetimes, where the specification of initial 
data on a slice that reaches spatial infinity is sufficient to evolve the entire 
interior. Such an approach to initial data is not useful in the asymptotically 
AdS case, particularly in problems that are relevant to AdS/CFT: for these 
problems, the time-like boundary must be included as part of the spacetime.

\subsection{Asymptotically AdS$_5$ Spacetimes}\label{subsection:aads_spacetimes}

An asymptotically AdS$_5$ (AAdS$_5$) spacetime is one that shares the boundary 
structure of AdS$_5$. In particular, it must have $SO(4,2)$ as its asymptotic 
symmetry group. To write down the asymptotics of the fields in such a spacetime, 
let us decompose the metric by writing
\begin{equation}\label{eqn:generalmetricdecomposition}
g_{\mu \nu} = \hat{g}_{\mu \nu} + h_{\mu \nu}
\end{equation}
where $h_{\mu \nu}$ is the deviation of the full metric $g_{\mu \nu}$ from the 
metric $\hat{g}_{\mu \nu}$ of AdS$_5$. For an explicit form of the metric that 
we evolve, the reader may wish to briefly skip to~(\ref{eqn:metric_asymptotics}).

The matter-free asymptotics of $h_{\mu \nu}$ corresponding to an AAdS$_5$ 
spacetime were found in~\cite{Henneaux:1985tv}, by requiring that these 
deviations $h_{\mu \nu}$ satisfy boundary conditions at spatial infinity that 
are invariant under the $SO(4,2)$ symmetry group of AdS$_5$. 
The resulting boundary conditions are consistent with the Fefferman-Graham
construction~\cite{fefferman} (see Appendix ~\ref{app:boundary_operator}). 
To obtain these boundary conditions, \cite{Henneaux:1985tv} begins by noting that a 
typical generator $\xi^\mu$ of $SO(4,2)$ has asymptotics 
\begin{eqnarray}
\xi^m &=& \mathcal{O}(1) \nonumber \\
\xi^r &=& \mathcal{O}(r) \nonumber \\
\xi^m_{,r} &=& \mathcal{O}(r^{-3}) \nonumber \\
\xi^r_{,r} &=& \mathcal{O}(1),
\end{eqnarray}
where the index $m$ denotes the non-radial coordinates $(t,\chi,\theta,\phi)$.
Since we need to ensure that $SO(4,2)$ is an asymptotic symmetry, we are 
interested in satisfying the Killing equation in an asymptotic sense
\begin{equation}\label{eqn:asymptotickillingeqn}
\mathcal{L}_\xi g_{\mu \nu} = \mathcal{O}(h_{\mu \nu}),
\end{equation}
so that for all generators $\xi^\mu$ of $SO(4,2)$, the Lie derivative of each 
metric component along $\xi^\mu$ approaches zero with the appropriate power 
of $r$ as $r\rightarrow\infty$. In other words, we are looking for an asymptotic 
form of the metric deviation $h_{\mu \nu} \sim r^{p_{\mu \nu}}$, such that this 
asymptotic form is preserved by the coordinate transformations that correspond to 
the generators $\xi^\mu$. Direct calculation reveals that (\ref{eqn:asymptotickillingeqn}) 
holds when $p_{r r}=-D-1$, $p_{r m}=-D$, $p_{m n}=-D+3$. Since we are interested 
in the case of $D=5$ dimensions, the vacuum boundary conditions read
\begin{eqnarray}\label{eqn:metricbcs}
h_{r r} &=& f_{r r}( t,\chi,\theta,\phi ) \frac{1}{r^6} + \mathcal{O}(r^{-7}) 
\nonumber \\
h_{r m} &=& f_{r m}( t,\chi,\theta,\phi ) \frac{1}{r^5} + \mathcal{O}(r^{-6}) 
\nonumber \\
h_{m n} &=& f_{m n}( t,\chi,\theta,\phi ) \frac{1}{r^2} + \mathcal{O}(r^{-3}).
\end{eqnarray}

These boundary conditions are valid for vacuum AAdS$_5$ spacetimes, and as we shall see 
in the next section, continue to hold for spacetimes containing localized matter distributions 
with sufficiently rapid fall-off near the boundary. For more general matter 
distributions, fall-off conditions can be obtained via different methods; see for 
example~\cite{deHaro:2000xn,Henneaux:2006hk}. The most general set of fall-off 
conditions were found in~\cite{Papadimitriou:2005ii}: these continue to hold for 
spacetimes with fewer restrictions on boundary conformal structure and boundary topology.

\subsection{Scalar Fields in Asymptotically AdS$_5$ Spacetimes}\label{subsection:scalar_fields_in_asymptotically_ads5_spacetimes}

As we will be coupling matter to gravity, it is important to know how the 
presence of matter alters the vacuum boundary conditions (\ref{eqn:metricbcs}). 
To understand just the asymptotics, it suffices to consider a static 
spherically symmetric scalar $\phi = \phi(r)$ of mass $m$, for which the 
Klein-Gordon equation 
\begin{equation}\label{eqn:kgeqn}
\Box \phi = m^2 \phi
\end{equation}
takes the form
\begin{equation}\label{eqn:kgeqn_spherical}
\left[ \frac{r D}{L^2} + \frac{D-2}{r} + \hat{g}^{rr} \frac{\partial}{\partial r} \right] 
\frac{\partial \phi}{\partial r} = m^2 \phi.
\end{equation}
The static ansatz $\phi(r) \sim r^{-\Delta}$ yields a quadratic equation for 
$\Delta$ whose solutions are the powers of allowed fall-offs. The asymptotics 
for a scalar field of mass $m$ are thus described by
\begin{equation}\label{eqn:scalarbcs}
\phi = \frac{A}{r^{d-\Delta}} + \frac{B}{r^{\Delta}},
\end{equation}
\begin{equation}\label{eqn:scalarpower}
\Delta =   \frac{d}{2} + \sqrt{ \left( \frac{d}{2}\right)^2 + L^2 m^2},
\end{equation}
which gives $\Delta = 2 + \sqrt{4+m^2}$ in $D \equiv d+1 = 5$ dimensions. The usual 
vacuum/localized-matter boundary conditions given in (\ref{eqn:metricbcs}) 
can accomodate a scalar field with a vanishing $A$ branch, $A=0$. We show in 
Appendix \ref{app:effect_of_scalar_backreaction_on_metric_falloff} that this 
case does not alter the boundary conditions (\ref{eqn:metricbcs}), because 
the scalar field falls off sufficiently quickly near the boundary. This also 
suggests why gravitational wave perturbations are consistent 
with~(\ref{eqn:metricbcs}), since their propagation characteristics are 
``morally'' equivalent to that of a massless scalar field, i.e. having
$\Delta=4$ for $B$ branch solutions.

The situation becomes richer for nonlocalized-matter boundary conditions with 
a non-zero $A$ branch, $A \ne 0$. In this case, the backreaction of the scalar field 
onto the geometry begins to alter the metric asymptotics. Scenarios studied in this 
paper require only the usual vacuum/localized-matter boundary conditions 
(\ref{eqn:metricbcs}); systems which require nonlocalized-matter boundary conditions 
will be addressed in a future work.

\subsection{The Stress Energy Tensor of the Dual CFT}\label{subsection:the_set_of_the_dual_cft}

Here we briefly mention the most relevant entry in the AdS/CFT dictionary for 
this study, namely the one that enables us to extract the expectation value 
$\left< T_{\mu \nu} \right>_{\text{CFT}}$ of the CFT stress energy tensor from 
the asymptotic behavior of the metric:\footnote{The full expression for the boundary 
stress tensor includes a factor of $1/G$, where $G$ is Newton's constant, corresponding 
to a scaling as $N^2$ in the large $N$ gauge theory dual to $AdS_5$.  We have omitted 
the factor of $1/G$ from (\ref{eqn:cftsetexpectation}) in keeping with our standard 
convention that $G=1$.  When quoting explicit numerical results for the boundary stress 
tensor, we will also set $L=1$.  By doing so we are restricting to a specific count of 
degrees of freedom in the boundary theory; but since the scaling of $\left< T_{\mu\nu} \right>$ 
is straightforward, there is no significant loss of generality.}
\begin{equation}\label{eqn:cftsetexpectation}
\left< T_{\mu \nu} \right>_{\text{CFT}}=\underset{q \rightarrow 0}{\lim}{\frac{1}{q^2} {}^{(q)}T_{\mu \nu}}.
\end{equation}
Here, $q=1-\rho$ (though more generally $q$ is a smooth positive scalar with a 
simple zero at the AAdS boundary), and ${}^{(q)} T_{\mu \nu}$ is the Brown-York 
quasi-local stress tensor~\cite{Brown:1992br}. For AAdS$_5$ spacetimes, the 
quasi-local stress tensor defined on a $q={\rm const.}$ time-like hypersurface 
$\partial M_q$ was constructed in \cite{Balasubramanian:1999re}, and is given by 
\begin{equation}\label{eqn:quasiset}
{}^{(q)} T^0_{\mu \nu} = \frac{1}{8 \pi} \left( {}^{(q)}\Theta_{\mu \nu} - {}^{(q)}\Theta \Sigma_{\mu \nu} 
- \frac{3}{L} \Sigma_{\mu \nu} + {}^{(q)} G_{\mu \nu} \frac{L}{2} \right).
\end{equation}
Here, ${}^{(q)}\Theta_{\mu \nu} = -{\Sigma ^\alpha}_\mu {\Sigma ^\beta}_\nu \nabla_{(\alpha} S_{\beta)}$ 
is the extrinsic curvature of the time-like surface $\partial M_q$, $S^\mu$ is a 
space-like, outward pointing unit vector normal to the surface $\partial M_q$, 
$\Sigma_{\mu \nu}\equiv g_{\mu\nu} - S_\mu S_\nu $ is the induced 4-metric on 
$\partial M_q$, $\nabla_\alpha$ is the covariant derivative 
operator, and ${}^{(q)} G_{\mu \nu}$ is the Einstein tensor associated with 
$\Sigma_{\mu \nu}$. The last two terms in (\ref{eqn:quasiset}) are counterterms 
designed to exactly cancel the divergent boundary behavior of the first two 
terms of~(\ref{eqn:quasiset}) evaluated in pure AdS$_5$.

A feature of the stress tensor (\ref{eqn:quasiset}) is that it is non-vanishing 
even when the geometry is that of pure AdS$_5$. This non-vanishing piece was 
already noticed in~\cite{Balasubramanian:1999re}, and was correctly identified 
as the contribution from the Casimir energy of the boundary CFT: this CFT is 
defined on a manifold with topology $\mathbb{R} \times S^3$, and so can have a 
non-vanishing vacuum energy. Since this Casimir contribution is 
non-dynamical, we consider it as part of our background vacuum and simply 
subtract it from (\ref{eqn:quasiset}), obtaining 
\begin{equation}\label{eqn:quasiset_subtracted}
{}^{(q)} T_{\mu \nu} = {}^{(q)} T^0_{\mu \nu} - {}^{(q)} t_{\mu \nu}.
\end{equation}
Setting $L=1$, the non-zero components of the Casimir 
contribution ${}^{(q)} t_{\mu \nu}$ are ${}^{(q)} t_{tt} = 3 q^2 / (64\pi)$, 
${}^{(q)} t_{\chi\chi} = q^2 / (64\pi)$, 
${}^{(q)} t_{\theta\theta} = q^2 \sin^2\chi / (64\pi)$, and 
${}^{(q)} t_{\phi\phi} = {}^{(q)} t_{\theta\theta} \sin^2\theta$.

\section{Initial Data}\label{section:initial_data}

Initial data for Cauchy evolution of the Einstein field equations 
(\ref{eqn:efe}) is not freely specifiable, but is subject to a set of $D$ 
constraint equations: the Hamiltonian constraint and the $D-1$ non-trivial 
components of the momentum constraints. There are many conceivable ways of 
finding initial data that are consistent with these constraints; here, we 
adapt to AAdS$_5$ spacetimes the traditional Arnowitt-Deser-Misner 
(ADM)~\cite{Arnowitt:1962hi}-based conformal decomposition approach often used 
in asymptotically flat spacetimes. See~\cite{Gourgoulhon:2007tn} for a recent review. 
To simplify this first study, we restrict initial data to a moment of time symmetry, 
and we use a scalar field to source non-trivial deviations from pure vacuum AdS$_5$.
 
Specifying time symmetric initial data is equivalent to demanding that the 
extrinsic curvature of the initial $t=0$ slice $\Sigma_{t=0}$ vanishes. The 
extrinsic curvature of a constant $t$ slice $\Sigma_{t}$ of the spacetime is 
defined as
\begin{eqnarray}\label{eqn:extrinsiccurvature}
K_{\mu \nu} &=& - {\gamma_{\mu}}^{\alpha} {\gamma_{\nu}}^{\beta} 
\nabla_{(\alpha} n_{\beta)} 
\nonumber \\
&=& -\frac{1}{2} \mathcal{L}_n \gamma_{\mu \nu},
\end{eqnarray}
where 
\begin{equation}\label{eqn:n}
n_\mu=-\alpha \partial_\mu t
\end{equation}
is the unit time-like one-form normal to the $\Sigma_{t}$ slice, $\alpha$ is 
the so-called lapse function, and $\gamma_{\mu \nu}$ is the 4-metric induced on 
$\Sigma_{t}$, expressible in local coordinates as 
\begin{equation}\label{eqn:gamma}
\gamma_{\mu \nu} = g_{\mu \nu} + n_\mu n_\nu.
\end{equation}

The momentum constraints are the $D-1$ equations
\begin{equation}\label{eqn:momentumconstraint}
D_{\nu} K^{\mu \nu} - \gamma^{\mu \nu} D_{\nu} K = 8 \pi j^{\mu},
\end{equation}
where $D_\mu=\gamma_\mu{}^\nu\nabla_\nu$ is the derivative operator intrinsic to 
$\Sigma_{t}$, and  
\begin{equation}
j^\mu= -T_{\alpha \beta} n^\alpha \gamma^{\mu \beta} 
\end{equation}
is the momentum of any matter in the spacetime. Time symmetry $K_{\mu\nu}|_{t=0}=0$
requires that in order for~(\ref{eqn:momentumconstraint}) to be satisfied, the 
momentum density $j^{\mu}$ must vanish everywhere on $\Sigma_{t=0}$. 
This leaves us with only the Hamiltonian constraint; the solution of this equation is 
the subject of the next few sections.

\subsection{Hamiltonian Constraint}\label{subsection:hamiltonian_constraint}

The Hamiltonian constraint is a single equation
\begin{equation}\label{eqn:admform_gen}
{}^{(4)}R + K^2 - K_{\mu \nu} K^{\mu \nu} - 2\Lambda_5 = 16 \pi \rho_E,
\end{equation}
where ${}^{(4)}R$ is the Ricci scalar of the geometry intrinsic to the slice 
$\Sigma_{t}$, and 
\begin{equation}\label{eqn:rho_E}
\rho_E=T_{\mu \nu} n^\mu n^\nu
\end{equation}
is the energy density on the slice. At a moment of time 
symmetry~(\ref{eqn:admform_gen}) simplifies to
\begin{equation}\label{eqn:admform}
{}^{(4)}R - 2\Lambda_5 = 16 \pi \rho_E.
\end{equation}
Following the conformal approach, we will solve this equation by requiring that 
our spatial 4-metric $\gamma_{\mu \nu}$ be conformal to the 4-metric 
$\hat{\gamma}_{\mu \nu}$ of a spatial slice of vacuum AdS$_5$ 
\begin{eqnarray}\label{eqn:conformalfactor}
\gamma_{\mu \nu} &=& \zeta^2 \hat{\gamma}_{\mu \nu} \nonumber \\
\gamma^{\mu \nu} &=& \zeta^{-2} \hat{\gamma}^{\mu \nu},
\end{eqnarray}
for some positive smooth function $\zeta$ on $\Sigma_{t=0}$ with boundary 
condition $\left. \zeta \right|_{\partial \Sigma}=1$. The Hamiltonian constraint 
(\ref{eqn:admform}) can then be expressed as 
\begin{equation}\label{eqn:conformalform}
{}^{(4)}\hat{R} - 6 \zeta^{-1} \hat{\gamma}^{\alpha \beta} 
 \hat{D}_\alpha \hat{D}_\beta \zeta - \zeta^2 2 \Lambda_5 = 16 \pi \zeta^2 \rho_E,
\end{equation}
where $\hat{D}_\alpha$ is the covariant derivative operator compatible with 
$\hat{\gamma}_{\mu \nu}$, and ${}^{(4)}\hat{R}$ is the corresponding Ricci 
scalar. ${}^{(4)}\hat{R}$ is readily computed from the spatial part of the 
AdS$_5$ metric (\ref{eqn:originalmetric}), giving a constant 
${}^{(4)}\hat{R} = -12/L^2 = 2 \Lambda_5$. This lets us rewrite 
(\ref{eqn:conformalform}) as 
\begin{equation}\label{eqn:conformalform2}
\hat{\gamma}^{\alpha \beta} \hat{D}_\alpha \hat{D}_\beta \zeta - \frac{1}{3} 
\Lambda_5 \zeta + \frac{1}{3} \left( \Lambda_5 + 8 \pi \rho_E \right) \zeta^3 
= 0.
\end{equation}
Notice that the Hamiltonian constraint (\ref{eqn:conformalform2}) does not contain 
the lapse function $\alpha$; this is consistent with the understanding that unlike 
a specification of the lapse $\alpha$, which encodes the manner in which data evolves 
away from the initial slice, the Hamiltonian constraint may only set data intrinsic 
to the initial slice itself. Furthermore, by restricting our attention to conformally 
AdS initial data, we have written this Hamiltonian constraint as a non-linear elliptic 
equation for the conformal factor $\zeta$. This equation can be solved once we specify 
the matter energy density $\rho_E$, which we discuss in the following section. 

\subsection{Hamiltonian Constraint with Scalar Matter}\label{scalarid}

In this section, we write down an explicit form for the matter source term in 
(\ref{eqn:conformalform2}), constructed from scalar field initial data. 
Scalar fields are particularly convenient for our purposes, since their energy 
density constitutes a parameter with which we can tune the initial data. 
Considering cases of incrementally larger energy density then allows us to 
approach the dynamical, strong field regime in a controlled fashion. 

The Lagrangian density of a scalar field $\phi$ with a potential $V(\phi)$ is 
given by
\begin{equation}\label{eqn:reallagrangian}
\mathcal{L}(\phi,\partial_\mu \phi ) = -\frac{1}{2} g^{\alpha \beta} 
\partial_{\alpha} \phi \partial_{\beta} \phi - V(\phi).
\end{equation}
Varying the action constructed from (\ref{eqn:reallagrangian}) with respect to 
the metric $g_{\alpha \beta}$ gives an energy-momentum tensor
\begin{equation}\label{eqn:realenergymomentum}
T_{\mu \nu} = \partial_\mu \phi \partial_\nu \phi - g_{\mu \nu} 
\left( \frac{1}{2} g^{\alpha \beta} \partial_{\alpha} \phi \partial_{\beta} 
\phi + V(\phi) \right).
\end{equation}
Substituting~(\ref{eqn:n}), (\ref{eqn:gamma}), (\ref{eqn:conformalfactor}), and 
(\ref{eqn:realenergymomentum}) into (\ref{eqn:rho_E}), then using the restriction 
of time-symmetry, which for the scalar field amounts to setting 
$\partial_t \phi|_{t=0}=0$, we obtain 
\begin{equation}\label{eqn:timesymmetryenergydensity}
\rho_E = \zeta^{-2} \hat{\gamma}^{ij} \partial_{i} \phi \partial_{j} \phi 
+ V(\phi).
\end{equation}
The Hamiltonian constraint thus takes the form
\begin{eqnarray}\label{eqn:conformalform3}
\hat{\gamma}^{kl} \hat{D}_k \hat{D}_l \zeta - \frac{1}{3} (\Lambda_5  - 
8\pi \hat{\gamma}^{ij} \partial_{i} \phi \partial_{j} \phi) \zeta \nonumber \\
+ \frac{1}{3} \left( \Lambda_5 + 8 \pi V(\phi) \right) \zeta^3 =0 .
\end{eqnarray}
The choice of $\phi$ on the spatial slice is completely arbitrary. For the tests 
and quasi-normal mode study described here, we restrict ourselves to free, massless 
fields i.e. $V(\phi)=0$. For the spatial profile of these fields, we use the following 
$5$-parameter generalized Gaussian function:
\begin{equation}\label{eqn:generalizedgaussian}
\phi(\rho,\chi) = q^4 (1+\rho)^4 A_0 \exp\left( -\frac{(R(\rho,\chi)-R_0)^2}{\delta^2} \right)
\end{equation}
where
\begin{eqnarray}
R(\rho,\chi) &=& \sqrt{\frac{x(\rho,\chi)^2}{{w_x}^2} + \frac{y(\rho,\chi)^2}{{w_y}^2}} \nonumber \\
x(\rho,\chi) &=& \rho \cos\chi \nonumber \\
y(\rho,\chi) &=& \rho \sin\chi.\label{eqn:generalizedgaussian_b}
\end{eqnarray}
Here, $A_0$ is the maximum amplitude, $R_0$ fixes the radial position of the 
maximum, $\delta$ sets the overall compactness of the profile, and $w_x$,$w_y$ 
can be adjusted to set the relative compactness of the profile in the $x$,$y$ 
directions. The $q^4$ factor ensures that this profile has the correct fall-off
for a massless scalar field, consistent with~(\ref{eqn:scalarbcs}) 
and~(\ref{eqn:scalarpower})\footnote{The $q^4$ prefactor on the right side of 
(\ref{eqn:generalizedgaussian}) means that we are not deforming the CFT by the 
dimension $4$ operator dual to the massless scalar $\phi$.}; this is supplemented 
by a $(1+\rho)^4$ factor to maintain the original Gaussian profile's even character 
near the origin. 

\subsection{Choice of Coordinates in Terms of Lapse and Shift on the 
Initial Slice}\label{sec:gauge_t0_ij}

To complete the specification of initial data, we must choose coordinates 
on the initial slice, i.e. the form in which we wish to represent our 
metric $\gamma_{\mu\nu}$. This specification fixes the remaining 
coordinate degrees of freedom, and within the ADM decomposition, it amounts to a 
choice of the lapse function $\alpha$ and shift vector $\beta^i$, defined via
\begin{equation}
g_{\mu \nu} dx^\mu dx^\nu = -\alpha^2 dt^2 +
\gamma_{ij}(dx^i + \beta^i dt)(dx^j +\beta^j dt).
\end{equation}
Here, the latin indices $i,j$ only sum over the spatial coordinates.

For spatial coordinates, we choose the compactified $x^i=(\rho,\chi,\theta,\phi)$ 
as defined in the discussion preceding~(\ref{eqn:compactifiedmetric}). To understand 
our choice of initial shift vector $\beta^i$ and an initial lapse function $\alpha$, 
we must first understand a few subtleties concerning gauge choices in asymptotically 
AdS spacetimes, and so we defer this discussion until Sec.~\ref{sec:gauge_t0}. In this 
later section, we will recast our choice of $\beta^i$,$\alpha$ in terms of an 
equivalent choice of $\bar{g}_{t\nu}$, $\partial_t \bar{g}_{t\nu}$ at $t=0$.

\section{Evolution}\label{section:evolution}
In this section we describe the ingredients we have found necessary to achieve
stable evolution of AAdS$_5$ spacetimes within the GH formalism. The solutions 
in this initial study preserve an $SO(3)$ symmetry, which is sufficiently general 
to be physically relevant, as well as capture many of the problems and issues that 
need to be resolved for stable evolution. Chief among these are 
(a) decomposing the metric into a form that analytically factors out the AdS 
divergences, and dividing out sufficient powers of $1-\rho$ from what remains, 
allowing us to set boundary conditions that impose the desired leading-order 
deviation from AdS, and 
(b) imposing an asymptotic gauge in terms of the source functions $H^\mu$ of 
the GH formalism that is consistent with the desired fall-off. 
These two issues are in fact intimately related in AAdS spacetimes---asking for 
coordinates where the leading-order metric deviations are non-singular in the 
approach to the boundary, together with a choice of the form of the background 
(singular) AdS metric, completely fixes any residual gauge freedom.

The next few sections are structured as follows: in Sec.~\ref{sec:harmonic} 
we review the GH formalism, in Sec.~\ref{sec:evo_vars} we describe the 
metric decomposition that we use, and in Sec.~\ref{sec:gauge} we look at the 
asymptotic form of the field equations and derive the relevant gauge conditions 
for GH evolution. In Sec.~\ref{sec:gauge_t0}, we show how to choose 
$\left. \bar{g}_{t\nu} \right|_{t=0}$, $\left. \partial_t \bar{g}_{t\nu} \right|_{t=0}$ 
on the initial time slice, such that they are compatible with these gauge conditions.

\subsection{The Generalized Harmonic Formalism}\label{sec:harmonic}

Here we give an overview of the generalized harmonic (GH) formalism with 
constraint damping; for more details see~\cite{2002APS..APRC12004G,
Pretorius:2004jg,Lindblom:2005qh,Pretorius:2007nq}.
The GH formalism is based on coordinates $x^\mu$ that are chosen so that each 
coordinate satisfies a scalar wave equation with source function 
$H^\mu$:~\footnote{As can be seen from (\ref{eqn:ghcondition}) $H^\mu$ is not 
a vector in the sense of its properties under a coordinate transformation, 
rather it transforms as the trace of the metric connection. One can introduce
additional geometric structure in the form of a background metric and
connection to write the GH formalism in terms of ``standard'' tensorial
objects. However, in a numerical evolution one must always choose a 
concrete coordinate system, and hence the resulting equations that are 
eventually discretized are the same regardless of the extra mathematical
structure introduced at the formal level.}
\begin{equation}\label{eqn:ghcondition}
\square x^{\mu} = \frac{1}{\sqrt{-g}}\frac{\partial}{\partial x^\alpha} 
\left( \sqrt{-g} g^{\alpha \mu} \right) = -g^{\alpha \beta} \Gamma^\mu_{\alpha \beta} \equiv  H^{\mu},
\end{equation}
where $g$ is the determinant of the metric, and $\Gamma^\mu_{\alpha \beta}$ are 
the Christoffel symbols. To see why this has proven to be so useful for Cauchy 
evolution, let us begin by rewriting the field equations (\ref{eqn:efe}) in 
trace-reversed form 
\begin{equation}\label{eqn:efe_tr}
R_{\mu \nu} = \bar{T}_{\mu \nu},
\end{equation}
where 
\begin{equation}
\bar{T}_{\mu \nu} = \frac{2}{3} \Lambda_5 g_{\mu \nu} + 8\pi \left( T_{\mu \nu} 
- \frac{1}{3} {T^\alpha}_\alpha g_{\mu \nu} \right)
\end{equation}
When viewed as a set of second-order differential equations for the metric 
$g_{\mu \nu}$, the field equations in the form (\ref{eqn:efe_tr}) do not have 
any well-defined mathematical character (namely hyperbolic, elliptic or 
parabolic), and in fact are ill-posed. Fixing this character requires choosing 
a coordinate system. A well-known way to arrive at a set of strongly hyperbolic 
equations is to impose {\em harmonic coordinates}, namely 
(\ref{eqn:ghcondition}) with $H^\mu=0$. Specifically, this condition 
(and its gradient) can be substituted into the field equations to yield a wave 
equation for the principal part of each metric element, 
$g^{\alpha\beta} \partial_\alpha \partial_\beta g_{\mu \nu} + ... = 0$,
where the ellipses denote lower order terms.

One potential problem with harmonic coordinates, in particular in a highly
dynamical, strong-field spacetime evolved via a Cauchy scheme, is that beginning
from a well-defined initial data surface $t={\rm const.}$ which is everywhere 
space-like, there is no guarantee that $t$, subject to the harmonic condition, 
will remain time-like throughout the spacetime as evolution proceeds. If $t$ 
becomes null or space-like at a point, standard numerical techniques will break 
down. A solution to this, first suggested in~\cite{2002APS..APRC12004G}, is to 
make use of source functions (as originally introduced 
in~\cite{1985CMaPh.100..525F}). Note that {\em any} spacetime in {\em any}
coordinate system can be written in GH form; the corresponding source functions 
are simply obtained by evaluating the definition (\ref{eqn:ghcondition}). Thus, 
trivially, if there is a well-behaved, non-singular coordinate chart that covers 
a given spacetime, then there is a GH description of it. The difficulty in a 
Cauchy evolution is that this chart is not known {\em a-priori}, and the source 
functions $H^\mu$ must be treated as independent dynamical fields. Finding a 
well-behaved coordinate chart then amounts to supplementing the Einstein field
equations with a set of evolution equations for $H^\mu$, which can now be 
considered to encode the coordinate degrees of freedom in our description of the 
spacetime. 

The field equations in GH form thus consist of the Einstein equations 
(\ref{eqn:efe_tr}), brought into hyperbolic form via the imposition of 
(\ref{eqn:ghcondition})
\begin{eqnarray}\label{eqn:efe_gh}
& & - \frac{1}{2} g^{\alpha \beta} g_{\mu \nu, \alpha \beta} - 
g^{\alpha \beta}{}_{,(\mu} g_{\nu) \alpha, \beta} \nonumber \\
& & - H_{(\mu, \nu)} + H_\alpha \Gamma^\alpha_{\mu \nu} - \Gamma^\alpha_{\beta 
\mu} \Gamma^\beta_{\alpha \nu} \nonumber \\
&= & \frac{2}{3} \Lambda_5 g_{\mu \nu} + 8\pi \left( T_{\mu \nu} - 
\frac{1}{3} {T^\alpha}_\alpha g_{\mu \nu} \right),
\end{eqnarray}
together with the relevant evolution equations for the matter (here the 
Klein-Gordon equation (\ref{eqn:kgeqn})) and a set of equations for the source 
functions, which we write symbolically as
\begin{equation}\label{eqn:guage}
\mathcal L^\mu [{H^\mu}] = 0.
\end{equation}

Even though $H^{\mu}$ are now treated as independent functions, we are only 
interested in the {\em subset} of solutions to the expanded 
system~(\ref{eqn:efe_gh}),(\ref{eqn:guage}) that satisify the GH 
constraints~(\ref{eqn:ghcondition}). Introducing
\begin{equation}\label{eqn:ghconstraint}
C^{\mu} \equiv H^{\mu} - \square x^{\mu},
\end{equation}
we thus seek solutions to~(\ref{eqn:efe_gh}),(\ref{eqn:guage}) for which 
$C^{\mu}=0$. An equivalent way of obtaining (\ref{eqn:efe_gh}) 
from~(\ref{eqn:efe_tr}) is to subtract $\nabla_{(\mu} C_{\nu)}$ from 
$R_{\mu \nu}$,
so that
\begin{equation}\label{eqn:efe_tr_gh}
R_{\mu \nu} - \nabla_{(\mu} C_{\nu)} - \bar{T}_{\mu \nu} = 0.
\end{equation}
The effect of this subtraction becomes obvious when we rewrite the Ricci tensor 
explicitly in terms of $\square x^{\mu}$, so    
$R_{\mu \nu} = -\frac{1}{2} g^{\alpha \beta} g_{\mu \nu, \alpha \beta} - 
{g^{\alpha \beta}}_{,(\mu} g_{\nu) \alpha, \beta} - \nabla_{(\mu} \square x_{\nu)} 
- {\Gamma^\alpha}_{\beta \mu} {\Gamma^\beta}_{\alpha \nu}$.
We see that the subtraction of $\nabla_{(\mu} C_{\nu)}$ is simply designed to 
replace the $\nabla_{(\mu} \square x_{\nu)}$ term in $R_{\mu \nu}$ by an equivalent 
$\nabla_{(\mu} H_{\nu)}$ term. We also see that a solution of the Einstein field 
equations (\ref{eqn:efe_tr}) is also a solution of (\ref{eqn:efe_tr_gh}), as long 
as the constraints $C^\mu=0$ are satisfied. 

For a Cauchy evolution of the system~(\ref{eqn:efe_gh}),(\ref{eqn:guage}), we 
need to specify initial data in the form
\begin{equation}\label{eqn:g_id}
g_{\mu\nu}|_{t=0}, \ \ \ \partial_t g_{\mu\nu}|_{t=0},
\end{equation}
subject to the constraints
\begin{equation}\label{eqn:ghconstraintid}
C^\mu|_{t=0} = 0, \ \ \ \partial_t C^\mu|_{t=0} = 0.
\end{equation}
One can show (see for e.g.~\cite{Lindblom:2005qh}) that if 
(\ref{eqn:ghconstraintid}) is satisfied, then the ADM Hamiltonian and momentum 
constraints will be satisfied at $t=0$. Conversely, if the ADM constraints are 
satisfied at $t=0$ together with $C^\mu|_{t=0}=0$ (this latter condition is 
satisfied trivially computing $H^\mu|_{t=0}$ by substituting (\ref{eqn:g_id})
into~(\ref{eqn:ghcondition})), then $\partial_t C^\mu|_{t=0}=0$. Thus, our 
initial data method described in the previous section will produce data 
consistent with~(\ref{eqn:ghconstraintid}). Furthermore, using a contraction 
of the second Bianchi identity $\nabla^\mu R_{\mu \nu} = \nabla_\nu R/2$, one 
can show that $C^\mu$ satisfies the following hyperbolic equation:
\begin{equation}\label{eqn:ghconstrainteqn}
\square C_\nu = - C^\mu \nabla_{(\mu} C_{\nu)} - C^\mu \bar{T}_{\mu \nu}.
\end{equation} 
Thus, if we imagine (analytically) solving~(\ref{eqn:efe_gh}),(\ref{eqn:guage}) 
using initial data satisfying (\ref{eqn:ghconstraintid})  supplemented with 
boundary conditions consistent with $C^\mu=0$ on the boundary for all time, then 
(\ref{eqn:ghconstrainteqn}) implies that $C^\mu$ will remain zero in the 
interior for all time.

At the level of the discretized equations, however, $C^\mu$ is only zero up to 
truncation error. This is not {\em a priori} problematic: numerically one 
only ever gets a solution approximating the continuum solution to within 
truncation error. However, experience with asymptotically-flat simulations 
suggest that in some strong-field spacetimes, 
equation~(\ref{eqn:ghconstrainteqn}) for $C^\mu$ admits exponentially growing 
solutions (the so-called ``constraint-violating modes''). At any practical 
resolution, this severely limits the amount of physical time for which an 
accurate solution to the desired $C^\mu=0$ Einstein equations can be obtained. 
In asymptotically flat spacetimes, supplementing the GH harmonic 
equations with {\em constraint-damping} terms as introduced 
in~\cite{Gundlach:2005eh} suppresses these unwanted solutions. Anticipating 
similar problems in AAdS spacetimes, and that constraint damping will similarly 
help, we add the same terms to (\ref{eqn:efe_gh}), and arrive at the final 
form of our evolution equations
\begin{eqnarray}\label{eqn:efe_gh_modified}
&-& \frac{1}{2} g^{\alpha \beta} g_{\mu \nu, \alpha \beta} - 
{g^{\alpha \beta}}_{,(\mu} g_{\nu) \alpha, \beta} \nonumber \\
&-& H_{(\mu, \nu)} + H_\alpha {\Gamma^\alpha}_{\mu \nu} - {\Gamma^\alpha}_{\beta 
\mu} {\Gamma^\beta}_{\alpha \nu} \nonumber \\
&-& \kappa \left( 2 n_{(\mu} C_{\nu)} - (1+P) g_{\mu \nu} n^\alpha 
C_\alpha \right) \nonumber \\
&=&   \frac{2}{3} \Lambda_5 g_{\mu \nu} + 8\pi \left( T_{\mu \nu} - 
\frac{1}{3} {T^\alpha}_\alpha g_{\mu \nu} \right).
\end{eqnarray}
Here, the unit time-like one-form $n_\mu$ is defined as in (\ref{eqn:n}), and
the constraint damping parameters $\kappa \in \left( -\infty,0 \right]$ and
$P \in \left[ -1,0 \right]$ are arbitrary constants. In all simulations 
described here, we use $\kappa=-10$ and $P=-1$. \footnote{We did not perform 
any systematic survey by varying $\kappa$ or $P$, though with a little 
experimentation found the exact value of $\kappa$ was not too important to 
achieve effective constraint damping, though we found that it was important 
to keep $P$ close to $-1$.}

Note that the new terms are homogeneous in $C_\mu$, and hence do not alter any 
of the properties discussed above that relate solutions of the Einstein 
evolution equations and ADM constraints with those of the corresponding GH 
equations, with the exception that the constraint propagation 
equation~(\ref{eqn:ghconstrainteqn}) picks up additional terms, again 
homogeneous in $C_\mu$ (see for example \cite{Gundlach:2005eh}).

\subsection{Evolution Variables and Boundary Conditions}\label{sec:evo_vars}

The boundary is crucial for evolution in asymptotically AdS$_5$ spacetimes. To 
find the most natural variables to evolve in this setting, we first need to gain 
some intuition on how the fields behave near the boundary at $\rho=1$. To begin, 
let us again use (\ref{eqn:generalmetricdecomposition}) to decompose the metric 
$g_{\mu \nu}$ into a pure AdS$_5$ piece $\hat{g}_{\mu \nu}$ and a deviation 
$h_{\mu \nu}$. From the point of view of the evolution 
equations~(\ref{eqn:efe_gh_modified}), this decomposition allows us to 
analytically eliminate a subset of asymptotically singular terms representing 
the pure AdS$_5$ solution. From the point of view of the boundary 
conditions~(\ref{eqn:metricbcs}), this decomposition guides our choice of 
variables that are most suitable for Cauchy evolution.

Our choice of evolution variables is largely motivated by considerations 
similar to those that arise when numerically solving hyperbolic partial 
differential equations (PDEs) whose domain includes a formally singular 
boundary, where one seeks solutions that remain regular at the boundary 
(see for example~\cite{Garfinkle:2000hd}, where this method was introduced to 
study the evolution of gravitational waves in axisymmetry in a domain that includes 
the axis of symmetry). Before listing the variables that we use to represent the 
metric deviation in AAdS spacetime, let us first colloquially describe the 
reasoning behind their definitions. We will not {\em prove} that the following is a 
correct (or complete) characterization of the AAdS boundary behavior of our 
coupled system of 
PDEs~(\ref{eqn:kgeqn}),(\ref{eqn:guage}),(\ref{eqn:efe_gh_modified});
rather we will take the empirical approach that if by using this regularization 
scheme we are able to obtain stable, convergent numerical solutions, then this 
strongly suggests that the regularization is consistent, at least for the set of 
initial data considered. Though note that in~\cite{Ishibashi:2004wx} a complete
characterization of boundary conditions consistent with stable evolution for 
{\em linearized} gravitational perturbations on an AdS {\em background} was 
given. As discussed in Appendix \ref{app:boundary_operator}, the boundary 
conditions we describe below are consistent with the Friedrich self-adjoint 
extension of the operator describing the scalar sector of gravitational 
perturbations. Thus, insofar as the linear problem guides the full non-linear 
problem, we can have some confidence that the following prescription is 
well-posed.

To illustrate, consider a function $f(t,q,\chi,\theta,\phi)$ that can be 
expanded in a power series in $q$, where the boundary is located at $q=0$:
\begin{equation}\label{eqn:f_example}
f=f_0 + f_1 q + f_2 q^2 + f_3 q^3 + ...
\end{equation}
Here, $f_0,f_1,f_2,f_3,...$ are functions of $t,\chi,\theta,\phi$. Now suppose 
that for a regular solution (which in our case means a solution consistent with 
the desired AAdS fall-off) the first $n$ terms of the RHS are required to be 
zero, and that the $n$-plus-first term describes the leading-order behavior for 
the particular physical solution of interest. At first glance, this would 
suggest that we need to supply $n+1$ boundary conditions. However, if $f$ 
satisfies a hyperbolic PDE with ``standard'' characteristic structure i.e. 
with an inward and outward propagating mode in each spatial direction, then we 
are usually only free to choose $1$ boundary condition, effectively fixing the mode 
propagating into the domain. Furthermore, at a singular boundary where we demand 
regularity, one is often not free to choose even this ingoing mode---the outgoing 
mode together with regularity completely fixes it. As proposed in~\cite{Garfinkle:2000hd}, a 
solution is to {\em define} a new evolution variable $\bar{f}$ via 
\begin{equation}
f(t,q,\chi,\theta,\phi) \equiv \bar{f} (t,q,\chi,\theta,\phi) q^{n-1},
\end{equation}
and demand that $\bar{f}$ satisfy a Dirichlet condition 
$\bar{f}(t,q=0,\chi,\theta,\phi)=0$ at the boundary. Plugging this into 
(\ref{eqn:f_example}) gives 
\begin{equation}\label{eqn:f_example_b}
\bar{f}= ...  + f_{n-2} q^{-1} + f_{n-1} + f_n q + f_{n+1} q^2 + ...
\end{equation}
One can see that if we choose regular initial data for $\bar{f}$ that has 
$\bar{f}(t=0,q=0)=0$, this eliminates all the components of $f$ with undesired 
fall-off at $t=0$. This will give a regular solution (analytically) for all time 
if the differential equation for $f$ admits a unique solution consistent with 
the desired fall-off. Note that by factoring out $n-1$ powers of $q$, we have 
assumed that the nature of the boundary is such that we are {\em not} free to 
specify the leading-order behavior encoded in $f_n$ as a formal boundary 
condition, i.e. initial data together with evolution should uniquely determine it 
(if this were not the case, one could factor out an additional power of $q$, and 
explicity set $f_n$). Again, as described in more detail in 
Appendix~\ref{app:boundary_operator}, this is consistent with the analysis 
of~\cite{Ishibashi:2004wx} on admissible boundary conditions for linearized 
gravitational perturbations of AdS. This is also the picture that arises in more 
formal derivations of the fluid/gravity correspondence in terms of derivative 
expansions of the field equations, where demanding normalizability at the 
boundary and regularity in the bulk (outside of black hole singularities) 
effectively constrains the gravitational dynamics of the bulk to have as many 
``degrees of freedom'' as the dual boundary fluid dynamics. For a review, see 
for example~\cite{Hubeny:2011hd}.

Applying the above reasoning to our metric fields $g_{\mu \nu}$, we construct 
regularized metric variables $\bar{g}_{\mu \nu}$ that asymptotically fall-off 
as $\bar{g}_{\mu \nu} \sim q$:
\begin{eqnarray}\label{eqn:metric_asymptotics}
g_{t t} &=& \hat{g}_{t t} + q (1+\rho) \bar{g}_{t t} \nonumber \\
g_{t \rho} &=& \hat{g}_{t \rho} + q^2(1+\rho)^2 \bar{g}_{t \rho} \nonumber \\
g_{t \chi} &=& \hat{g}_{t \chi} + q (1+\rho) \bar{g}_{t \chi} \nonumber \\
g_{\rho \rho} &=& \hat{g}_{\rho \rho} +  q (1+\rho) \bar{g}_{\rho \rho} \nonumber \\
g_{\rho \chi} &=& \hat{g}_{\rho \chi} + q^2(1+\rho)^2 \bar{g}_{\rho \chi} \nonumber \\
g_{\chi \chi} &=& \hat{g}_{\chi \chi} + q (1+\rho) \bar{g}_{\chi \chi} \nonumber \\
g_{\theta \theta} &=& \hat{g}_{\theta \theta} + q (1+\rho) (\rho^2 \sin^2 \chi) \bar{g}_{\psi} \nonumber \\
g_{\phi \phi} &=& \hat{g}_{\phi \phi} + q (1+\rho) (\rho^2 \sin^2 \chi \sin^2 \theta) \bar{g}_{\psi}
\end{eqnarray}

The term in the metric $g_{\mu \nu}$ that is conformal to $S^2$ can be kept 
track of by a single variable $\bar{g}_{\psi}$, since we are considering 
solutions that preserve an $SO(3)$ symmetry that acts to rotate this $S^2$.
The extra factors $\rho^2\sin^2\chi$ and $\rho^2\sin^2\chi\sin^2\theta$ that appear 
in (\ref{eqn:metric_asymptotics}) are included to ensure regularity at the 
origin $\rho=0$, for reasons similar to those discussed above. Using the 
regularized variables defined in (\ref{eqn:metric_asymptotics}), the boundary 
conditions (\ref{eqn:metricbcs}) can be fully captured by a simple set of 
Dirichlet boundary conditions at spatial infinity: 
\begin{eqnarray}\label{eqn:metric_bcs}
\left. \bar{g}_{t t} \right|_{\rho=1} &=& 0 \nonumber \\
\left. \bar{g}_{t \rho} \right|_{\rho=1} &=& 0 \nonumber \\
\left. \bar{g}_{t \chi} \right|_{\rho=1} &=& 0 \nonumber \\
\left. \bar{g}_{\rho \rho} \right|_{\rho=1} &=& 0 \nonumber \\
\left. \bar{g}_{\rho \chi} \right|_{\rho=1} &=& 0 \nonumber \\
\left. \bar{g}_{\chi \chi} \right|_{\rho=1} &=& 0 \nonumber \\
\left. \bar{g}_{\psi} \right|_{\rho=1}    &=& 0. 
\end{eqnarray}

In polar-like coordinates, the Taylor expansion of tensorial quantities are 
typically either even or odd in $\rho$ about $\rho=0$; the extra factors of 
$(1+\rho)$ in (\ref{eqn:metric_asymptotics}) are to ensure that $g_{\mu\nu}$ and
$\bar{g}_{\mu \nu}$ have the same even/odd character in the limit 
$\rho\rightarrow 0$. We use standard results for the origin 
regularity conditions, which in our context read:
\begin{eqnarray}\label{eqn:metric_orireg}
\left. \partial_\rho \bar{g}_{t t} \right|_{\rho=0}       &=& 0 \nonumber \\
\left. \partial_\rho \bar{g}_{t \chi} \right|_{\rho=0}    &=& 0 \nonumber \\
\left. \partial_\rho \bar{g}_{\rho \rho} \right|_{\rho=0} &=& 0 \nonumber \\
\left. \partial_\rho \bar{g}_{\chi \chi} \right|_{\rho=0} &=& 0 \nonumber \\
\left. \partial_\rho \bar{g}_{\psi} \right|_{\rho=0}      &=& 0 \nonumber \\
\left. \bar{g}_{t \rho} \right|_{\rho=0}                  &=& 0 \nonumber \\
\left. \bar{g}_{\rho \chi} \right|_{\rho=0}               &=& 0.
\end{eqnarray}
Similar regularity conditions apply on axis:
\begin{eqnarray}\label{eqn:metric_axireg}
\left. \partial_\chi \bar{g}_{t t} \right|_{\chi=0,\pi}       &=& 0 \nonumber \\
\left. \partial_\chi \bar{g}_{t \rho} \right|_{\chi=0,\pi}    &=& 0 \nonumber \\
\left. \partial_\chi \bar{g}_{\rho \rho} \right|_{\chi=0,\pi} &=& 0 \nonumber \\
\left. \partial_\chi \bar{g}_{\chi \chi} \right|_{\chi=0,\pi} &=& 0 \nonumber \\
\left. \partial_\chi \bar{g}_{\psi} \right|_{\chi=0,\pi}      &=& 0 \nonumber \\
\left. \bar{g}_{t \chi} \right|_{\chi=0,\pi}                  &=& 0 \nonumber \\
\left. \bar{g}_{\rho \chi} \right|_{\chi=0,\pi}               &=& 0.
\end{eqnarray}

Elementary flatness imposes an additional relation among the metric variables 
on the axis, ensuring that no conical singularities arise there. Given our axial 
Killing vector $\partial_\phi$, with norm-squared $\xi = g_{\phi \phi}$, 
this statement is made precise by the condition~\cite{Stephani:2003tx}
\begin{equation}\label{eqn:elementaryflatness}
\left. \frac{g^{\mu \nu} \partial_\mu \xi \partial_\nu \xi}{4 \xi} \right|_{\chi=0,\pi} = 1,
\end{equation}
which evaluates to the following in terms of our regularized metric variables:
\begin{equation}\label{eqn:metric_ef}
\left. \bar{g}_{\chi \chi} \right|_{\chi=0,\pi} = \rho^2 \left. \bar{g}_{\psi} \right|_{\chi=0,\pi}.
\end{equation}
We ensure that this relationship is satisfied by explicitly 
setting $\bar{g}_{\chi \chi}$ in terms of $\bar{g}_{\psi}$ on the axis as 
dictated by~(\ref{eqn:metric_ef}), instead of applying the regularity condition 
for $\bar{g}_{\chi \chi}$ as displayed in~(\ref{eqn:metric_axireg}).

To find the appropriate regularized source function variables $\bar{H}_{\mu}$, 
we insert the above expressions~(\ref{eqn:metric_asymptotics}) for the metric 
components into the definition~(\ref{eqn:ghcondition}) to find how the source
functions deviate from their AdS$_5$ values $\hat{H}_\mu$ near the boundary.
Factoring out the appropriate powers of $q$ so that $\hat{H}_\mu(q=0)=0$, and
adding powers of $(1+\rho)$ to maintain the same expansions near the origin, we 
obtain 
\begin{eqnarray}\label{eqn:source_asymptotics}
H_{t} &=& \hat{H}_{t} + q^3 (1+\rho)^3 \bar{H}_{t} \nonumber \\
H_{\rho} &=& \hat{H}_{\rho} + q^2 (1+\rho)^2 \bar{H}_{\rho} \nonumber \\
H_{\chi} &=& \hat{H}_{\chi} + q^3 (1+\rho)^3 \bar{H}_{\chi} 
\end{eqnarray}
with boundary conditions:
\begin{eqnarray}\label{eqn:source_bcs}
\left. \bar{H}_t \right|_{\rho=1}       &=& 0 \nonumber \\
\left. \bar{H}_\rho \right|_{\rho=1}       &=& 0 \nonumber \\
\left. \bar{H}_\chi \right|_{\rho=1}       &=& 0 \nonumber \\
\end{eqnarray}
origin regularity conditions:
\begin{eqnarray}\label{eqn:source_orireg}
\left. \partial_\rho \bar{H}_{t} \right|_{\rho=0}       &=& 0 \nonumber \\
\left. \partial_\rho \bar{H}_{\chi} \right|_{\rho=0}    &=& 0 \nonumber \\
\left. \bar{H}_{\rho} \right|_{\rho=0}                  &=& 0.
\end{eqnarray}
and axis regularity conditions:
\begin{eqnarray}\label{eqn:source_axireg}
\left. \partial_\chi \bar{H}_{t} \right|_{\chi=0,\pi}       &=& 0 \nonumber \\
\left. \partial_\chi \bar{H}_{\rho} \right|_{\chi=0,\pi}    &=& 0 \nonumber \\
\left. \bar{H}_{\chi} \right|_{\chi=0,\pi}                  &=& 0.
\end{eqnarray}
We also need a regularized massless scalar field variable $\bar{\phi}$ that
asymptotically falls off as $\bar{\phi} \sim q$, so we let
\begin{equation}\label{eqn:scalar_asymptotics}
\phi = q^3 (1+\rho)^3 \bar{\phi}.
\end{equation}
with boundary condition:
\begin{eqnarray}\label{eqn:scalar_bcs}
\left. \bar{\phi} \right|_{\rho=1}        &=& 0
\end{eqnarray}
origin regularity condition:
\begin{eqnarray}\label{eqn:scalar_orireg}
\left. \partial_\rho \bar{\phi} \right|_{\rho=0}       &=& 0. \nonumber \\
\end{eqnarray}
and axis regularity condition:
\begin{eqnarray}\label{eqn:scalar_axireg}
\left. \partial_\chi \bar{\phi} \right|_{\chi=0,\pi}       &=& 0. \nonumber \\
\end{eqnarray}

\subsection{Gauge Choice}\label{sec:gauge}

Choosing generalized harmonic gauge conditions in AAdS$_5$ spacetimes is 
somewhat more subtle than in the usual asymptotically flat case. Roughly 
speaking, it turns out that it is not enough to simply demand that the metric 
and source functions satisfy the requisite rates of fall-off approaching the 
boundary as indicated in~(\ref{eqn:metric_asymptotics}) and 
(\ref{eqn:source_asymptotics}); rather, there are further restrictions amongst 
the leading-order behavior of certain fields that need to be explicitly enforced, 
so that the requisite fall-off is preserved during evolution.

To show this more clearly, we expand the regularized metric variables 
$\bar{g}_{\mu \nu}$, source functions $\bar{H}_{\mu}$, and scalar field 
$\bar{\phi}$ in power series about $q=0$
\begin{equation}\label{eqn:qexpmetric}
\bar{g}_{\mu \nu} = \bar{g}_{(1) \mu \nu}( t,\chi,\theta,\phi ) q + 
\bar{g}_{(2) \mu \nu}( t,\chi,\theta,\phi ) q^2 + \mathcal{O}(q^3) 
\end{equation}
\begin{equation}\label{eqn:qexpsourcefunctions}
\bar{H}_{\mu} = \bar{H}_{(1) \mu}( t,\chi,\theta,\phi ) q + 
\bar{H}_{(2) \mu}( t,\chi,\theta,\phi ) q^2 + \mathcal{O}(q^3) 
\end{equation}
\begin{equation}\label{eqn:qexpscalar}
\bar{\phi} = \bar{\phi}_{(1)}( t,\chi,\theta,\phi ) q + 
\bar{\phi}_{(2)}( t,\chi,\theta,\phi ) q^2 + \mathcal{O}(q^3) 
\end{equation}
and substitute these expressions into the field 
equations~(\ref{eqn:efe_gh_modified}). Since the GH form of the field equations 
results in PDEs where the principle part of each equation is a wave operator 
acting on the metric (this fact guides our numerical solution method), we will 
schematically write this perturbative expansion of the field equations for the 
leading component $\bar{g}_{(1) \mu \nu}$ of $\bar{g}_{\mu \nu}$ as follows
\begin{equation}\label{eqn:schem_wave}
\tilde{\square} \bar{g}_{(1) \mu \nu} = ...
\end{equation}
where we use the symbol $\tilde{\square}$ to denote a wave-like operator active 
within the $(t,\chi,\theta,\phi$) subspace, and containing terms of the form 
$c_0 \cdot \partial^2 \bar{g}_{(1) \mu \nu}/\partial t^2 - 
c_1 \cdot \partial^2 \bar{g}_{(1) \mu \nu}/\partial \chi^2 + ...$. Here, 
$c_0,c_1,...$ are coefficient functions that are in general different for each 
component of the field equations, but are regular and finite on the boundary. 
Their particular form is unimportant here, as we are interested in highlighting 
the leading-order terms sourcing the wave-like equation on the RHS of 
(\ref{eqn:schem_wave}). 

In this notation, the field equations near $q=0$ read
\begin{widetext}
\begin{eqnarray}\label{eqn:nearbdyefes}
\tilde{\square} \bar{g}_{(1) tt} &=& 
   ( -8\bar{g}_{(1) \rho \rho} + 4 \bar{H}_{(1) \rho} )q^{-2} + 
   \mathcal{O}(q^{-1}) \nonumber \\
\tilde{\square} \bar{g}_{(1) t\rho} &=& 
   ( -60\bar{g}_{(1) t \rho} - 8\cot\chi\bar{g}_{(1) t \chi} + 
     24\bar{H}_{(1) t} - \bar{g}_{(1) tt,t} \nonumber \\
     &&
     + 2\bar{g}_{(1) t \chi,\chi} + 2\bar{g}_{(1) \rho\rho,t} - \bar{g}_{(1) \chi\chi,t} - 
     2\bar{g}_{(1) \psi,t} - 2\bar{H}_{(1) \rho,t} )q^{-2} 
   + \mathcal{O}(q^{-1}) \nonumber \\
\tilde{\square} \bar{g}_{(1) t\chi} &=& 
   \mathcal{O}(q^{-1}) \nonumber \\
\tilde{\square} \bar{g}_{(1) \rho\rho} &=& 
   ( -8\bar{g}_{(1) t t} -24\bar{g}_{(1) \rho \rho} + 
     8\bar{g}_{(1) \chi \chi}  +16\bar{g}_{(1) \psi} + 16\bar{H}_{(1) \rho} )q^{-2} 
   + \mathcal{O}(q^{-1}) \nonumber \\
\tilde{\square} \bar{g}_{(1) \rho\chi} &=& 
   ( -60\bar{g}_{(1) \rho \chi} - 8\cot\chi\bar{g}_{(1) \chi \chi} 
     + 8\cot\chi\bar{g}_{(1) \psi} + 24\bar{H}_{(1) \chi} \nonumber \\
     &&
     + \bar{g}_{(1) tt,\chi} - 2\bar{g}_{(1) t\chi,t} + 
     2\bar{g}_{(1) \rho\rho,\chi} + \bar{g}_{(1) \chi\chi,\chi} - 
     2 \bar{g}_{(1) \psi,\chi} - 2\bar{H}_{(1) \rho,\chi} )q^{-2} 
   + \mathcal{O}(q^{-1}) \nonumber \\
\tilde{\square} \bar{g}_{(1) \chi\chi} &=& 
   ( 8\bar{g}_{(1) \rho \rho} - 4\bar{H}_{(1) \rho} )q^{-2} 
   + \mathcal{O}(q^{-1}) \nonumber \\
\tilde{\square} \bar{g}_{(1) \psi} &=& 
   \sin^2\chi( 8\bar{g}_{(1) \rho \rho} -4 \bar{H}_{(1) \rho} )q^{-2} 
   + \mathcal{O}(q^{-1}),
\end{eqnarray}
\end{widetext}
and again we emphasize that $\tilde{\square}$ is used to schematically denote 
a regular wave operator, though its specific form is in general different for 
each equation. For reference, we also list the leading-order behavior of the GH 
constraints (\ref{eqn:ghconstraint}): 
\begin{widetext}
\begin{eqnarray}\label{eqn:nearbdyC}
C_t    &=& ( 20 \bar{g}_{(1) t \rho} - 8 \bar{H}_{(1) t} + \bar{g}_{(1) tt,t} 
             - 2 \bar{g}_{(1) t \chi,\chi} + \bar{g}_{(1) \rho\rho,t} 
             + \bar{g}_{(1) \chi\chi,t} + 2 \bar{g}_{(1) \psi,t} )q^4
           + \mathcal{O}(q^5) \nonumber \\
C_\rho &=& ( 8 \bar{g}_{(1) t t} + 8 \bar{g}_{(1) \rho\rho} 
             - 8 \bar{g}_{(1) \chi\chi} - 16 \bar{g}_{(1) \psi} 
             - 8 \bar{H}_{(1) \rho} )q^3
           + \mathcal{O}(q^4) \nonumber \\
C_\chi &=& ( 20 \bar{g}_{(1) \rho\chi} - 8 \bar{H}_{(1) \chi} - \bar{g}_{(1) tt,\chi} 
             + 2 \bar{g}_{(1) t \chi,t} + \bar{g}_{(1) \rho\rho,\chi} 
             - \bar{g}_{(1) \chi\chi,\chi} + 2 \bar{g}_{(1) \psi,\chi} )q^4
           + \mathcal{O}(q^5). \nonumber \\
\end{eqnarray}
\end{widetext}

In order for the evolution to be consistent with each metric component's desired 
fall-off indicated in (\ref{eqn:metric_asymptotics}), the leading-order 
``source'' on the right-hand side of each equation in (\ref{eqn:nearbdyefes}) 
must scale as $q^0$ due to the Dirichlet boundary 
conditions~(\ref{eqn:metric_bcs}). This implies that all terms of order $q^{-2}$ 
and $q^{-1}$ must vanish, and it would thus appear as though there were an 
additional hierarchy of constraints that need to be satisfied before the 
leading-order dynamics in the $q^0$ term becomes manifest (note that these 
constraints are {\em not} simply the GH constraints (\ref{eqn:nearbdyC})). 
This is, in part, an artifact of having decomposed the field equations in a 
near-boundary expansion such as (\ref{eqn:qexpmetric}): when solving the full
equations consistently (i.e., with a Cauchy evolution scheme and initial data 
satisfying the constraints, along with stable, consistent boundary conditions 
as discussed in the previous sections), one usually expects that the evolution 
will ``conspire'' to preserve what appears as constraints in the perturbative 
expansion. However, there are two potential complications in the AAdS case, 
highlighted in the above by the fact that the leading-order power in the 
expansions {\em diverge} as $q^{-2}$. 

First, in the GH method, one is free to choose $\bar{H}_\mu$ as the gauge, and 
though the structure of the field equations guarantee that the resultant 
solution will be consistent, there is {\em no} guarantee that a given choice of 
$\bar{H}_\mu$ will preserve the desired asympotic fall-off of the 
metric~(\ref{eqn:metric_asymptotics}). Case in point, suppose that we started 
with some initial value $\bar{H}_\mu(t=0)$, and wanted to evolve to a gauge that 
in time becomes harmonic with respect to pure AdS$_5$, namely 
$\bar{H}_\mu(t\gg 0)=0$ so that $H_\mu(t\gg 0)=\hat{H}_\mu$ 
in~(\ref{eqn:source_asymptotics}). Then~(\ref{eqn:nearbdyefes}) for 
$\bar{g}_{(1) tt}$ immediately tells us that either $\bar{g}_{(1) \rho \rho}$ 
must evolve to $0$, or  $\bar{g}_{(1) tt}$ ceases to remain regular. Whichever 
scenario unfolds, this choice of gauge leads to a representation of the metric 
that is not consistent with the fall-off assumed 
in~(\ref{eqn:metric_asymptotics}), and as discussed in 
Sec.~\ref{sec:evo_vars}, it would generically be difficult (or even impossible) to 
come up with a numerical scheme to stably evolve such a situation.

Second, the form of the equations in (\ref{eqn:nearbdyefes}) {\em imply} that 
regularity requires a delicate cancellation between terms in the near-boundary 
limit. Thus any truncation error introduced by a numerical scheme must be 
sufficiently small to effectively scale by some high power of $q$ in the limit. 
In a typical finite difference scheme, the closest point to the boundary will be 
$q=h$ , where $h$ is the mesh-spacing there. Naively then, 
from~(\ref{eqn:nearbdyefes}) one would expect to need a discretization scheme 
that has local convergence order $n+2$ to obtain global $n^{th}$ order convergence 
of the solution. Fortunately, this naive argument apparently does not hold with our 
code and the situations we have explored to date: we observe global second-order 
convergence using second-order accurate finite difference discretization. We 
are not sure why this is so; again, it could simply be that the near-boundary 
expansion is giving a misleading picture of the nature of evolution of the full 
set of equations, or it could be due to the particular asymptotic gauge choice 
we use, described in the remainder of this section.

We have experimented with a small handful of gauge choices which were unstable, 
including evolution to harmonic w.r.t AdS ($\bar{H}_\mu(t > 0) \rightarrow 0$, 
though as argued above one expects problems with this), and a fixed gauge 
($\partial{\bar{H}_\mu}/\partial t =0$, and we note that our initial data is 
fully consistent with all the ``constraints'' in~(\ref{eqn:nearbdyefes}) and 
(\ref{eqn:nearbdyC})). When a numerical solution is unstable, it is often 
difficult to isolate the source: the gauge could be inconsistent in the sense 
discussed above, the discretization scheme could be unstable for the particular 
set of equations, there is a bug, etc. So here we simply list the asymptotic 
gauge condition we have empirically found to be stable (see 
Sec.~\ref{sec:num_source} for an explicit expression of the particular source 
functions used in this study), which in the notation introduced above is:
\begin{eqnarray}\label{eqn:nearbdygauge}
\bar{H}_{(1) t}    &=& \frac{5}{2} \bar{g}_{(1) t \rho} \nonumber \\
\bar{H}_{(1) \rho} &=& 2 \bar{g}_{(1) \rho\rho} \nonumber \\
\bar{H}_{(1) \chi} &=& \frac{5}{2} \bar{g}_{(1) \rho\chi}.
\end{eqnarray}
This choice was in part motivated by the asymptotic form of the field equations, 
in that these conditions, in conjuction with the GH 
constraints~(\ref{eqn:nearbdyC}) explicitly eliminate many of the order $q^{-2}$ 
terms that appear in~(\ref{eqn:nearbdyefes}). However, this is almost certainly
not a unique choice for stability, and one can anticipate that modifications 
would be required if, for example, the code is used to explore situations with 
less symmetry, or to study scenarios where the boundary theory is ``deformed'' 
in a manner that alters the leading-order AAdS asymptotics.

\subsection{Initializing the 5-metric at $t=0$}\label{sec:gauge_t0}

Our goal in this section is to describe a choice 
of $\left. \bar{g}_{t\nu} \right|_{t=0}$, $\left. \partial_t \bar{g}_{t\nu} \right|_{t=0}$ such 
that the source functions at $t=0$, as evaluated via~(\ref{eqn:ghcondition}), are compatible with 
our target gauge~(\ref{eqn:nearbdygauge}). (The spatial components of
the initial data $\left. \bar{g}_{ij} \right|_{t=0}$ and $\left. \partial_t \bar{g}_{ij} \right|_{t=0}$
come from the solution to the constraint equations described in Sec.~\ref{section:initial_data}).
In the notation of the power-series expansion about $q=0$ 
performed in the previous section, the leading-order coefficients of~(\ref{eqn:ghcondition}) 
evaluated using~(\ref{eqn:metric_asymptotics}) read 
\begin{eqnarray}\label{eqn:nearbdygauge_t0}
\bar{H}_{(1) t}    &=& \frac{5}{2}\bar{g}_{(1) t \rho} + \frac{1}{8}\bar{g}_{(1) tt,t} 
                     - \frac{1}{4}\bar{g}_{(1) t \chi,\chi} + \frac{1}{8}\bar{g}_{(1) \rho\rho,t} \nonumber \\
                   &&
                     + \frac{1}{8}\bar{g}_{(1) \chi\chi,t} + \frac{1}{4}\bar{g}_{(1) \psi,t} \nonumber \\
\bar{H}_{(1) \rho} &=& \bar{g}_{(1) t t} + \bar{g}_{(1) \rho\rho} 
                     - \bar{g}_{(1) \chi\chi} - 2 \bar{g}_{(1) \psi} \nonumber \\
\bar{H}_{(1) \chi} &=& \frac{5}{2}\bar{g}_{(1) \rho\chi} - \frac{1}{8}\bar{g}_{(1) tt,\chi} 
                     + \frac{1}{4}\bar{g}_{(1) t \chi,t} + \frac{1}{8}\bar{g}_{(1) \rho\rho,\chi} \nonumber \\
                   &&
                     - \frac{1}{8}\bar{g}_{(1) \chi\chi,\chi} + \frac{1}{4}\bar{g}_{(1) \psi,\chi}.
\end{eqnarray} 

Inspection of~(\ref{eqn:nearbdygauge}) and~(\ref{eqn:nearbdygauge_t0}), together with our 
choice of time symmetry ($\left. \partial_t \bar{g}_{ij} \right|_{t=0} = 0$), reveals that the
following is required to leading order in the approach to $\rho=1:$
\begin{eqnarray}
\left. \bar{g}_{t t} \right|_{t=0}               &=& \left. \left( \bar{g}_{\rho\rho} + \bar{g}_{\chi\chi} + 2\bar{g}_{\psi} \right) \right|_{t=0}  \label{eqn:gauget0_1} \\
\left. \partial_t \bar{g}_{t t} \right|_{t=0}    &=& 0 \label{eqn:gauget0_2} \\
\left. \partial_t \bar{g}_{t \chi} \right|_{t=0} &=& \left. \partial_\chi \bar{g}_{\chi\chi} \right|_{t=0}. \label{eqn:gauget0_3}
\end{eqnarray}
In practice we impose (\ref{eqn:gauget0_2}) and (\ref{eqn:gauget0_3}) over the entire computational
domain, and impose ($\ref{eqn:gauget0_1}$) in an asymptotic sense: we set 
$\left. \bar{g}_{t t} \right|_{t=0} =0 $ for $\rho<\rho_0$, where $\rho_0<1$ is
a user-specified parameter, and smoothly transition to ($\ref{eqn:gauget0_1}$) as 
$\rho\rightarrow 1$. In this way, we keep the form of $\left. \bar{g}_{t t} \right|_{t=0}$ 
as simple as possible in the interior. The relations~(\ref{eqn:nearbdygauge}) 
and~(\ref{eqn:nearbdygauge_t0}) leave the remaining metric variables 
$\left. \bar{g}_{t \rho} \right|_{t=0}$, 
$\left. \partial_t \bar{g}_{t \rho} \right|_{t=0}$ and  
$\left. \bar{g}_{t \chi} \right|_{t=0}$ unconstrained, and we 
again take the simplest choice and set them to zero. In ADM language, 
these conditions amount to a choice of initial lapse, shift and their 
first time derivatives.

\section{Numerical Scheme}\label{section:numerical_scheme}

The primary challenge in dealing with AAdS spacetimes lies in the nature of the
boundary, which must largely be dealt with analytically, as discussed in detail 
in the previous section. Once this is done, the numerical discretization and 
solution method is rather conventional. The methods we use are very similar to 
those described 
in~\cite{Pretorius:2004jg}, and so we simply list them in 
Sec.~\ref{sec:num_methods}, referring the reader to~\cite{Pretorius:2004jg} for 
the details. In Sec.~\ref{sec:num_source}, we describe the full (interior) form 
of the source functions that we use (in the previous section we only described 
their asymptotic behavior), and in Sec.~\ref{sec:conv_tests} we show some 
convergence results from a representative solution.

\subsection{Numerical Solution Methods}\label{sec:num_methods}

All equations and boundary conditions are discretized using second-order 
accurate finite difference methods, the only non-trivial boundary conditions 
being the Neumann conditions for the origin ($\rho=0$), and the axis 
($\chi=0,\pi$) regularity conditions. The discretized Hamiltonian 
constraint~(\ref{eqn:conformalform3}) is solved using a full approximation 
storage (FAS) multigrid algorithm with v-cycling, and Newton-Gauss-Seidel 
iteration for the smoother. The evolution equations for the 
metric~(\ref{eqn:efe_gh_modified}) and scalar field~(\ref{eqn:kgeqn}) 
(with energy-momentum tensor~(\ref{eqn:realenergymomentum})), are discretized 
after substituting in the definitions for our regularized metric 
variables~(\ref{eqn:metric_asymptotics}), scalar 
field~(\ref{eqn:scalar_asymptotics}), and source 
functions~(\ref{eqn:source_asymptotics}). The corresponding 
regular fields $(\bar{g}_{\mu\nu},\bar{\phi})$ are solved for using an 
iterative, Newton-Gauss-Seidel relaxation procedure. In this study we have not 
explored any dynamical gauge evolution equations for $\bar{H}^\mu$, and simply 
set them to prescribed functions of the $\bar{g}_{\mu\nu}$ as outlined in 
the next section.

We use the excision method to solve for black hole spacetimes, which involves
{\em excising} a portion of the grid within the apparent horizon (AH), thus 
removing the geometric singularity from the computational domain. Due to the 
causal structure of the spacetime within the horizon, all physical characteristics 
of the equations flow out of the domain (i.e. into the excised region). Thus 
excision implies that the usual field equations are still solved on the excision 
surface, except that centered difference stencils are replaced with sideways 
stencils where appropriate, in order to reference information that is only from 
the outside of the excision surface (in other words, {\em no} boundary 
conditions are placed there). We search for the AH using a flow method; the 
excision surface is defined to be a surface with the same coordinate shape as 
the AH, but some fraction $1-\delta_{ex}$ smaller, where we typically use 
$\delta_{ex}\in [0.05,0.2]$. One issue with using polar-like coordinates is that 
it incurs a rather severe restriction on the Courant-Friedrichs-Lewy (CFL) 
factor $\lambda$ that defines the time step 
$\Delta t \equiv \lambda \min(\Delta \rho,\Delta \chi)$, where $\Delta \rho$ and 
$\Delta \chi$ are the mesh spacings in $\rho$ and $\chi$ respectively. Roughly 
speaking, with a uniform discretization in $\rho$ and $\chi$, the condition for 
stability is $\lambda < \rho_{min}$, where $\rho_{min}$ is the 
smallest non-zero coordinate within the discrete domain. Ostensibly 
this occurs next to the origin, where $\rho_{min}=\Delta \rho$, so in the limit 
of high resolution the time-step size can become prohibitively small. For the 
tests and results presented here, we sidestep this issue by only studying black 
hole spacetimes, where excision removes the origin from the computational domain.

Kreiss-Oliger dissipation~\cite{KO}, with reduction of order approaching 
boundaries as described in~\cite{Calabrese:2003vx}, is used to help damp 
unphysical high-frequency solution components that sometimes arise at grid 
boundaries, in particular the excision surface; we typically use a dissipation 
parameter of $\epsilon=0.35$.

We use the 
{\tt PAMR/AMRD}~\footnote{\tt http://laplace.physics.ubc.ca/Group/Software.html}
libraries to provide support for running in parallel on Linux computing 
clusters. The libraries also support adaptive mesh refinement (AMR), though all 
results described here are unigrid. 

\subsection{Source Functions}\label{sec:num_source}

As described in Sec.~\ref{sec:gauge}, we are not free to arbitrarily choose
the leading-order behavior of the source functions approaching the AdS boundary 
if subsequent evolution of the field equations is to preserve the desired 
asymptotic form for the metric~(\ref{eqn:metricbcs}). Specifically, we want to 
consider a class of gauges that satisfy the asymptotic 
conditions~(\ref{eqn:nearbdygauge}). 

A naive implementation of~(\ref{eqn:nearbdygauge}), wherein one 
would set $H^\mu(t>0)$ {\em everywhere} on the grid via 
$\bar{H}_t = 5/2 \bar{g}_{t \rho}$, 
$\bar{H}_\rho = 2 \bar{g}_{\rho \rho}$, 
$\bar{H}_\chi = 5/2 \bar{g}_{\rho \chi}$,  
would result in a discontinuous gauge at 
$t=0$, as our method of solving for the initial data in general gives a 
different form for $H^\mu(t=0)$ on the interior. This is rather common in GH evolution 
i.e. the initial data provides source functions that are different from that of the 
target gauge, and the usual way to deal with this is to construct a hybrid gauge 
that smoothly transitions in time from the initial gauge to the target gauge. 
The specific transition we use is as follows. Denote the source functions coming 
from the initial data by 
$\bar{H}^{(0)}_\mu \equiv \left. \bar{H}_\mu \right|_{t=0}$, and define the 
functions $F_\nu$ to be the asymptotic constraints trivially extended into the 
interior\footnote{Of course, a more general implementation could allow for an 
arbitrary gauge in the interior, though for the spacetimes we have evolved to-date 
this simple choice has worked well.}:
\begin{eqnarray}\label{eqn:nearbdygauge_F}
F_t (t,\rho,\chi) &\equiv& \frac{5}{2} \bar{g}_{t \rho} (t,\rho,\chi) \nonumber \\
F_\rho (t,\rho,\chi) &\equiv& 2 \bar{g}_{\rho \rho} (t,\rho,\chi) \nonumber \\
F_\chi (t,\rho,\chi) &\equiv& \frac{5}{2} \bar{g}_{\rho \chi} (t,\rho,\chi).
\end{eqnarray}
Then, we choose
\begin{eqnarray}\label{eqn:gauge_choice}
\bar{H}_\mu(t,\rho,\chi) &=& \bar{H}^{(0)}_\mu(t,\rho,\chi) \left[ g(t,\rho) \right] \nonumber\\
                         &+& F_\nu(t,\rho,\chi) \left[ 1-g(t,\rho) \right],
\end{eqnarray}
with 
\begin{equation}
g(t,\rho) = \exp(-z(t,\rho)^4),
\end{equation}
\begin{equation}
z(t,\rho) = \frac{t}{\xi_2 f(\rho) + \xi_1 [1-f(\rho)]},
\end{equation}
and
\begin{equation}
f(\rho)=
\left\{
\begin{array}{lll}
1                         & \rho \ge \rho_2 \\
1-y^3 ( 6y^2 - 15y + 10 ) & \rho_2 \ge \rho \ge \rho_1 \\
0                         & \hbox{otherwise}
\end{array}
\right.,
\end{equation}
where $y(\rho) = (\rho_2 - \rho)/(\rho_2 - \rho_1))$, and $\xi_1,\xi_2,\rho_1$ 
and $\rho_2$ are user-specified constants. The time-transition function $g$ is 
such that $g(0,\rho)=1$, $\dot{g}(0,\rho)=0$, $\ddot{g}(0,\rho)=0$, and 
$\underset{t\rightarrow \infty}{\lim}g(t,\rho)=0$; it is designed to 
give~(\ref{eqn:gauge_choice}) the correct initial and target values, and 
transition between the two in a continuous fashion. The function $f(\rho)$, on 
the other hand, is such that $f(\rho_1)=0$, $f'(\rho_1)=0$, $f''(\rho_1)=0$ and 
$f(\rho_2)=0$, $f'(\rho_2)=0$, $f''(\rho_2)=0$; it is designed to let the 
transition occur with characteristic time $\xi_1$ for radii $\rho < \rho_1$, 
interpolating to a characteristic time of $\xi_2$ for radii $\rho \ge \rho_2$. 
It is important near the boundary to reach the target gauge quickly, as this is 
where the delicate cancellations made possible by the 
gauge~(\ref{eqn:nearbdygauge}) are crucially needed. Accordingly, 
$\xi_2$ is generally set to a small number. On the other hand, in may not be 
desirable to have such rapid gauge dynamics in the interior, and $\xi_1$ thus 
allows us to independently control the characteristic gauge evolution time 
there. On a typical run, we set 
$\rho_1=0.0,\rho_2=0.95,\xi_1=0.1,\xi_2=0.0025$. 

\subsection{Convergence Tests}\label{sec:conv_tests}

To check the stability and consistency of our numerical solutions we employ a 
pair of standard convergence tests. Here we show convergence results from one
typical representative case, namely strong scalar field initial data with 
non-trivial $\chi$-dependence. Specifically, the initial data 
parameters~(\ref{eqn:generalizedgaussian}) used were 
$A_0=10.0$, $R_0=0.0$, $\delta=0.2$, $w_x=4.0$, $w_y=16.0$. The resulting 
evolution describes a highly deformed black hole which settles down to 
an AdS-Schwarzschild solution with outermost apparent horizon radius $r_h=5.0$. 
The physics of this solution will be discussed in Sec.~\ref{results}, 
together with additional tests showing conservation of the boundary 
stress energy tensor in Sec.~\ref{subsec:bdy_stress_tensor}.

First, in order to determine whether the evolution is stable and consistent, 
assuming that the solution admits a Richardson expansion  we compute the 
rate of convergence $Q(t,x^i)$ at each point on the grid for a given field 
\begin{equation}\label{eq:qconv}
Q(t,x^i)=\frac{1}{\ln(2)}\ln\left( \frac{f_{4h}(t,x^i)-f_{2h}(t,x^i)}{f_{2h}(t,x^i)-f_{h}(t,x^i)} \right).
\end{equation}
Here, $f_\Delta$ denotes one of $\bar{g}_{\mu\nu},\bar{H}_\mu,\bar{\phi}$ from a 
simulation with mesh spacing $\Delta$. Given that we use second-order accurate 
finite difference stencils, with $2:1$ refinement in $\Delta$ between 
successive resolutions, and similarly in the time-step $\Delta t$, since we 
keep the CFL factor at a constant $\lambda=0.2$, we expect $Q$ to asymptote to $Q=2$ 
in the limit $\Delta\rightarrow0$.

Second, to determine whether we are converging to the correct solution, i.e. 
to a solution of the Einstein field equations, we compute an independent 
residual of the field equations. This is obtained by taking the numerical 
solution and substituting it into a discretized version of 
$G_{\mu\nu} + \Lambda_5 g_{\mu\nu} - 8\pi T_{\mu\nu}$. Since the numerical 
solution was found solving the GH form of the field equations, we do not expect 
the independent residual to be exactly zero; rather, if the solution is correct 
the independent residual should be purely numerical truncation error, and hence 
converge to zero. Thus, we can compute a convergence factor for it by using only
two resolution results via
\begin{equation}\label{eq:qires}
Q_{EFE}(t,x^i)=\frac{1}{\ln(2)}\ln\left( \frac{f_{2h}(t,x^i)}{f_{h}(t,x^i)} \right).
\end{equation}
Here, $f_\Delta$ denotes a component of 
$G_{\mu\nu} + \Lambda_5 g_{\mu\nu} - 8\pi T_{\mu\nu}$. 
Again, given our second-order accurate finite difference stencils and with $2:1$ 
refinement in $\Delta$ between successive resolutions, we expect $Q$ to approach 
$Q=2$ as $\Delta\rightarrow0$.

\begin{figure}[h]
	\centering
	\includegraphics[width=4.0in, bb=0 0 500 240]{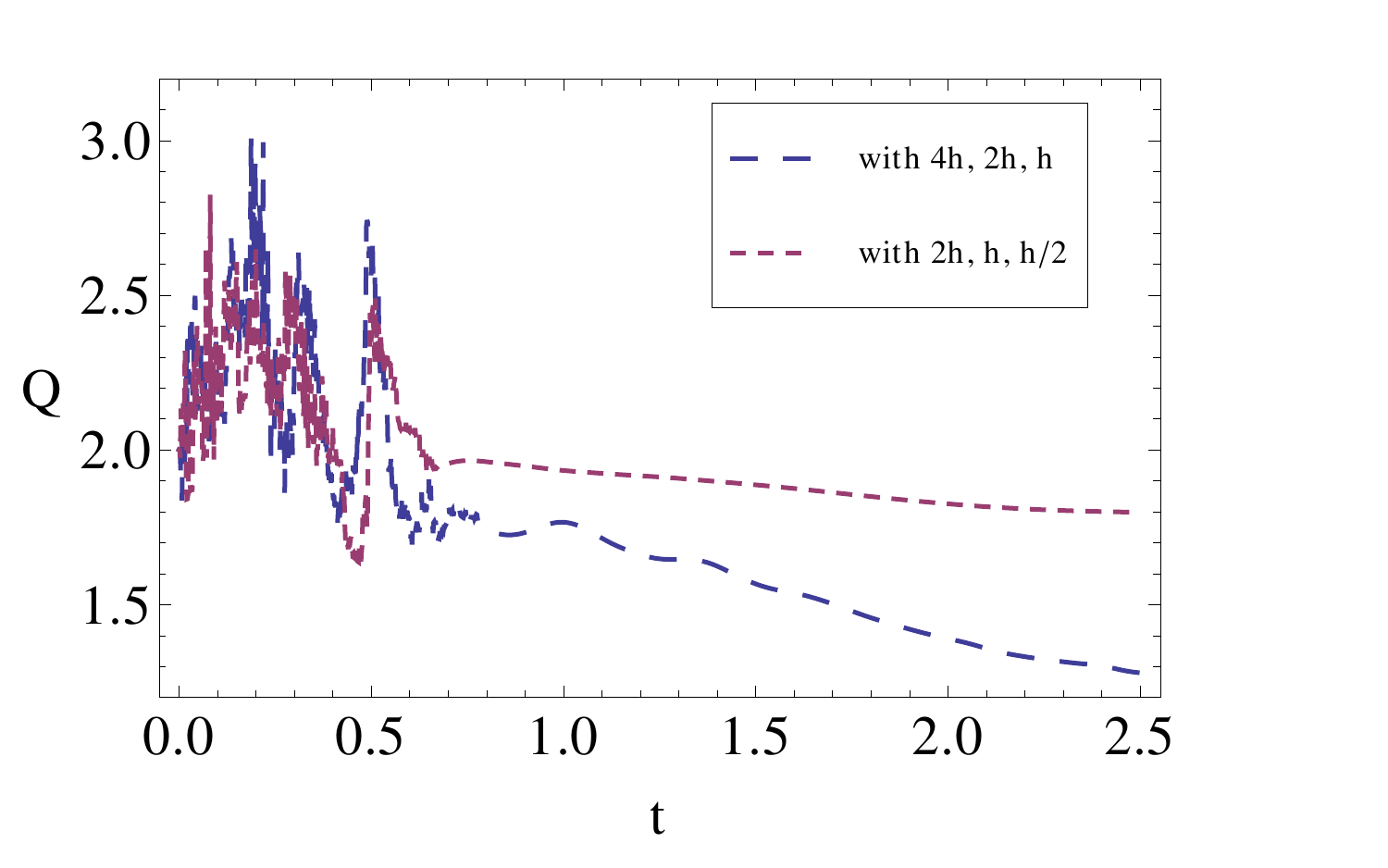}
\parbox{3.3in}{\caption{Convergence factors (\ref{eq:qconv}) for the 
$\bar{g}_{\rho\rho}$ grid function, constructed from a simulation run at 4 different 
resolutions; the highest resolution run has mesh spacing $h/2$.
Here the $L^2$-norm of the convergence factors are taken over the 
entire grid. The trends indicate that this grid function is converging to second-order.
Other grid functions exhibit similar trends.
	}\label{fig:qconv}}
\end{figure}

\begin{figure}[h]
	\centering
	\includegraphics[width=4.0in, bb=0 0 500 240]{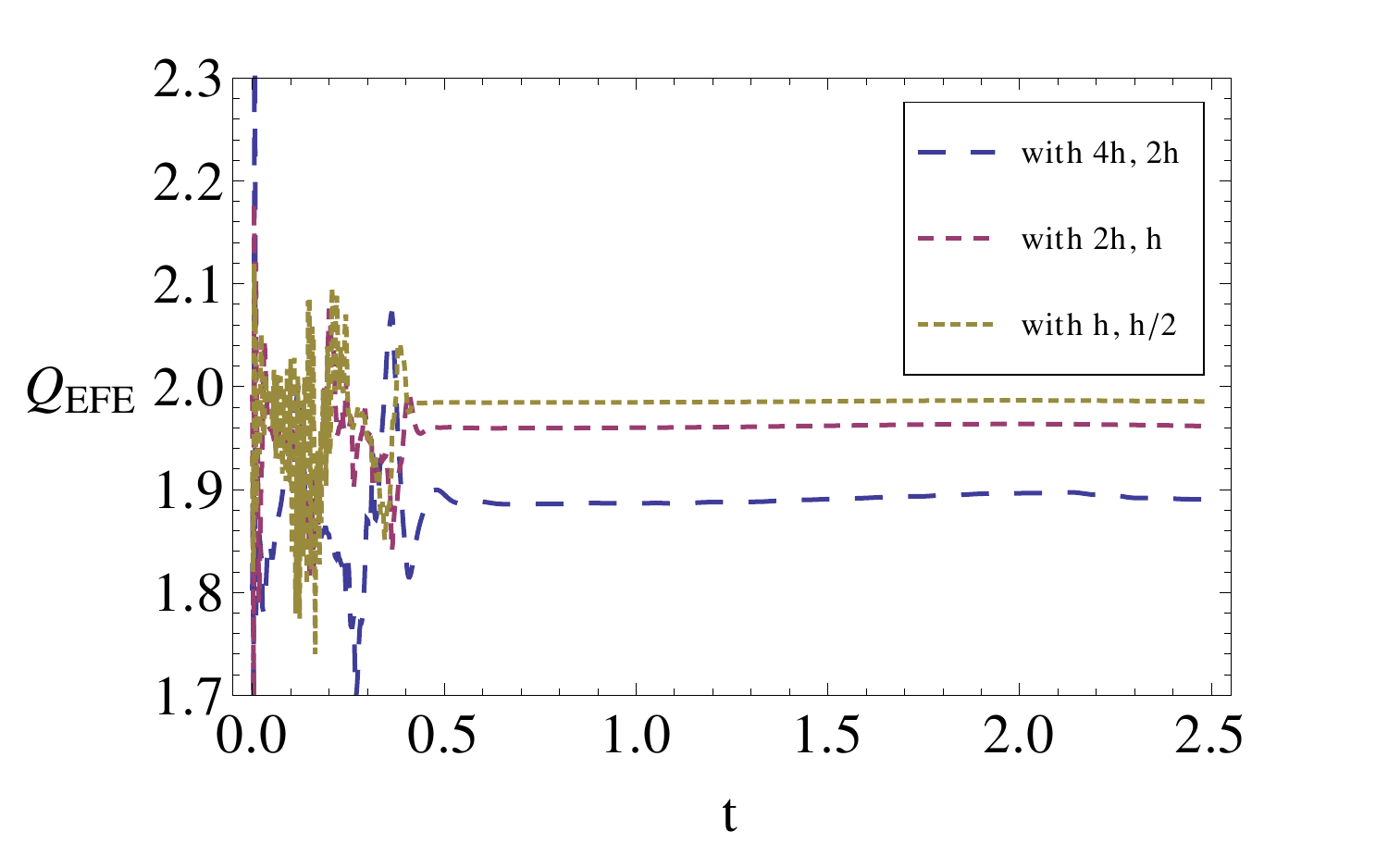}
\parbox{3.3in}{\caption{Convergence factors for the 
independent residual (\ref{eq:qires}), constructed from simulations run at 4 
different resolutions; the highest resolution run has mesh spacing $h/2$.
At each point on the grid an $L^\infty$ norm is taken over all components
of the independent residual, and what is shown here is then 
the $L^2$-norm of this over the entire grid. 
The trends in this plot indicate second-order convergence.
	}\label{fig:qires}}
\end{figure}

Figs.~\ref{fig:qconv} and~\ref{fig:qires} show $L^2$-norms of~(\ref{eq:qconv}) 
and~(\ref{eq:qires}) respectively, obtained from evolving the particular initial 
data described above. We used 4 different resolutions to help see the 
trends in the respective $Q$'s. At early times, the solution is not yet in the 
asymptotic scaling regime for the convergence factors, however that they 
are typically greater than $2$ indicates that the solution error is nevertheless 
small. Also, the trends going to higher resolution, in particular at later times, 
appear consistent with second-order convergence. 

The early-time deviations in Fig.~\ref{fig:qires} coincide with an initial 
transient associated with the gauge transition described 
by~(\ref{eqn:gauge_choice}), that emanates from the boundary and dissipates as 
it travels further into the interior. This transient is clearly seen as a 
temporary dip in the {\em point-wise} convergence factors for independent 
residuals near the boundary; taking the $L^2$-norm as was done in 
Fig.~\ref{fig:qires} to some extent masks this dip, though is still visible 
with the highest resolution curve in that plot. Nevertheless, this transient 
converges away in the sense that the region of the domain affected shrinks as 
resolution is increased. 

\section{Results}\label{results}

We now describe several early results obtained with this code. 
First, in Sec.~\ref{mass-amplitude}, we show solutions of the Hamiltonian 
constraint with a scalar field source~(\ref{eqn:conformalform3}), demonstrating 
that this approach is capable of producing initial data containing trapped 
surfaces. Then, in Sec.~\ref{sec:qnr} we show the evolution of initial data 
describing highly deformed black holes that subsequently shed their asymmetries 
via quasi-normal ringdown. A proper extraction of quasi-normal modes, and in particular 
a meaningful comparison with perturbative calculations where these modes can be 
defined, requires the identification of a reference background metric and a transformation
of the solution to a gauge consistent with that of the perturbative calculations.
We have made no attempt in this direction, besides matching the areal radius on the 
extraction sphere, and do in fact see what appear to be gauge modes. 
Nevertheless, we can extract the leading order linear quasi-normal
modes, and for the higher angular number modes, use a simple forced harmonic oscillator 
model to identify what appear to be non-linearly excited harmonics of the lower angular modes. 
In Sec.~\ref{subsec:bdy_stress_tensor} we discuss the extracted 
boundary stress energy tensor of these solutions, and in 
Sec.~\ref{section:hyd_desc_of_the_bdy_cft} analyse how well the CFT state can 
be described by hydrodynamics. We find that the extracted stress energy tensor is consistent,
essentially to within numerical truncation error, with that of a viscous, conformal fluid from 
$t=0$ onwards. In Sec.~\ref{section:passing_to_minkowski_space}, we transform these solutions onto 
a Minkowski piece of the boundary, and find an initial fluid geometry that resembles a 
Lorentz-flattened pancake. Defining the beam-line direction as the one along which the initial data 
is flattened, the evolution of this fluid exhibits both longitudinal and transverse flow 
relative to the beam-line, with most of the energy flowing along the longitudinal directions.

\subsection{Strong-field Solutions of the Hamiltonian Constraint}\label{mass-amplitude}

To generate black holes spacetimes, we choose initial data where the deviation 
from pure AdS$_5$ is sourced by a highly compact distribution of scalar field energy, 
with profile given by~(\ref{eqn:generalizedgaussian}). In this paper we only 
consider free, massless scalar fields, so $V(\phi)=0$ in (\ref{eqn:timesymmetryenergydensity}).
In this section, we begin by focusing on spherically symmetric initial data, 
so $w_x=w_y=1$ in (\ref{eqn:generalizedgaussian}).

\begin{figure}[h]
\begin{center}
\includegraphics[width=3.6in, bb=0 0 480 280]{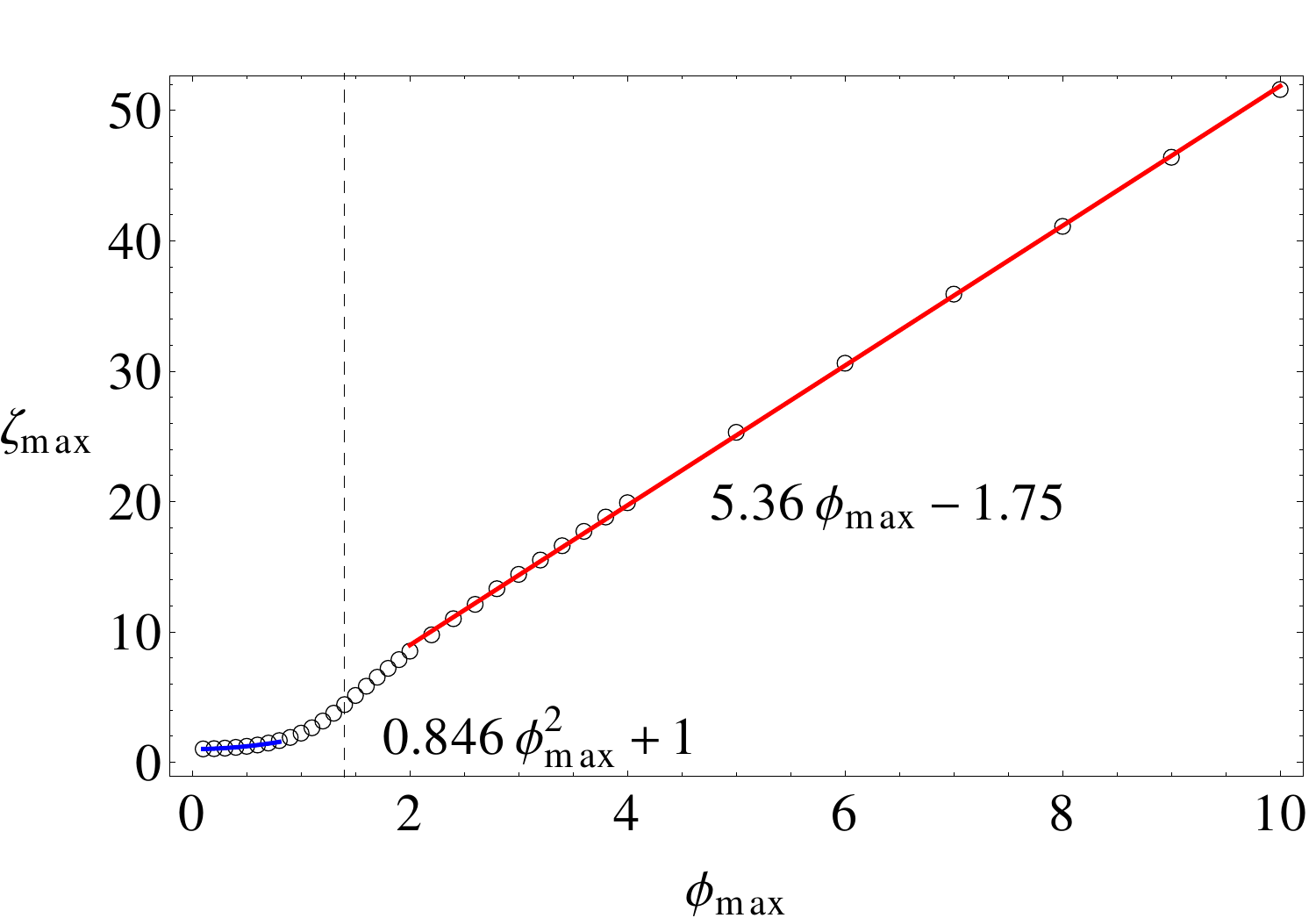}
\parbox{3.3in}{
\caption{Maximum conformal factor vs. maximum amplitude of 
initial scalar matter distribution, with Gaussian profile 
(\ref{eqn:timesymmetryenergydensity}) with
$A=\phi_{\text{max}}$, $\delta=0.2$ (with AdS scale $L=1$), 
$w_x=w_y=1$ and $R_0=0$. 
We differentiate between ``strong-field'' and ``weak-field'' data
based on whether there is a trapped surface present on the initial slice or 
not, respectively. (Though of course this distinction is somewhat arbitrary, 
particularly since subsequent evolution of weak-field data could 
eventually result in black hole formation, as argued in \cite{Bizon:2011gg} 
even for arbitrarily small amplitude initial data.) The value of $\phi_{max}$ 
beyond which trapped surfaces are found in the initial data is indicated by 
the dashed vertical line. The open circles denote numerical solutions,
while the solid lines are fits to the data, as shown.
	}\label{fig:conformalfactorvsamplitude}}
\end{center}
\end{figure}

Fig.~\ref{fig:conformalfactorvsamplitude} summarizes solutions of the 
Hamiltonian constraint~(\ref{eqn:conformalform3}) by plotting the maximum value 
of the conformal factor versus the maximum of the scalar field (both maxima  
occur at the origin of the domain for these initial data). Let us begin by contrasting 
the qualitative behavior of this plot with its counterpart in the asymptotically flat case, 
presented in~\cite{Pfeiffer:2005jf}. This earlier work employed the conformal 
thin sandwich method to solve the constraints. There, it was discovered that 
there exists a critical point above which no numerical solutions for the initial 
data could be found, and where the conformal factor diverges as the amplitude 
approaches the critical point. Even more surprisingly, it was found that 
generalizing to the extended conformal thin sandwich method gives rise to a 
branch point instead of a critical point, with solutions along the upper branch 
exhibiting non-uniqueness for any given set of initial data. 

In contrast, we find no divergent behavior in the conformal factor, nor any 
behavior suggesting the non-uniqueness of solutions $\phi_{max}$, at least in 
the regime where the cosmological length scale is relevant\footnote{We have not
investigated the limit where the characteristic scale in the initial data $\delta$
is much less than the cosmological scale $L$, where one might intuitively expect
$L$ to become irrelevant in governing the local nature of the solution,
and results consistent with the asymptotically flat case~\cite{Pfeiffer:2005jf}
may be recovered.}. Furthermore, the linear dependence of the maximum of the 
conformal factor versus amplitude for large amplitude data suggests that the 
AdS$_5$ Hamiltonian constraint admits conformal solutions for arbitrarily strong 
initial matter distributions. Of course, given the presence of the cosmological 
constant and its relevance on these scales, it is not too surprising that we find 
such qualitatively different behavior compared to the asymptotically flat case. From 
a more formal perspective, the local existence and uniqueness of solutions to 
non-linear elliptic PDEs can be understood by applying a maximum principle, 
where the signs and relative magnitudes of coefficients in the PDE are crucial; 
one can show that in this regard cosmological the constant in AAdS spacetimes ``helps''.

\begin{figure}
\centering
\includegraphics[width=3.6in, bb=0 0 480 280]{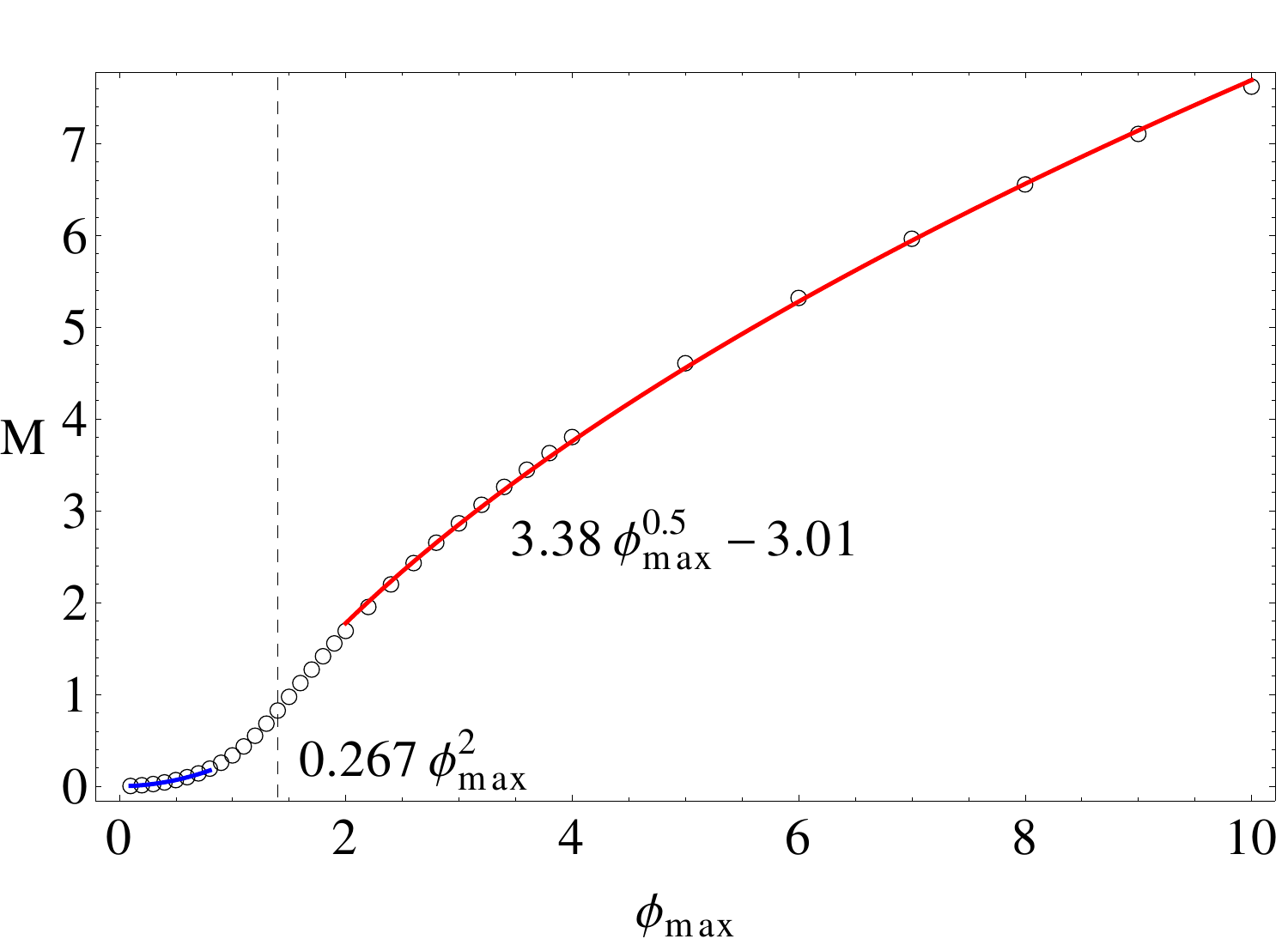}
\parbox{3.3in}{\caption{AdS mass vs. maximum amplitude of massless scalar 
matter, with Gaussian profile defined by $A=\phi_{\text{max}}$, $(x_0,y_0)=(0,0)$, $\delta=0.2$. 
Conventions are the same as those used in Fig.~\ref{fig:conformalfactorvsamplitude}.
In cases where trapped surfaces are present (to the right of the dashed
vertical line), an estimate of the mass based on the area of the apparent
horizon gives a value very close the asymptotic mass plotted here, though
systematically smaller (the two values would essentially be indistinguishable on
the scale of this figure, hence we only show the asymptotic mass to avoid
clutter).
	}\label{fig:massvsamplitude}}
\end{figure}

Fig.~\ref{fig:massvsamplitude} shows a plot of the conserved mass $M$ of the 
spacetime versus scalar field amplitude, computed from the quasi-local stress 
tensor~(\ref{eqn:quasiset_subtracted}) as follows. We take a spatial 
$t={\rm const.}$ (here $t=0$) slice in $\partial M_q$, with induced 3-metric 
$\sigma_{\mu \nu}$,  lapse $N$ and shift $N^i$ such that 
$\Sigma_{\mu \nu} dx^\mu dx^\nu = -N^2 dt^2 + \sigma_{ij}(dx^i + N^i dt)(dx^j +N^j dt)$, 
then compute 
\begin{equation}\label{eqn:adsmass}
M = \underset{q \rightarrow 0}{\lim} \int_{\Sigma} d^3 x \sqrt{\sigma} N ( {}^{(q)} T_{\mu \nu} u^\mu u^\nu )
\end{equation}
where $u^\mu$ is the time-like unit vector normal to $t={\rm const.}$
Note that for a vacuum AdS-Schwarzschild black hole, this prescription gives a 
result consistent with the usual definition of its mass from the analytic 
solution, namely $M = 3 \pi \left({r_0}^2/8 \right)$, where 
$r_0=r_H\sqrt{1+r_H^2/L^2}$, given a horizon with areal radius $r_H$.

\subsection{Quasinormal Ringing}\label{sec:qnr}

As a first application of our evolution scheme, we study the quasi-normal 
ringdown of an initially high distorted (i.e. non-spherical) black hole. 
Here we focus on the dynamics of the bulk, and in the following sections discuss 
the corresponding dynamics of the CFT boundary stress tensor. For some previous 
work on the subject see for 
example~\cite{Horowitz:1999jd,Cardoso:2003cj,Kovtun:2005ev,Janik:2006gp,
Teaney:2006nc,Friess:2006kw}. For a review of black hole quasi-normal modes in 
AdS see~\cite{Berti:2009kk}. Again, we use large-amplitude scalar field data to
create an initial slice of the spacetime containing a trapped surface, and here 
introduce a non-trivial $\chi$-dependence by adjusting the shape of the profile 
through the parameters $w_x$ and $w_y$ as defined 
in~(\ref{eqn:generalizedgaussian}). Specifically, unless otherwise stated we 
choose $R_0=0.0$, $\delta=0.2$, $w_x=4.0$, $w_y=16.0$, and vary the amplitude
$A_0$ to control the size of the resultant black hole. 

\begin{figure}[h]
        \centering
        \includegraphics[width=4.3in, bb=0 0 500 240]{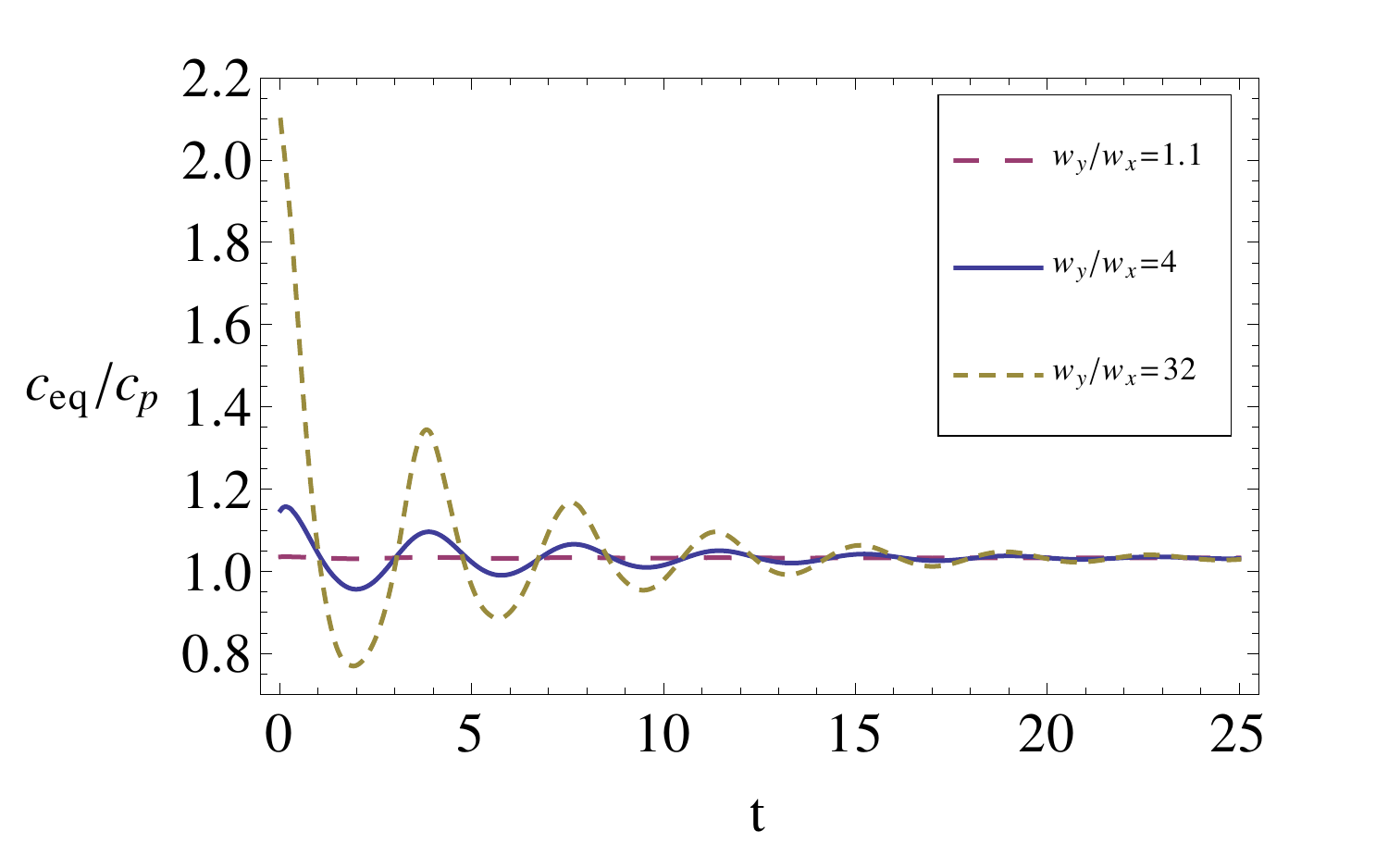}
\parbox{3.3in}{\caption{The ratio $c_{eq}/c_p$ of proper equatorial to polar
circumference of the apparent horizon versus time. $w_y/w_x$ denotes
parameters in the initial data describing the scalar field profile (\ref{eqn:generalizedgaussian}),
and $w_y/w_x=4$ is the canonical case studied in this paper, though for interest
we also show examples representing a weaker and stronger initial asymmetry. In all cases, the 
final black hole has radius $r_h\approx 5$. Note that a geometric sphere has 
$c_{eq}/c_p=1$, whereas a geometric disk has $c_{eq}/c_p=\pi/2$. Thus, even the 
$w_y/w_x=4$ case is initially a rather large deformation of the $3$-sphere; the 
curve for the more extreme case $w_y/w_x= 32$ implies that at early times the 
Gaussian curvature is negative over at least some regions of the horizon.
        }\label{fig:ceqcp}}
\end{figure}

First, to demonstrate that we are looking at relatively large perturbations of 
the spherical black hole, in Fig.~\ref{fig:ceqcp} we plot the ratio of the 
equatorial $c_{eq}$ to polar $c_p$ proper circumferences of the apparent horizon 
versus time\footnote{Note that the apparent horizon is to some extent slicing 
dependent. However, given the symmetries in our problem, in particular that 
$t=0$ is a moment of time-symmetry (where we see the largest deformation of the 
horizon), it is difficult to imagine how the intrinsic geometry of the apparent 
horizon does not give a good indication of the relative magnitude of the 
spacetime perturbation. Nevertheless, it would be interesting to compare to the 
event horizon, though finding event horizons are more complicated and we leave 
that to a future study.}. A geometric sphere has $c_{eq}/c_p=1$, whereas a 
geometric disk $c_{eq}/c_p=\pi/2$. The $w_y/w_x=4$ case studied in detail here 
has $c_{eq}/c_p(t=0) \approx 1.2$, so a fairly sizeable deformation. For
certain quasi-normal mode and hydrodynamic extraction results we also show
data from the $w_y/w_x=32$ case, which has a very large initial $c_{eq}/c_p(t=0) \approx 2.2$.
 
To extract the quasi-normal modes, the first step would be to transform
our metric to a gauge consistent with standard perturbative calculations
for quasi-normal modes. This is a rather non-trivial step in general,
and we do not do so here, hoping that our asymptotic gauge choice
is sufficiently close to that of the perturbative calculations that
artifacts introduced are small. As we shall see,
this seems to hold to good approximation, though we do find modes
that appear to be pure gauge.

We begin by projecting our field 
variables onto the spherical harmonics on $S^3$; see~\cite{Avery:2000tx}. We 
first need the standard spherical harmonics on $S^2$:
\begin{equation}
Y_{lm}(\theta,\phi) = \sqrt{\frac{2l+1}{4\pi} \frac{(l-m)!}{(l+m)!}} 
P^m_l(\cos\theta) \exp(i m \phi)
\end{equation}
where the associated Legendre polynomials are defined as:
\begin{equation}
P^m_l(x) = \frac{(-1)^m}{2^l l!} \frac{d^{(l+m)}}{dx^{(l+m)}} (x^2-1)^l.
\end{equation}
The scalar spherical harmonics on $S^3$ are then:
\begin{eqnarray}
\mathbb{S}(klm) 
&=& (-1)^k i^l (2l)!! \sqrt{\frac{2(k+1)(k-l)}{\pi(k+l+1)!}} \nonumber \\
& & \times C^{l+1}_{k-l}(\cos\chi) \sin^l(\chi) Y_{lm}(\theta,\phi) 
\end{eqnarray}
where the Gegenbauer polynomials are defined as:
\begin{eqnarray}
C^a_k(x) 
&=& \frac{(-2)^k}{k!} \frac{\Gamma(k+a) \Gamma(k+2a)}{\Gamma(a) \Gamma(2k+2a)} 
(1-x^2)^{-a+1/2} \nonumber \\
& & \times \frac{d^k}{dx^k} (1-x^2)^{k+a-1/2}.
\end{eqnarray}
Our restriction to solutions that preserve a symmetry in $\theta,\phi$ mandates 
that we have $l=0,m=0$. 

\begin{figure}[h]
	\centering
	\includegraphics[width=4.0in, bb=0 0 520 290]{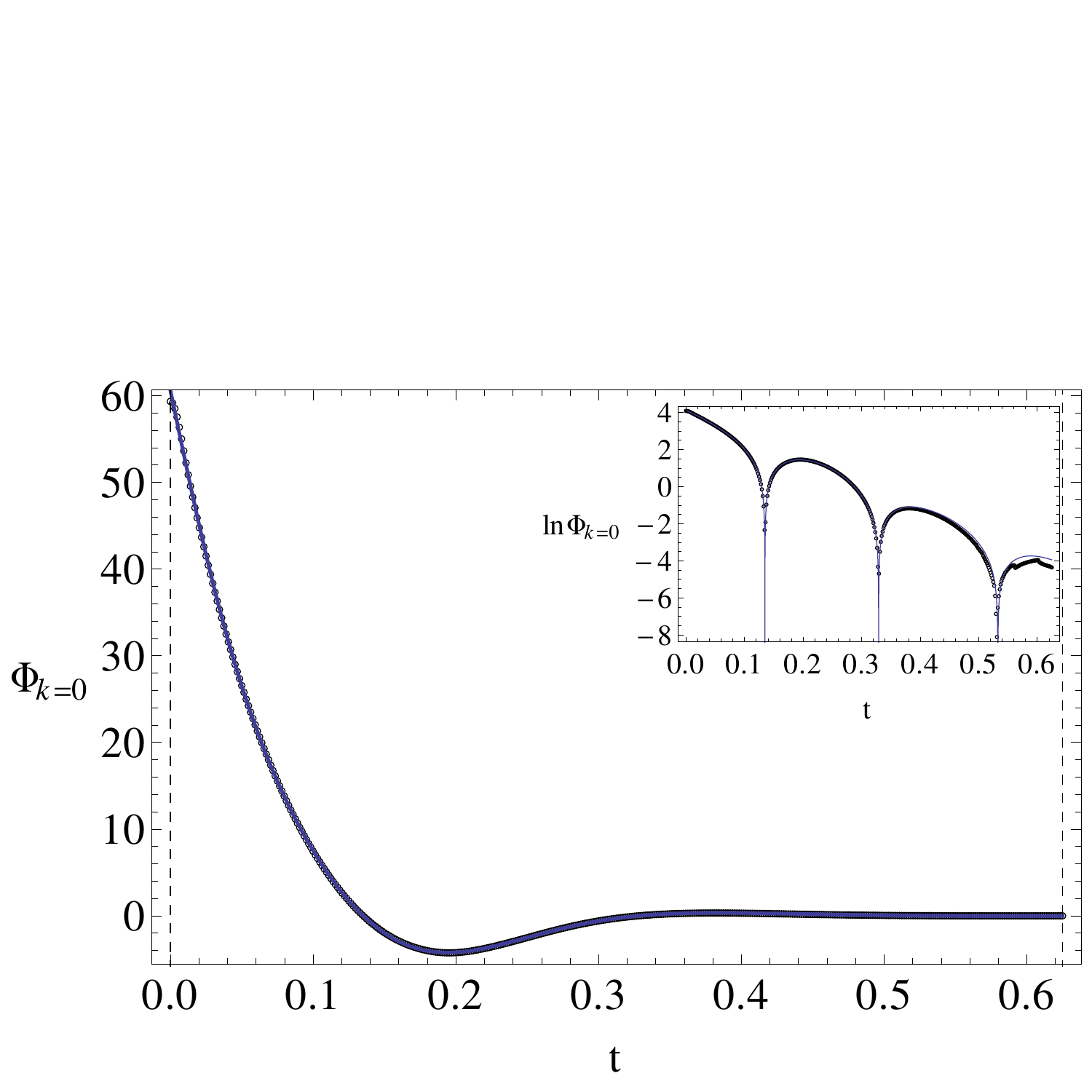}
\parbox{3.3in}{\caption{The leading-order behavior of the scalar field variable 
$\bar{\phi}$ near the boundary $q=1$, projected onto the $3$-sphere by 
$\Phi_{k=0}=\int d\Omega_3 (\bar{\phi}/q) \mathbb{S}(000)$, 
and plotted over a global time interval of $t\in[0,0.62]$ (open circles). A fit 
(solid line) $\Phi_{k=0}^{fit}$ is extracted using the data inbetween the dashed 
vertical lines. The inset shows a logarithmic plot of the data and the fit over the 
full global time interval. 
	}\label{fig:scalarqnms_k0}}
\end{figure}

For perturbations of the scalar field, these $\mathbb{S}(klm)$ scalar harmonics 
exist for all $k \ge 0$. Fig.~\ref{fig:scalarqnms_k0} shows the scalar field projected 
onto the 3-sphere ($k=0$), for a simulation with initial $w_y/w_x=4$
and whose final state black hole has horizon radius $r_h = 5$. The 
quasi-normal mode frequencies extracted from the scalar field are collected in 
Appendix~\ref{app:tables_of_scalar_qnm_frequencies}, under 
Table~\ref{tab:scalarqnms_fastmodes_n0} for the $n=0$ fundamental mode and under 
Table~\ref{tab:scalarqnms_fastmodes_n1} for the $n=1$ first overtone. This was done for 
several $r_h$ cases, wherein each pair of fundamental and first overtone was 
found by a simultaneous non-linear least-squares fit to two damped sinusoids, one 
corresponding to the $n=0$ fundamental and the other corresponding to the $n=1$ overtone. 
Each sinusoid is described by four parameters\footnote{These parameters are amplitude $A$, 
phase $\varphi$, imaginary frequency $\omega_i$, and real frequency $\omega_r$, so the fitting 
functions take the form $A \exp(-\omega_i t) \cos(\omega_r t + \varphi)$.}. The $n=0$ fundamental 
frequencies $\omega_r$ and $\omega_i$ scale linearly with $r_h$, and their dependence on $k$ shows 
that $\omega_r$ increases with $k$ and that $\omega_i$ decreases with $k$. The $n=1$ first 
overtone frequencies $\omega_r$ and $\omega_i$ also scale linearly with $r_h$, 
as with the $n=0$ frequencies; however, the dependence on $k$ differs from the $n=0$ fundamental 
case in that the $\omega_r$ and $\omega_i$ both increase with $k$. To make 
contact with earlier work on quasi-normal modes in AdS, we note that the frequencies extracted 
for the $n=0$ fundamental are a close match to those found in the seminal study \cite{Horowitz:1999jd} 
on this subject. 

\begin{figure}[h]
	\centering
	\includegraphics[width=4.0in, bb=0 0 520 290]{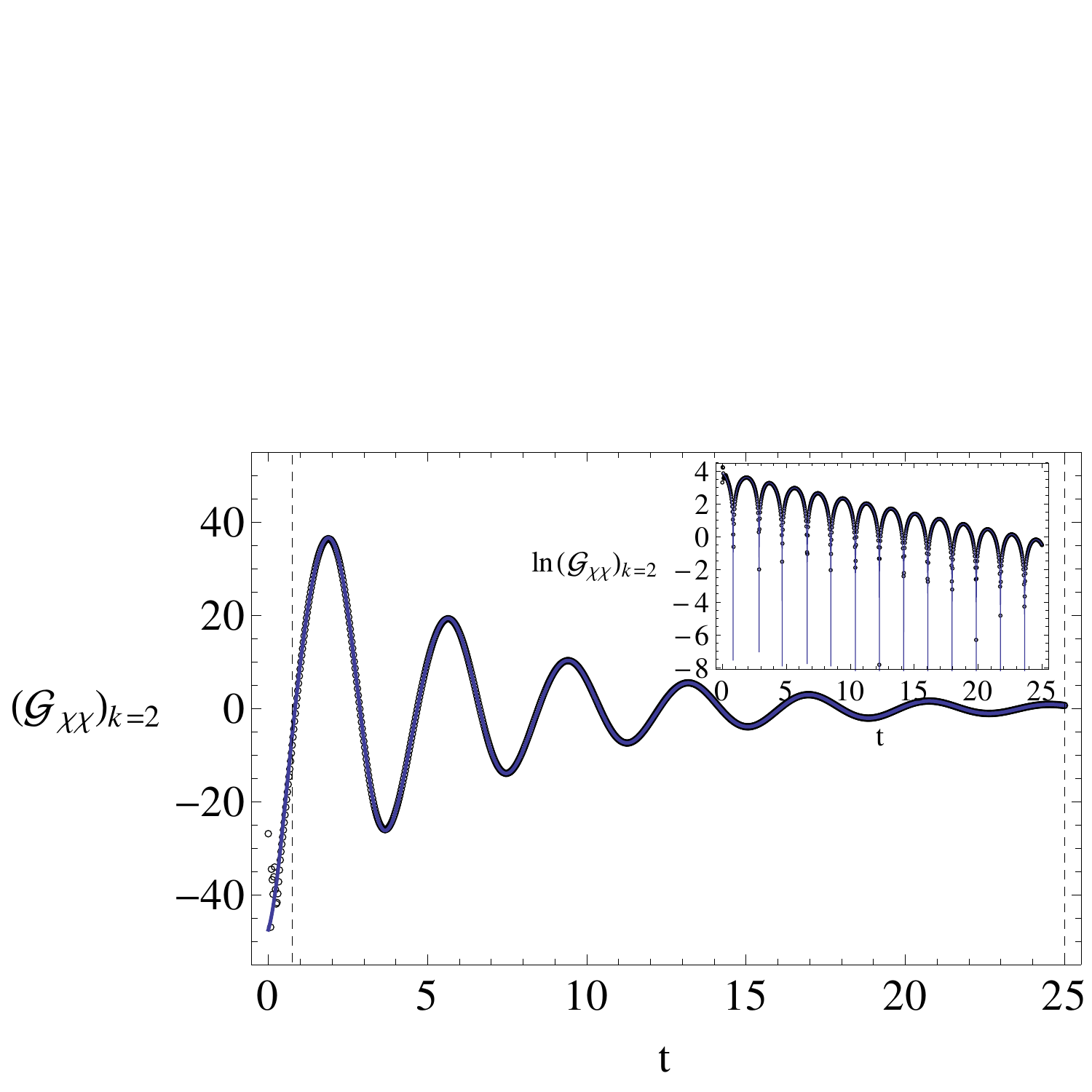}
\parbox{3.3in}{\caption{The leading-order behavior of the metric variable 
$\bar{g}_{\chi\chi}$ near the boundary $q=0$, projected onto the $\mathbb{S}(200)$ 
scalar harmonic by 
$(\mathcal{G}_{\chi\chi})_{k=2}=\int d\Omega_3 (\bar{g}_{\chi\chi}/q) \mathbb{S}(200)$, 
plotted over a global time interval of $t\in[0,8\pi]$ (open circles). A fit (solid line)
$(\mathcal{G}_{\chi\chi})_{k=2}^{fit}$ is extracted using the data inbetween the dashed 
vertical lines. The inset shows a logarithmic plot of the data and the fit over the full 
global time interval. Other metric variables show similar behavior.
	}\label{fig:gravqnms_k2}}
\end{figure}

\begin{table}[h]
\begin{center}
\begin{tabular}{|ll|l|}
\hline
&fund. 		&k=2					\\
\hline
&$\frac{r_h}{L}$=12.2 	&$(1.640 \pm 0.007)-i(0.902 \pm 0.067)\frac{L}{r_h}$	\\
&$\frac{r_h}{L}$=11.3 	&$(1.640 \pm 0.005)-i(0.895 \pm 0.058)\frac{L}{r_h}$	\\
&$\frac{r_h}{L}$=10.5 	&$(1.641 \pm 0.002)-i(0.876 \pm 0.039)\frac{L}{r_h}$	\\
&$\frac{r_h}{L}$=9.0 	&$(1.643 \pm 0.003)-i(0.864 \pm 0.028)\frac{L}{r_h}$	\\
&$\frac{r_h}{L}$=6.5 	&$(1.650 \pm 0.0007)-i(0.841 \pm 0.010)\frac{L}{r_h}$	\\
&$\frac{r_h}{L}$=5.0 	&$(1.661 \pm 0.0006)-i(0.837 \pm 0.005)\frac{L}{r_h}$	\\
&$\frac{r_h}{L}$=4.5 	&$(1.666 \pm 0.0006)-i(0.823 \pm 0.004)\frac{L}{r_h}$	\\
&$\frac{r_h}{L}$=4.0 	&$(1.675 \pm 0.0005)-i(0.823 \pm 0.003)\frac{L}{r_h}$	\\
&$\frac{r_h}{L}$=3.3 	&$(1.692 \pm 0.0004)-i(0.808 \pm 0.002)\frac{L}{r_h}$	\\
\hline
\end{tabular}
\end{center}
\centering
\parbox{3.3in}{\caption{The fundamental ($n=0$) quasi-normal mode frequencies $\omega_r - i\omega_i$ 
extracted from the metric variable $\bar{g}_{\chi\chi}$. These are shown for various 
horizon radii $r_h$ of the final state AdS-Schwarzschild black hole and for 
$SO(4)$ quantum numbers $k=2,l=0,m=0$. Uncertainties are estimated from convergence
studies. 
	}\label{tab:gravqnms_slowmodes_n0}}
\end{table}

\begin{table}[h]
\begin{center}
\begin{tabular}{|ll|l|}
\hline
&n=1  		&k=2					\\
\hline
&$\frac{r_h}{L}$=12.2 	&$(3.497 \pm 0.131)\frac{r_h}{L}-i(2.087 \pm 0.142)\frac{r_h}{L}$	\\
&$\frac{r_h}{L}$=11.3 	&$(3.320 \pm 0.156)\frac{r_h}{L}-i(2.185 \pm 0.174)\frac{r_h}{L}$	\\
&$\frac{r_h}{L}$=10.5 	&$(3.378 \pm 0.339)\frac{r_h}{L}-i(2.120 \pm 0.368)\frac{r_h}{L}$	\\
&$\frac{r_h}{L}$=9.0 	&$(3.256 \pm 0.078)\frac{r_h}{L}-i(2.100 \pm 0.125)\frac{r_h}{L}$	\\
&$\frac{r_h}{L}$=6.5 	&$(3.329 \pm 0.178)\frac{r_h}{L}-i(2.091 \pm 0.091)\frac{r_h}{L}$	\\
&$\frac{r_h}{L}$=5.0 	&$(3.103 \pm 0.264)\frac{r_h}{L}-i(2.626 \pm 0.495)\frac{r_h}{L}$	\\
&$\frac{r_h}{L}$=4.5 	&$(3.217 \pm 0.084)\frac{r_h}{L}-i(2.718 \pm 0.130)\frac{r_h}{L}$	\\
&$\frac{r_h}{L}$=4.0 	&$(3.274 \pm 0.051)\frac{r_h}{L}-i(2.688 \pm 0.070)\frac{r_h}{L}$	\\
&$\frac{r_h}{L}$=3.3 	&$(3.318 \pm 0.059)\frac{r_h}{L}-i(2.914 \pm 0.307)\frac{r_h}{L}$	\\
\hline
\end{tabular}
\end{center}
\centering
\parbox{3.3in}{\caption{The first overtone ($n=1$) quasi-normal mode frequencies $\omega_r - i\omega_i$ 
extracted from the metric variable $\bar{g}_{\chi\chi}$. These are shown for various 
horizon radii $r_h$ of the final state AdS-Schwarzschild black hole and for 
$SO(4)$ quantum numbers $k=2,l=0,m=0$. Uncertainties are estimated from convergence
studies. 
	}\label{tab:gravqnms_slowmodes_n1}}
\end{table}

For the metric, the $\mathbb{S}(klm)$ scalar harmonics are restricted by 
$0 \le l \le k$. The $k=1$ mode is not associated with any physical degrees of 
freedom \cite{Kodama:2003jz}, and these can be ignored; indeed our choice of 
initial data where the scalar field profile is symmetric about $\chi=\pi/2$ prevent 
these from being excited\footnote{In addition to the scalar harmonics, the metric also 
in general admits vector $\mathbb{V}(klm)$ ($1 \le l \le k$) and tensor harmonics 
$\mathbb{T}(klm)$ ($2 \le l \le k$), but our $SO(3)$ symmetry precludes the 
excitation of these modes: the vector and tensor harmonics require $\theta$ 
dependence (i.e. $l \ne 0$) that our symmetry does not allow.}. The $k=0$ mode 
corresponds to a perturbation of a black hole that is 
itself spherically-symmetric in all the $S^3$ angles $\chi,\theta,\phi$. 
Thus, as a consequence of Birkhoff's theorem, we might also expect to ignore 
this $k=0$ case as we have with $k=1$. However, despite this reasoning, we {\em see} 
non-trivial dynamics in the data's projection onto the $k=0$ scalar harmonic. 
To understand why, recall that we {\em have not} transformed 
our metric to a gauge that is consistent with the standard one assumed by the 
perturbative calculations for quasi-normal modes. Consequently, the metric components 
that we extract generally contain both physical quasi-normal modes as well as contributions 
from gauge modes that are introduced by our ``non-standard'' choice of time-slicing, and 
of the $r$ and $\chi$ coordinates on each slice. This gauge contribution is 
of course present in the data for all $k$, and is most obvious in the $k=0$ projection 
since it contains no physical quasi-normal modes. For $k \ge 2$, this 
gauge contribution manifests itself as a decaying mode (i.e. a mode whose 
frequency has a negligibly small real part). In the following, we consider the 
$k \ge 2$ modes of the metric, taking into account this gauge contribution.

Fig.~\ref{fig:gravqnms_k2} shows a representative metric variable projected onto 
the $\mathbb{S}(200)$ ($k=2$) scalar harmonic, for a simulation with 
final state horizon radius $r_h = 5$ and initial asymmetry $w_y/w_x=4$. The quasi-normal 
mode frequencies extracted from this representative metric variable are displayed 
in Table~\ref{tab:gravqnms_slowmodes_n0}. This analysis was repeated for several 
simulations with varying $r_h$ but fixed $w_y/w_x=4$. Fits to damped sinusoids 
yield $n=0$ fundamental frequencies with imaginary parts $\omega_i$ that scale as 
$\sim 1/r_h$, and real parts $\omega_r$ that are largely insensitive to these changes 
in $r_h$. 

Obtaining the $\mathbb{S}(200)$ harmonic's $n=1$ overtone from the metric variables 
is more difficult, largely because it decays much more quickly than the $n=0$ fundamental. 
To obtain good fits, we focus on an early-time segment of the data, and substract 
the $n=0$ fit, then fit to the remainder. The quasi-normal frequencies extracted in 
this way are tabulated in Table~\ref{tab:gravqnms_slowmodes_n1}. This table shows 
that the $n=1$ fundamental frequencies $\omega_r$ and $\omega_i$ both scale linearly 
with $r_h$. To compare with earlier work, notice that these extracted frequencies for 
the $n=1$ overtone are a close match to the first set of ``fast-modes'' found 
in \cite{Friess:2006kw}, and that the extracted frequencies for the $n=0$ fundamental 
closely match the low-lying ``slow-modes'' found in the same study. 
 
Discrepancies with the linear quasi-normal mode description start to appear in the next 
highest scalar harmonic $\mathbb{S}(400)$ ($k=4$): direct fitting yields a dominant 
frequency that does not match \cite{Friess:2006kw}. The $k=4$ fundamental mode 
with frequency $\omega_4 \approx 2.949 + i3.428/r_h$ is present, but is 
overshadowed by a mode with frequency $\omega^{dbl}_4\approx 3.312 + i1.627/r_h$, 
which is close to double that of the $k=2$ fundamental $\omega_2 \approx 1.652 + i0.826/r_h$ 
that appears in Table~\ref{tab:gravqnms_slowmodes_n0}. We expect that this
frequency-doubling in the $k=4$ modes arises from a non-linear mode-coupling, 
which we attempt to model as a damped harmonic oscillator driven at double 
the frequency of the $k=2$ fundamental mode\footnote{This is reasonable
if one considers the form the field equations take, perturbing about 
a black hole solution to second-order. }. Given a $k=2$ fundamental mode 
$\Psi_2(t)=A_2 \exp\left( -i\omega_2 t \right)$, 
the $k=4$ mode $\Psi_4(t)$ in this simple model satisfies
\begin{equation}\label{eqn:nonlinear_model}
\partial_t^2 \Psi_4 + k_4^2 \Psi_4 
- \lambda_4 \partial_t \Psi_4 = B \exp\left( -2i\omega_2 t \right),
\end{equation}
$\lambda_4 = \omega_{4i}$, $k_4^2 = (\omega_{4r})^2 + (\omega_{4i})^2$ (with
the $r$ and $i$ subscripts denoting the real and imaginary
components of the corresponding number respectively), and
we expect the driving amplitude $B$ to scale as $B \sim {(A_2)}^2$.

\begin{table*}[!t]
\begin{center}
\begin{tabular}{|ll|l|l|l|l|l|l|}
\hline
&$\frac{w_y}{w_x}$  
&$A_4$                       
&$A^{dbl}_4$                     
&$A^{g}_4$      
&$\omega^{g}_{4i}$ 
&$\Delta\varphi$   
&$\frac{A^{dbl}_4}{(A_2)^2} \times 10^3$       \\
\hline
&$1.1$      
&$(1.49 \pm 0.09)\times 10^{-3}$   
&$(6.44\pm 0.02)\times 10^{-3}$  
&$(4.40\pm 0.03)\times 10^{-4}$  
&$(1.23\pm 0.23) \frac{L}{r_h}$ 
&$0.832\pm 0.011$  
&$4.03\pm 0.02$ \\ 

&$4$        
&$2.75 \pm 0.01$                   
&$9.62\pm 0.01$                  
&$0.508\pm 0.030$                
&$(1.26\pm 0.04) \frac{L}{r_h}$ 
&$0.873\pm 0.032$  
&$4.00\pm 0.02$ \\ 

&$32$       
&$(5.39 \pm 0.04)\times 10^1$      
&$(1.16\pm 0.01)\times 10^{2}$   
&$1.59\pm 0.06$                  
&$(1.16\pm 0.01) \frac{L}{r_h}$ 
&$0.894\pm 0.066$  
&$4.60\pm 0.02$ \\ 
\hline
\end{tabular}
\end{center}
\centering
\parbox{6.2in}{\caption{Extracted parameters from the $k=4$ fit corresponding 
to Fig.~\ref{fig:gravqnms_k4} at $w_y/w_x=4$, as well as for the largest and 
smallest $w_y/w_x$ cases considered. 
The fit includes the fundamental $k=4$ mode with amplitude $A_4$,
a mode of amplitude $A^{dbl}_4$ that has twice the frequency
of the $k=2$ fundamental mode of amplitude $A_2$, and a
putative gauge mode $A^{g}_4 \exp(-\omega^{g}_{4i} t)$. The model for the mode coupling
discussed in the text predicts a phase difference between 
the frequency-doubled $k=4$ mode and its $k=2$ fundamental mode source of 
$\Delta\varphi \sim 0.842$, and that
$A^{dbl}_4$ scales as the square $A_2$; these two quantities are shown in
the last two columns of the table, and indicate the solution is reasonably
consistent with the model.
Uncertainties are estimated via convergence.
}\label{tab:nonlinear_modemixing_model}}
\end{table*}

The solution of~(\ref{eqn:nonlinear_model}) is a sum of the $k=4$ fundamental mode and 
the driven mode
\begin{equation}\label{eqn:QNM_fund_k4}
\Psi_4(t) = A_4 \exp\left( -i\omega_4 t \right) + A^{dbl}_4 \exp\left( -2i\omega_2 t \right),
\end{equation}
where $A_4$ is a constant depending on the initial data, and $A^{dbl}_4$
is a (complex) constant depending upon the other parameters of the model via
\begin{equation}\label{dbl_k4_amp}
A^{dbl}_4 = (\tilde{\omega}_r + i\tilde{\omega}_i)^{-1}B,
\end{equation} 
where we have introduced for notational convenience
$\tilde{\omega}_i = 4\omega_{2r}(2 \omega_{2i} - \omega_{4i})$
and 
$\tilde{\omega}_r = (\omega_{4r})^2 + (\omega_{4i})^2 
- 4\left( (\omega_{2r})^2-(\omega_{2i})^2 
+ \omega_{4i}\omega_{2i} \right)$. 

\begin{figure}[h]
	\centering
	\includegraphics[width=4.0in, bb=0 0 520 290]{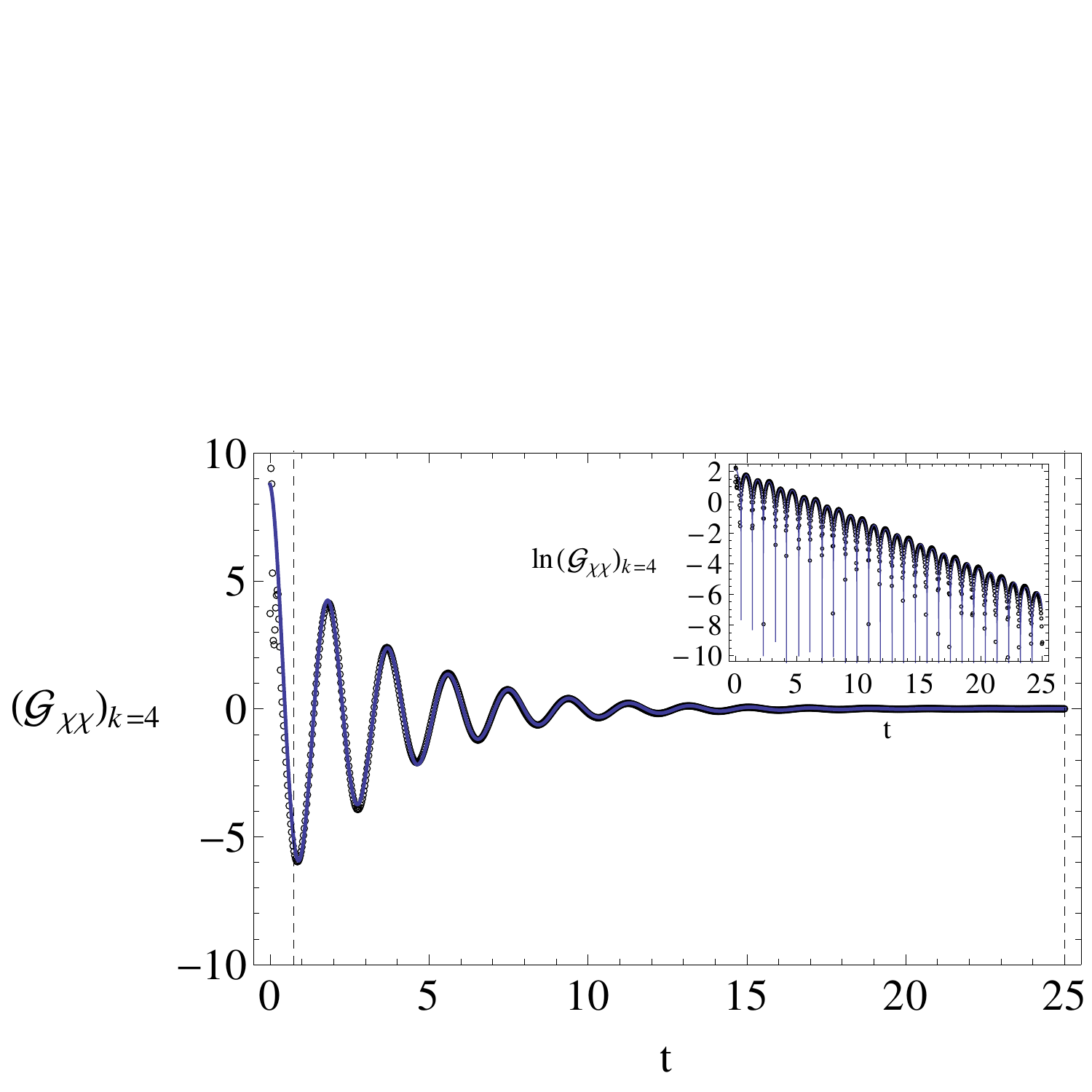}
\parbox{3.3in}{\caption{The leading-order behavior of the metric variable 
$\bar{g}_{\chi\chi}$ near the boundary $q=0$, projected onto the $\mathbb{S}(400)$ 
scalar harmonic by 
$(\mathcal{G}_{\chi\chi})_{k=4}=\int d\Omega_3 (\bar{g}_{\chi\chi}/q) \mathbb{S}(400)$, 
plotted over a global time interval of $t\in[0,8\pi]$ (open circles). A fit (solid line)
$(\mathcal{G}_{\chi\chi})_{k=4}^{fit}$ is extracted using the data inbetween the dashed 
vertical lines. This is a fit to a gauge mode, and two damped sinusoids with fixed frequencies 
$\omega_4 \approx 2.949 + i3.428/r_h$ and 
$2 \ \omega_2 \approx 2 \times ( 1.652 + i 0.826/r_h) $ (see Table ~\ref{tab:nonlinear_modemixing_model}
for the corresponding parameters of the fit).
The inset shows a logarithmic plot of the data and the fit over the full global 
time interval. Other metric variables show similar behavior.
	}\label{fig:gravqnms_k4}}
\end{figure}

Our strategy is to compare the predictions of this simplified model of non-linear 
mode-mixing to the parameters extracted from the fits. In particular
(\ref{dbl_k4_amp}) gives us quantitative predictions for the relative amplitude and
phase difference between the frequency-doubled
$k=4$ mode and the $k=2$ fundamental mode that sources it: that the phase difference
should be $\Delta\varphi = \arctan(\tilde{\omega}_i/\tilde{\omega}_r)$, and that 
$A^{dbl}_4$ should scale as $({A_2})^2$.
To test these predictions, we fit the $k=4$ data to two damped sinusoids simultaneously
(in addition to the decaying gauge mode), though 
{\em fixing} their frequencies to $\omega_4$ and 
$2\omega_2$. We can then check if the extracted amplitude and phase of the frequency doubled
mode matches the model prediction.  The results, presented
in Table~\ref{tab:nonlinear_modemixing_model} for three values of $w_y/w_x$,
and an example fit is shown in 
Fig.~\ref{fig:gravqnms_k4} for the $w_y/w_x=4$ case, show good agreement with the model.

\begin{figure}[h]
        \centering
        \includegraphics[width=4.6in, bb=0 0 500 240]{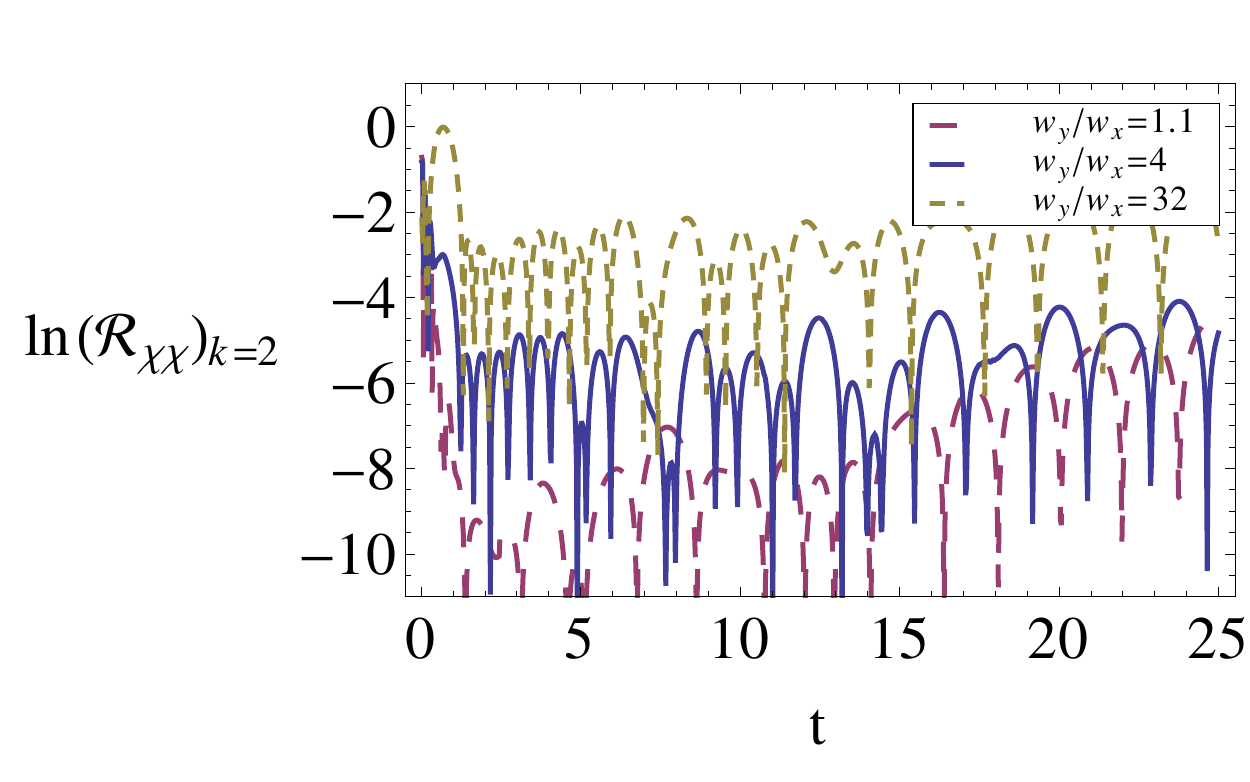}
        \includegraphics[width=4.6in, bb=0 0 500 240]{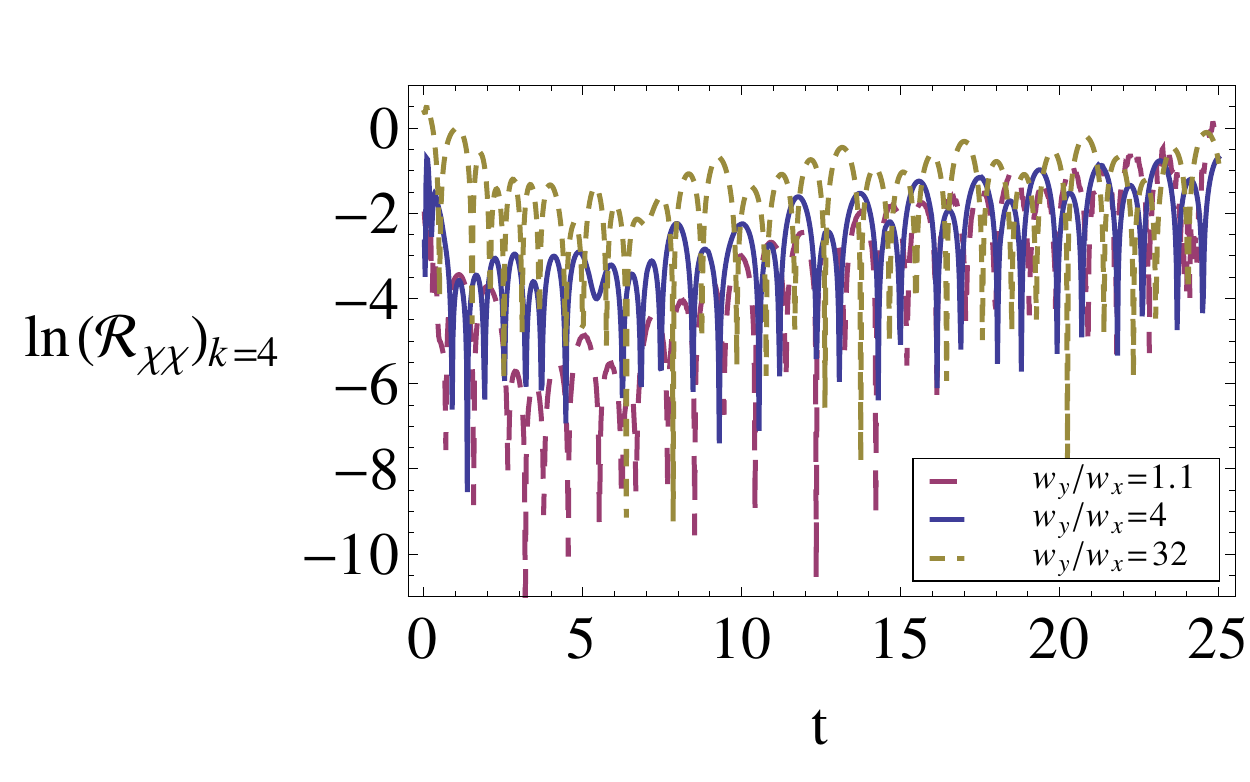}
\parbox{3.3in}{\caption{The normalized residuals for the $k=2$ and $k=4$ fits, corresponding 
to Fig.~\ref{fig:gravqnms_k2} and Fig.~\ref{fig:gravqnms_k4}, respectively, where
$(\mathcal{R}_{\chi\chi})_k = \left( (\mathcal{G}_{\chi\chi})_k 
- (\mathcal{G}_{\chi\chi})_k^{fit} \right) 
/ \left| (\mathcal{G}_{\chi\chi})_k^{fit}) \right|$. This quantity shows the extent 
to which the projection of the metric element $\bar{g}_{\chi\chi}$ onto the 
$\mathbb{S}(k00)$ harmonic {\em fails} to be described by the 
sum of the $k$ quasi-normal mode, gauge mode, and for the $k=4$
case the non-linear mode.
At times before $t \approx 1$, the residual becomes large going to smaller $t$ 
due to a growing phase offset between the 
data and the fit (see Fig.~\ref{fig:gravqnms_k2} for the $k=2$ case, and 
Fig.~\ref{fig:gravqnms_k4} for the $k=4$ case), and at very early times there is an 
additional contribution from the early-time transient. 
At late times there is a slight increase in the residual. 
This is largely a consequence of the fitting procedure and that we are
plotting a normalized residual : the fit is dominated by the data at early times
as the perturbation is decaying exponentially, hence a small phase
difference between the fit and data will result in a normalized
residual that grows with time.
        }\label{fig:qnmres}}
\end{figure}

\begin{figure}[h]
        \centering
        \includegraphics[width=4.6in, bb=0 0 500 240]{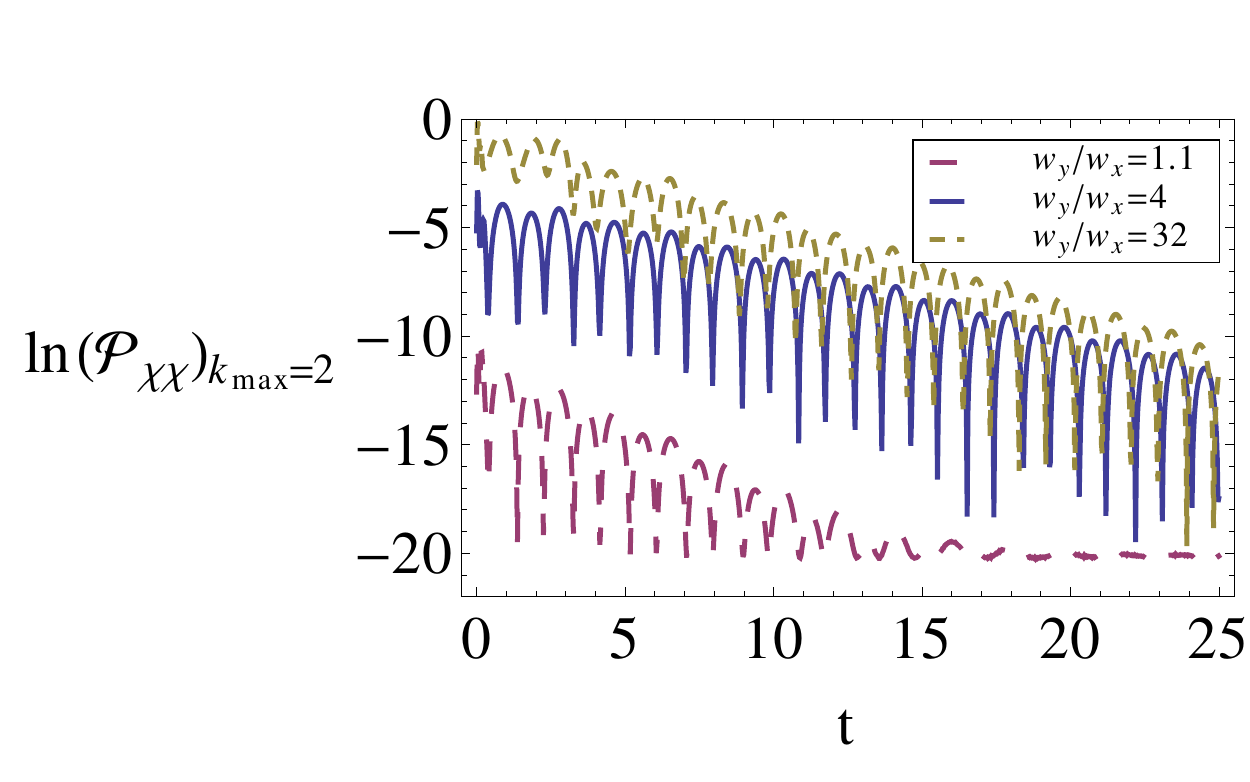}
        \includegraphics[width=4.6in, bb=0 0 500 240]{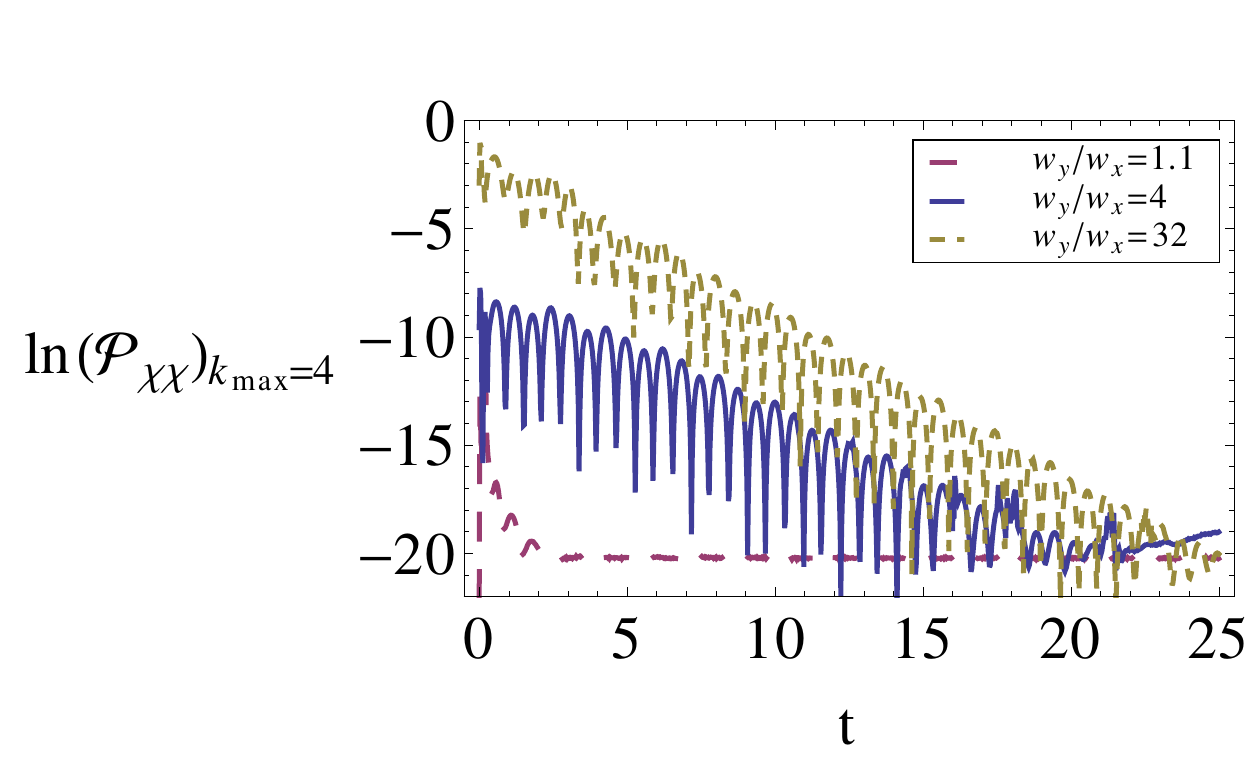}
\parbox{3.3in}{\caption{The normalized difference 
$(\mathcal{P}_{\chi\chi})_{k_{max}} = \left( \int d\Omega_3 (\bar{g}_{\chi\chi}/q)^2 - \sum_{k=0}^{k=k_{max}} (\mathcal{G}_{\chi\chi})_k^2 \right) / \left| \int d\Omega_3 (\bar{g}_{\chi\chi}/q)^2 \right|$. For the 
$w_y/w_x=4$ case, the contribution due to the $k>2$ ($k>4$) modes constitutes $\approx \%2$ ($\%0.1$) of 
the metric perturbation, and decreases with time (the metric modes decay faster with higher $k$). 
        }\label{fig:projres}}
\end{figure}

To quantify precisely how well the fits describe the actual data, we compute 
a residual that measures the difference between the data and the fits. This 
residual quantifies the part of the dynamics that we have not been able to fit 
to with linear quasi-normal modes supplemented by the pure decay gauge mode, 
and in the $k=4$ case, the mode arising from non-linear mode-mixing modeled by~(\ref{eqn:nonlinear_model}). 
The solid blue curve of the first panel of Fig.~\ref{fig:qnmres} depicts a normalized 
residual that is generated from the fit shown in Fig.~\ref{fig:gravqnms_k2}. 
To illustrate the dependence on $w_y/w_x$, this analysis is repeated for the largest 
and smallest $w_y/w_x$ we have considered. The solid blue curve shows that the fit to the 
$k=2$ fundamental quasi-normal mode for $t\gtrsim 1$ captures all but $\approx 1\%$ 
of the perturbation projected onto the $\mathbb{S}(200)$ harmonic, for the $w_y/w_x=4$ case. The 
other curves show that a similar fit does worse for larger $w_y/w_x$, which is consistent 
with the expectation that other unmodeled effects (i.e. those not accounted for by the linear 
quasi-normal modes and our simplified model of the gauge and non-linear effects) should become 
more significant for data with larger deformations. Similarly, the solid blue curve of the 
second panel in Fig.~\ref{fig:qnmres} depicts the normalized residual corresponding to the fit shown 
in Fig.~\ref{fig:gravqnms_k4}. It shows that for $t\gtrsim 1$, the fit to the fixed-frequency 
$k=4$ fundamental mode, the frequency-doubled $k=2$ mode and single gauge mode captures all 
but $\approx 5\%$ of the solution projected onto the $\mathbb{S}(400)$ harmonic
for $w_y/w_x=4$. Again, the corresponding residual grows with larger initial asymmetry.

Finally, to quantify how much of the full solution is not accounted for by the 
$\mathbb{S}(200)$ and $\mathbb{S}(400)$ harmonics that we discussed above, we look at 
the normalized difference between the square of the full solution integrated over the 3-sphere, 
and the sum of the squares of all projections onto the $\mathbb{S}(k00)$ harmonics for all 
$k \le k_{max}$. The result of this procedure is summarized in Fig.~\ref{fig:projres}. 

\subsection{Boundary Stress Tensor}\label{subsec:bdy_stress_tensor}

In terms of the regularized metric variables (\ref{eqn:metric_asymptotics}), 
and setting $L=1$, the stress energy tensor of the boundary CFT~(\ref{eqn:cftsetexpectation}) 
evaluates to 

\begin{eqnarray}\label{eqn:extractedset}
\left< T_{tt} \right>_{\text{CFT}} &=& \frac{1}{64\pi}( -24\bar{g}_{\rho \rho ,
\rho} - 32\bar{g}_{\chi \chi ,\rho} - 64\bar{g}_{\psi ,\rho} ) \nonumber \\
\left< T_{t\chi} \right>_{\text{CFT}} &=& \frac{1}{2\pi} ( -\bar{g}_{t \chi ,
\rho} ) \nonumber \\
\left< T_{\chi\chi} \right>_{\text{CFT}} &=& \frac{1}{64\pi}( -32\bar{g}_{t t ,
\rho} + 24\bar{g}_{\rho \rho ,\rho} + 64\bar{g}_{\psi ,\rho} ) \nonumber \\
\left< T_{\theta\theta} \right>_{\text{CFT}} &=& \frac{\sin^2\chi}{64\pi}( 
-32\bar{g}_{t t ,\rho} + 24\bar{g}_{\rho \rho ,\rho} + 32\bar{g}_{\chi \chi ,
\rho} \nonumber\\ &\ & + 32\bar{g}_{\psi ,\rho} ). \nonumber \\
\end{eqnarray}
and $\left< T_{\phi\phi} \right>_{\text{CFT}} = \sin^2\theta \left< T_{\theta\theta} \right>_{\text{CFT}}$. 
In the remainder of this section we discuss the CFT stress tensor extracted
from a representative numerical solution, namely the $r_h=5$ quasi-normal 
ringdown spacetime described in the previous section. 

\begin{figure}[!]
	\centering
	\includegraphics[width=4.0in, bb=0 0 540 240]{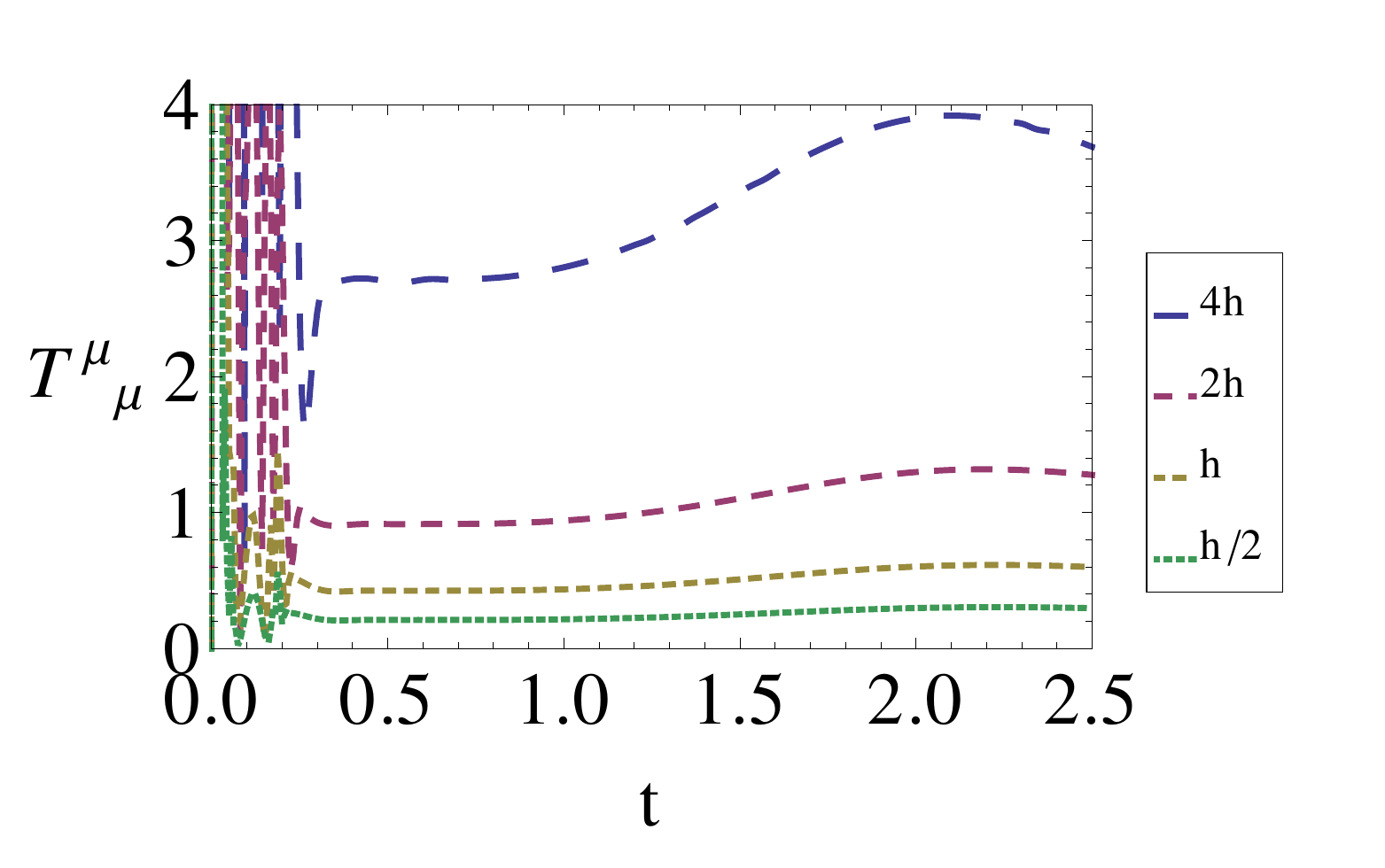}
\parbox{3.3in}{\caption{
The trace $\left< {T^\mu}_\mu \right>_{\text{CFT}}$ of 
the boundary stress tensor, constructed from a simulation run at 4 different 
resolutions, labeled by the mesh spacing relative to the highest resolution
run $h$. Here the $L^2$-norm of $\left< {T^\mu}_\mu \right>_{\text{CFT}}$ is 
taken over the entire grid, and the trends indicate convergence to 
a trace-free stress tensor with increasing resolution.
	}\label{fig:qstrace}}
\end{figure}

\begin{figure}[!]
	\centering
	\includegraphics[width=4.0in, bb=0 0 540 240]{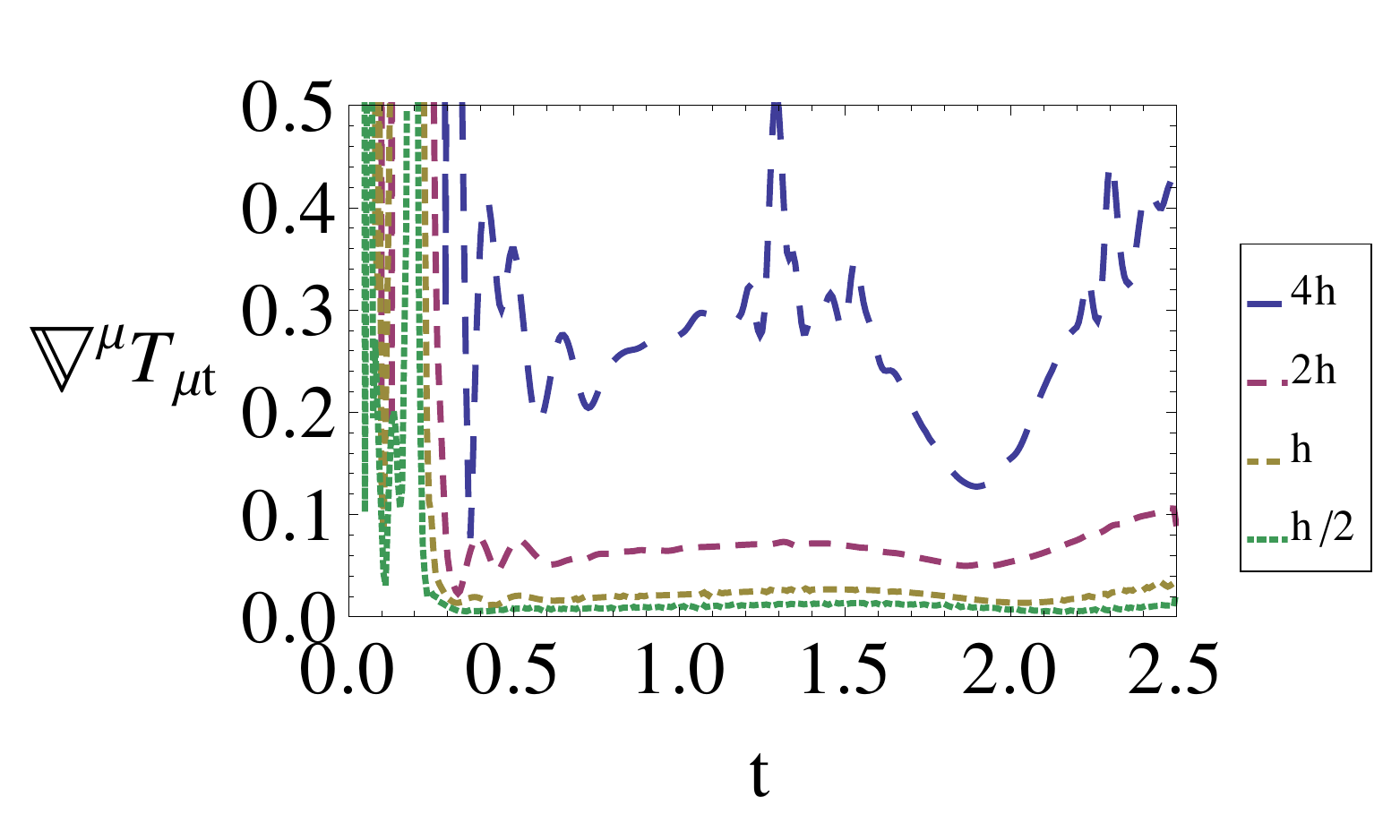}
	\includegraphics[width=4.0in, bb=0 0 540 240]{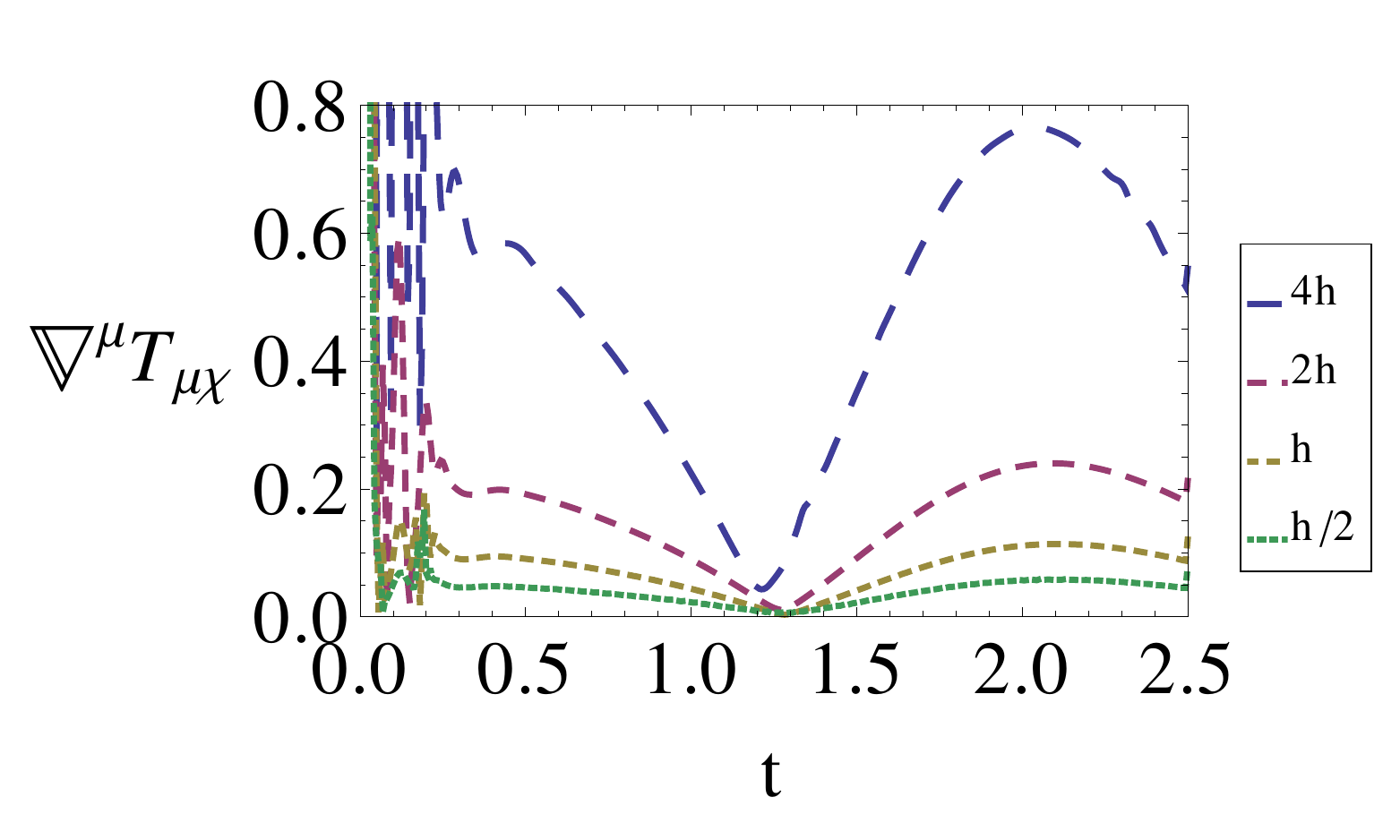}
\parbox{3.3in}{\caption{Two components of the divergence of the
stress tensor, $\nabla^\mu \left< T_{\mu t} \right>_{\text{CFT}}$ (top)
and $\nabla^\mu \left< T_{\mu \chi} \right>_{\text{CFT}}$ (bottom), constructed 
from a simulation run at 4 different resolutions, labeled by the mesh spacing 
relative to the highest resolution run $h/2$. The $L^2$-norm of the respective components
are taken over the entire grid, and the trends indicate convergence to
a conserved stress tensor with increasing resolution. 
	}\label{fig:qsdivt_qsdivy}}
\end{figure}

First, for a consistency check of expressions (\ref{eqn:extractedset}), we test 
whether (to within truncation error) the stress tensor is traceless and whether 
it is conserved with respect to the Levi-Civita connection on the 
$\mathbb{R} \times S^3$ boundary. Results for the trace and two non-trivial 
components of the divergence are displayed in Fig.~\ref{fig:qstrace} 
and~\ref{fig:qsdivt_qsdivy}, respectively. These plots demonstrate 
that as resolution is increased, we are indeed converging to a CFT stress tensor 
that is conserved and traceless, i.e. to matter that obeys the hydrodynamic 
equations of motion, and whose equation of state is consistent with conformal 
invariance. Note that the early-time transient related to 
the initial gauge dynamics discussed in Sec.~\ref{sec:conv_tests} is also 
visible in these plots, though as with the independent residual it also 
converges away in the sense that it occupies a smaller region of the spacetime 
domain with increased resolution.

\begin{figure}[!]
	\centering
	\includegraphics[width=2.4in]{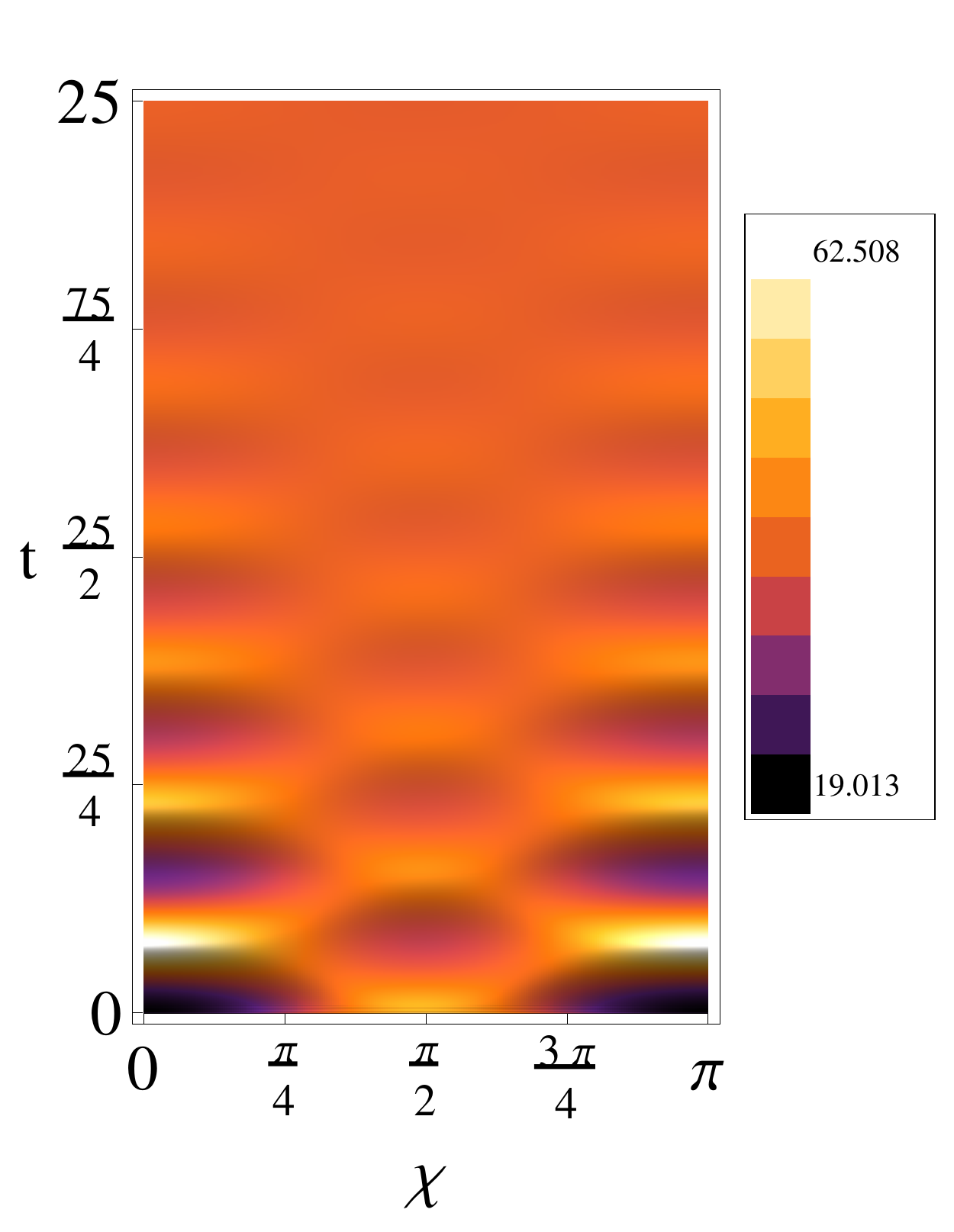}
	\includegraphics[width=2.4in]{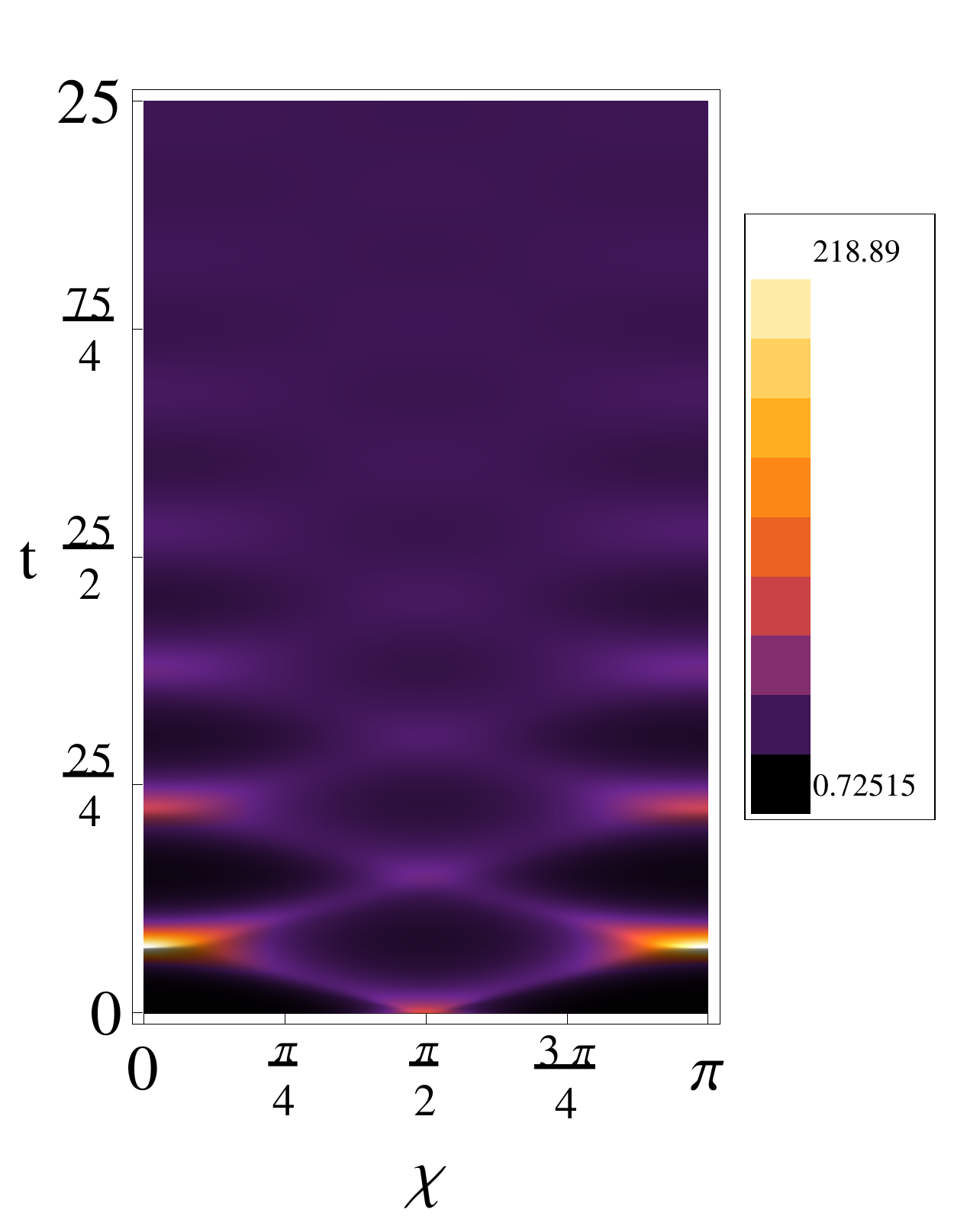}
\parbox{3.3in}{\caption{The energy density $\left< T_{tt} \right>_{\text{CFT}}$ 
of the boundary CFT, displayed on a ($t,\chi$) spacetime diagram, extracted from the 
$r_h=5$ quasi-normal black hole ringdown simulation described in Sec.~\ref{sec:qnr}. 
The first panel corresponds to the $w_y/w_x=4$ case, and the $w_y/w_x=32$ case in the 
second panel is included for comparison.
	}\label{fig:rh05_qstt}}
\end{figure}

\begin{figure}[!]
	\centering
	\includegraphics[width=2.4in]{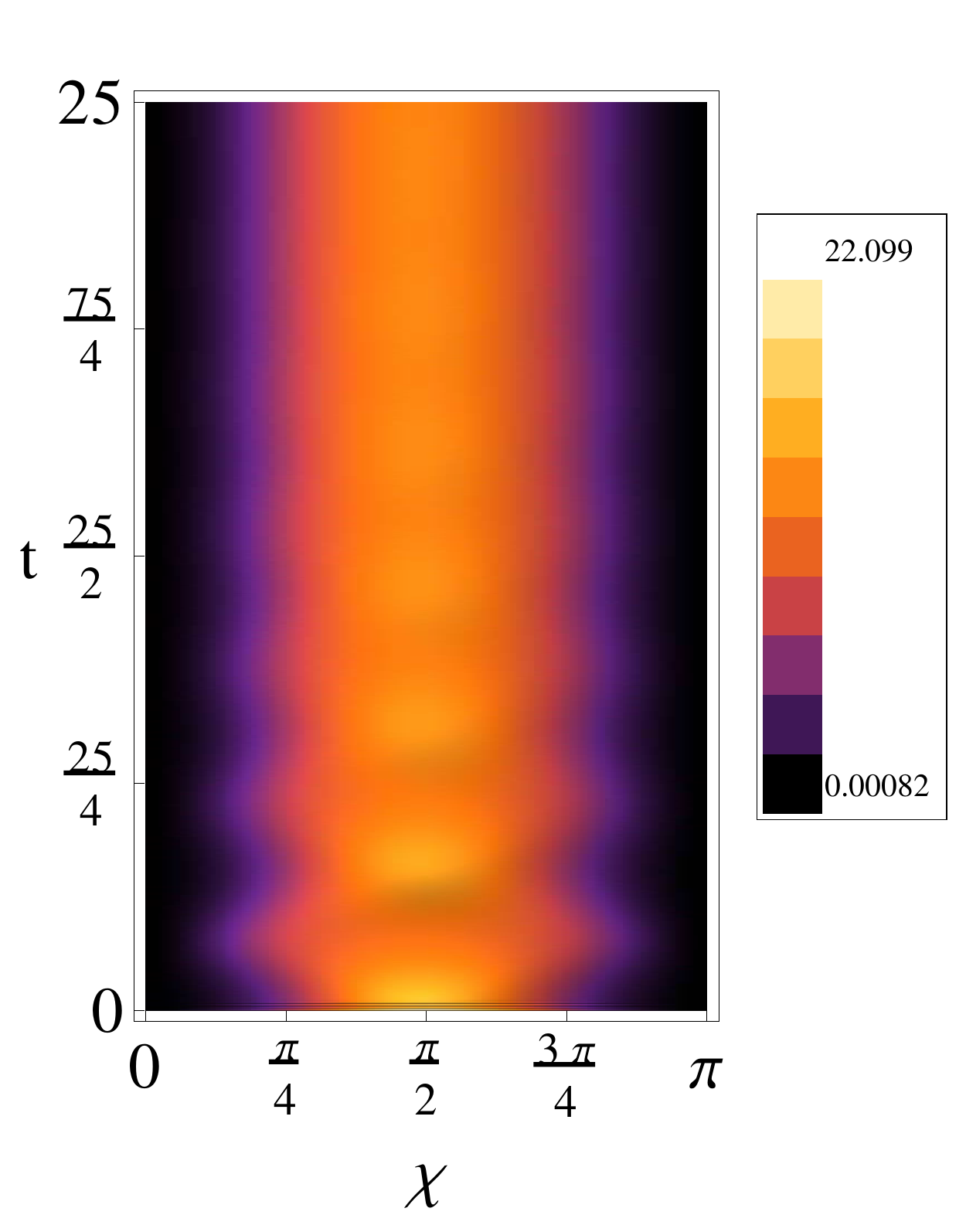}
	\includegraphics[width=2.4in]{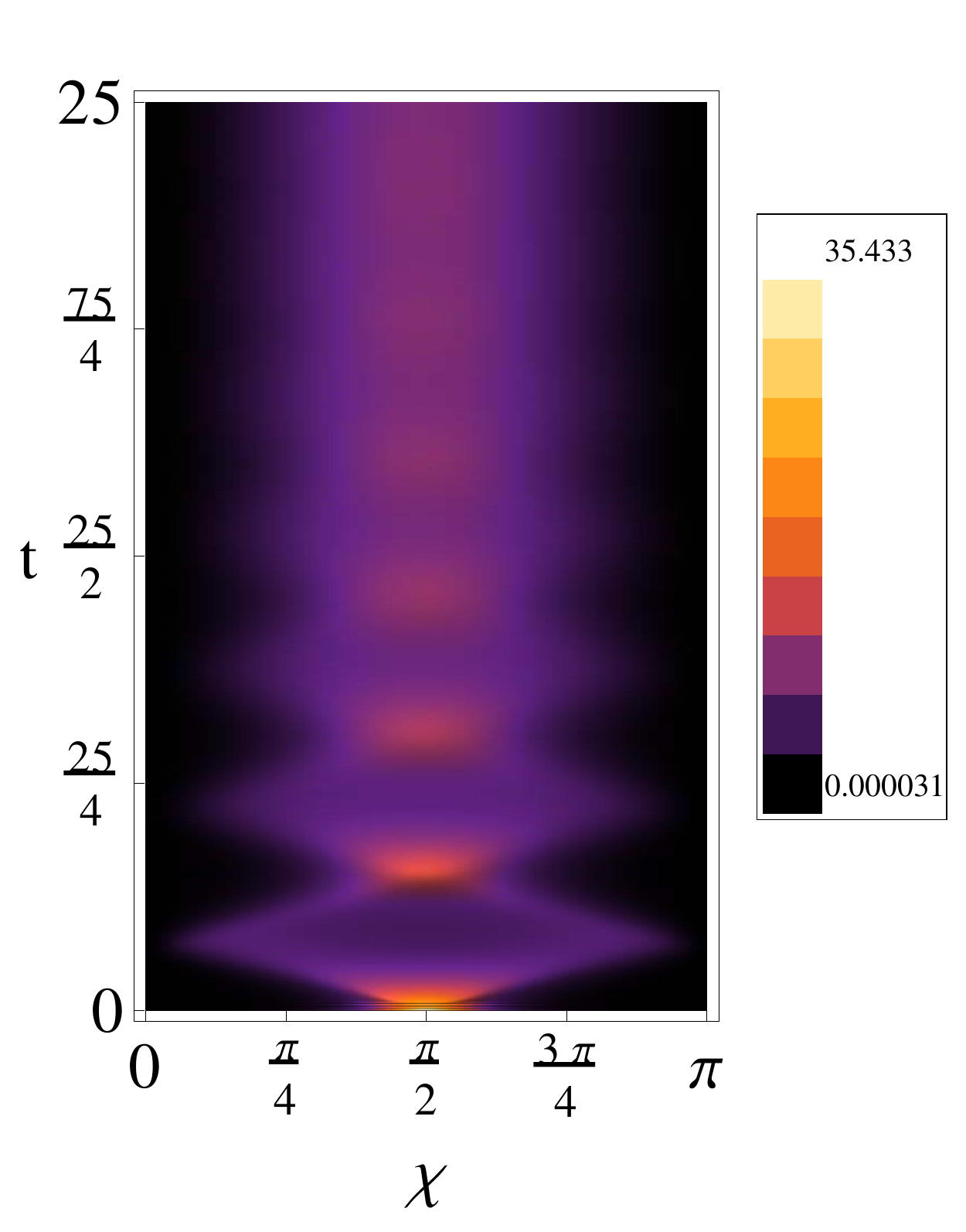}
\parbox{3.3in}{\caption{The $S^2$ component 
$\left< T_{\theta\theta} \right>_{\text{CFT}}$ of the boundary CFT stress tensor, 
displayed on a ($t,\chi$) spacetime diagram, extracted from the $r_h=5$  
quasi-normal black hole ringdown simulation described in Sec.~\ref{sec:qnr}. 
The first panel corresponds to the $w_y/w_x=4$ case, and the $w_y/w_x=32$ case in the 
second panel is included for comparison.
	}\label{fig:rh05_qspsi}}
\end{figure}

In Figs.~\ref{fig:rh05_qstt} and~\ref{fig:rh05_qspsi} we display representative 
components of the boundary CFT stress tensor on $\mathbb{R} \times S^3$ 
spacetime diagrams, specifically the energy density 
$\left< T_{tt} \right>_{\text{CFT}}$ and $S^2$ component of the pressure
$\left< T_{\theta\theta} \right>_{\text{CFT}}$ respectively. At $t=0$ the boundary 
state is clearly inhomogeneous, and in time evolves to a homogeneous state in 
a manner that mirrors the quasi-normal decay of the spacetime to a 
static black hole.

\subsection{Hydrodynamic Description of the Boundary CFT}\label{section:hyd_desc_of_the_bdy_cft}

Essentially all known physical systems at sufficiently high temperature, 
including those described by quantum field theories, exhibit hydrodynamic behavior 
once local thermodynamic equilibrium has been attained. 
In the gauge/gravity duality, stationary black holes are dual to equilibrium
thermal states, and studies have shown that perturbations of the bulk
spacetime manifest as hydrodynamic behavior in the boundary CFT. 
See~\cite{Hubeny:2011hd} for a recent review, and~\cite{Emparan:2011br} for a 
review of the ``blackfolds'' approach, which is similar in spirit but connects 
the dynamics of a perturbed black brane with that of an enclosing world-volume
via a derivative expansion of the Einstein field equations. Here, we have 
studied the evolution of initially highly distorted black holes, far from the 
perturbative regime, and so the question naturally arises: at what time does 
hydrodynamics become a good description of the boundary state? More precisely, 
we would like to know whether the extracted 
$\left< T_{\mu \nu} \right>_{\text{CFT}}$ on the $\mathbb{R} \times S^3$ 
boundary is that of the particular kind of fluid predicted by the duality, 
namely an $\mathcal{N}=4$ SYM fluid with equation of state $\epsilon=3P$ (where $\epsilon$ 
and $P$ are the energy density and isotropic pressure in the rest frame of the 
fluid, respectively), and transport coefficients that match those found 
via holographic methods~\cite{Loganayagam:2008is}.

In the previous section, we showed that convergence trends in the solution imply that 
in the continuum limit, $\left< T_{\mu \nu} \right>_{\text{CFT}}$ is conserved and
traceless. Thus, two necessary ingredients are already satisfied: conservation 
is required for the effective hydrodynamic variables to satisfy the Navier-Stokes 
equations, and for an isotropic fluid, tracelessness implies $\epsilon=3P$. In this 
section we show that one {\em can} map $\left< T_{\mu \nu} \right>_{\text{CFT}}$ to a 
corresponding set of hydrodynamic variables, and that essentially from the 
{\em initial time}, their behavior {\em is} consistent with that of 
an $\mathcal{N}=4$ SYM fluid, at least to within truncation error.
%  and modulo our ignorance of the higher-order transport coefficients of the fluid.
% {\bf HB: changed text, reflecting our new hydro extraction method}

In a low-energy effective description, the hydrodynamic stress tensor can be 
expressed as a velocity-gradient expansion, i.e. as a power series in the 
covariant gradients of a local fluid $4$-velocity one form $u_\nu$:
\begin{equation}\label{eqn:hydroset}
T_{\mu \nu} = (\epsilon + P) u_\mu u_\nu + P g_{\mu \nu} -2 \eta \sigma_{\mu \nu} + \Pi_{\mu \nu},
\end{equation}
where we have introduced the (symmetric, traceless) shear tensor 
\begin{equation}\label{eqn:shear}
\sigma^{\mu \nu} = \perp^{\mu \alpha} \perp^{\nu \beta} \nabla_{(\alpha} u_{\beta)} - \frac{1}{d-1} \nabla_\alpha u^\alpha \perp^{\mu \nu},
\end{equation}
and subsumed all higher-order terms under $\Pi_{\mu \nu}$. 
All covariant differentiation is performed with respect to the boundary metric 
$g_{\mu \nu} dx^\mu dx^\nu = -dt^2 + d\chi^2+\sin^2\chi(d\theta^2+\sin^2\theta d\phi^2)$,
and $\perp^{\mu \nu}$ in (\ref{eqn:shear}) is  
the projector onto the spatial slices orthogonal to the fluid $4$-velocity 
\begin{equation}\label{eqn:projector}
\perp^{\mu \nu} = g^{\mu \nu} + u^\mu u^\nu.
\end{equation}
In this sub-section, we will for now ignore the higher-order terms, i.e. we set 
$\Pi_{\mu \nu}=0$. 
In terms of mapping from stress tensor to hydrodynamic variables this should be a 
good approximation except in situations where the 
shear $\sigma_{\mu \nu}$ of the flow becomes small, which does occur 
periodically during the evolution of the distorted black holes discussed here.  
These higher order terms would lead to additional problems regarding the uniqueness 
of the mapping that we will now attempt. In the sub-section following this one, we will employ 
a different strategy and will instead extract a minimal subset, then, assuming an 
$\mathcal{N}=4$ SYM fluid, test for consistency with higher order transport coefficients 
(instead of trying to directly measure all these quantities, as we will now proceed to do).

Let us first identify a set of independent fluid variables which we can use to 
characterize the stress tensor. This set certainly includes $\epsilon$ and $P$, 
the energy density and isotropic pressure in the rest frame of the fluid, respectively. 
Given the $SO(3)$ symmetry of our solutions, the (unit) 4-velocity vector must 
take the form $u^\mu = \gamma (1,v, 0, 0)$ in $(t,\chi,\theta,\phi)$
coordinates, with $\gamma=1/\sqrt{1-v^2}$. This gives us a third fluid variable 
to add to our list: $v$, the $\chi$-coordinate velocity of the flow. As for the shear 
$\sigma_{\mu \nu}$, notice that it is symmetric, traceless and satisfies
the identity $u^\mu \sigma_{\mu \nu} = 0$. Together with the imposed $SO(3)$ 
symmetry, these imply that $\sigma_{\mu \nu}$ only has one degree of freedom in 
$d=4$ dimensions. Thus defining $\sigma_{\chi \chi} = \sigma$, one can 
straightforwardly show that the only other non-zero components of 
$\sigma_{\mu \nu}$ are:
\begin{equation}\label{eqn:shear_constraints}
\sigma_{t \chi} = -v \sigma, \hspace{+0.4cm} \sigma_{t t} = v^2 \sigma, \hspace{+0.4cm} \sigma_{\theta \theta} = \frac{\sigma_{\phi \phi}}{\sin^2\theta} = -\frac{\sin^2\chi}{2 \gamma^2} \sigma.
\end{equation}
Given how $\eta$ and 
$\sigma$ always appear as a product in (\ref{eqn:hydroset}), we 
treat the two as a single quantity $\eta\sigma$ for the purposes of extraction. 
Thus, ignoring the higher-order terms $\Pi_{\mu\nu}$, the variables 
$(\epsilon,P,v,\eta\sigma)$, each of which is a function of $(t,\chi)$ in general,
completely describes a conformal fluid flow on $\mathbb{R} \times S^3$ that 
preserves our $SO(3)$ symmetry.

On the gravity side, the quantities we measure are the components of the boundary stress 
tensor, which we denote $T_{\mu \nu} \equiv \left< T_{\mu \nu} \right>_{\text{CFT}}$ henceforth 
for the sake of brevity. In $SO(3)$ symmetry, there are $5$ non-zero stress tensor components, 
$4$ of which are independent. We can relate these $4$ to the fluid variables via (\ref{eqn:hydroset}):
\begin{eqnarray}
T_{tt}           &=& (\epsilon + P) \frac{1}{1-v^2} - P - 2 \eta\sigma v^2 \nonumber \\
T_{t\chi}        &=& -(\epsilon + P) \frac{v}{1-v^2} + 2 \eta\sigma v \nonumber \\
T_{\chi\chi}     &=& (\epsilon + P) \frac{v^2}{1-v^2} + P - 2 \eta\sigma \nonumber \\
T_{\theta\theta} &=& \sin^2\chi \left( P + \eta\sigma (1-v^2) \right)
\end{eqnarray}
where $T_{\phi\phi} = \sin^2\theta T_{\theta\theta}$. Inverting these relations in 
favor of the rest-frame hydrodynamic quantities $(\epsilon,P,v,\eta\sigma)$, and 
defining the auxiliary quantities
\begin{eqnarray}\label{eqn:hydro_aux_var}
\Xi      &\equiv& \sqrt{(T_{\chi\chi} + 2 T_{t\chi}+T_{tt}) (T_{\chi\chi} - 2 T_{t\chi}+T_{tt})} \nonumber \\
T_{\psi} &\equiv& \frac{T_{\theta\theta}}{\sin^2\chi},
\end{eqnarray}
we obtain
\begin{eqnarray}\label{eqn:hydroextraction}
\epsilon &=& \frac{1}{2} \left(T_{tt} -T_{\chi\chi} +\Xi \right) \nonumber \\
P &=& \frac{1}{6} \left(\Xi+T_{\chi\chi} + 4 T_{\psi} -T_{tt}\right) \nonumber \\
v &=& \frac{\Xi-T_{\chi\chi}-T_{tt}}{2 T_{t\chi}} \nonumber \\
\eta\sigma &=& \left[ -T_{\chi\chi}^3+T_{\chi\chi}^2 \left(-\Xi + T_{\psi} -2 T_{tt}\right) \right. 
\nonumber \\
&& + T_{\chi\chi} \left( T_{\psi} \left(\Xi+2T_{tt}\right) \right. 
\left. -T_{tt} \left(\Xi+T_{tt}\right) - 4 T_{t\chi}^2\right) 
\nonumber \\
&& + T_{\psi} \left(T_{tt} \left(\Xi +T_{tt}\right) + 4 T_{t\chi}^2 \right) 
\left. -2 T_{t\chi}^2 \Xi \right] / \left[ 6 \Xi \right],
\nonumber \\
\end{eqnarray}

\begin{figure}[!]
        \centering
        \includegraphics[width=2.4in]{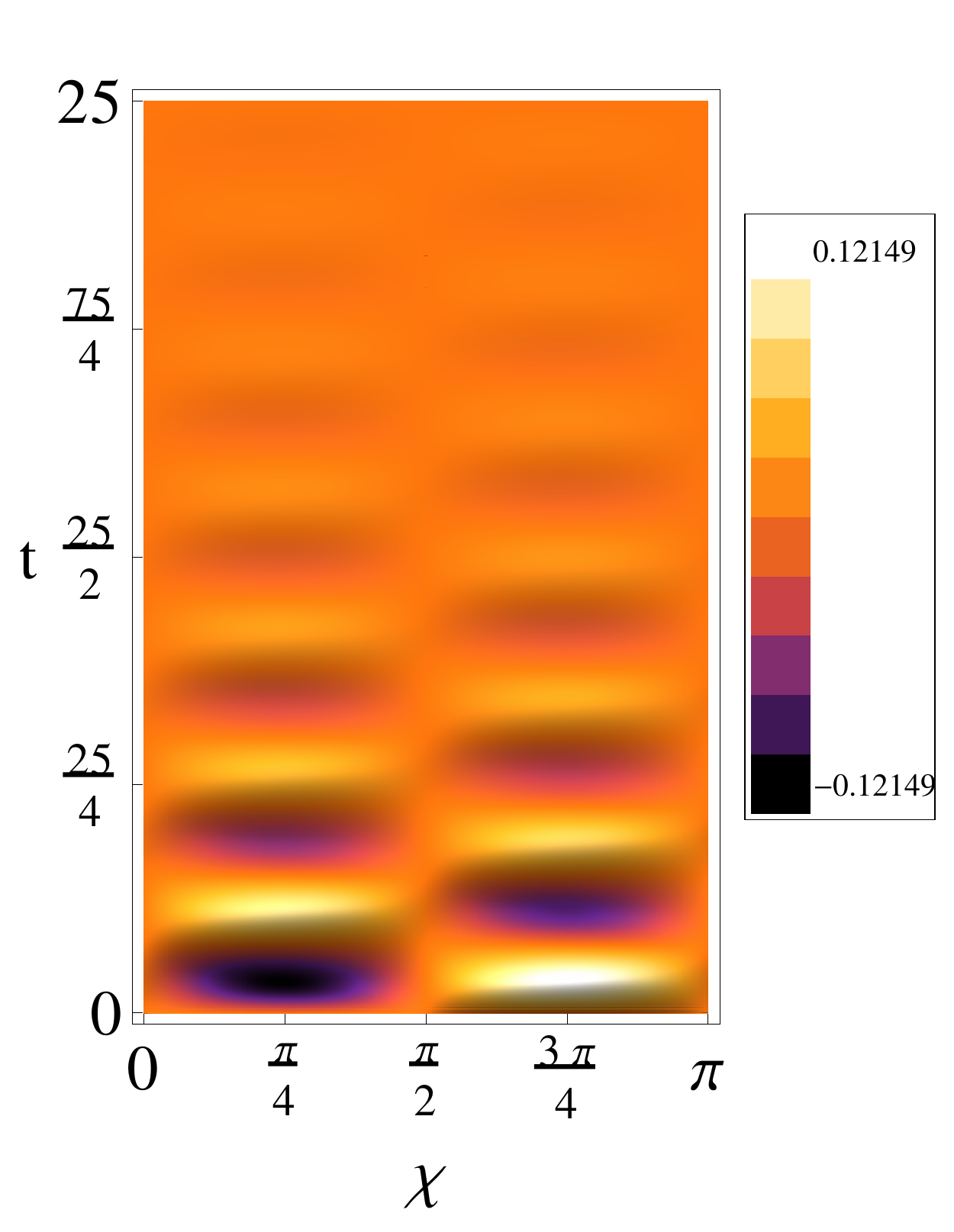}
        \includegraphics[width=2.4in]{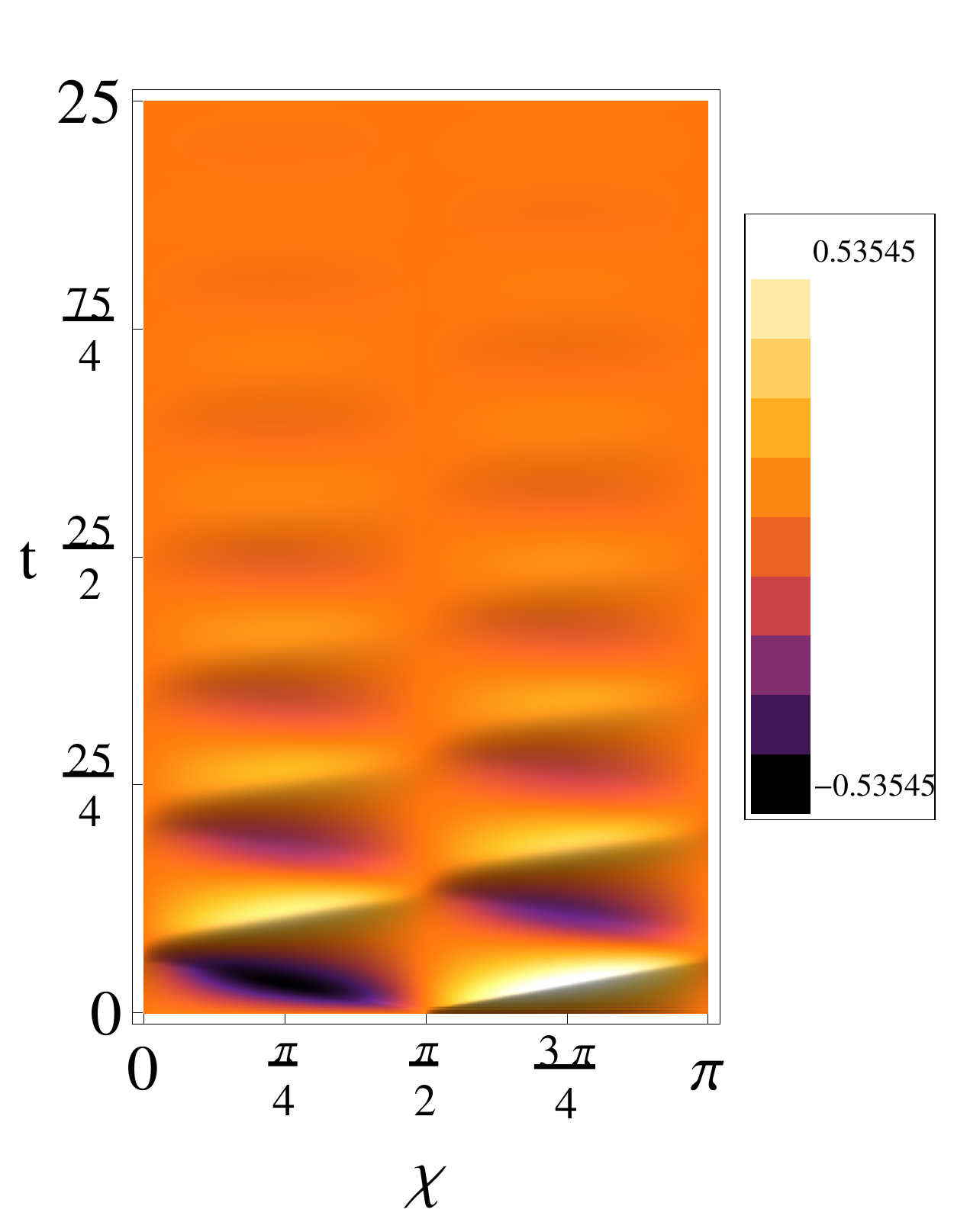}
\parbox{3.3in}{\caption{The velocity field $v$ of the fluid describing the boundary CFT, 
displayed on a $t,\chi$ spacetime diagram, extracted from 
the $r_h=5$ quasi-normal black hole ringdown simulation described in Sec.~\ref{sec:qnr}. 
The first panel corresponds to the $w_y/w_x=4$ case, and the $w_y/w_x=32$ case in the 
second panel is included for comparison.
        }\label{fig:rh05_qsv}}
\end{figure}

By itself, (\ref{eqn:hydroextraction}) is not profound: the stress-tensor
has 4 independent components, and we have simply mapped them to a new set
of variables ($\epsilon,P,v,\eta\sigma$). The crucial question though is whether 
these variables behave as those of a conformal fluid. The answer appears to be 
{\em yes}, even for these initially highly distorted black 
holes\footnote{A few of the explicit examples shown here correspond to a ``moderately'' 
distorted initial black hole, with $r_h=5$ and 
$c_{eq}/c_p(t=0) \approx 1.2$---see Fig.~\ref{fig:ceqcp}, though the same 
conclusions hold for the most distorted cases we have looked at to date 
($c_{eq}/c_p(t=0) \approx 2.2$), where the fluid velocity reaches a maximum 
$|v| \approx 0.54$---see Fig.~\ref{fig:rh05_qsv}.}.

\begin{figure}[!]
        \centering
        \includegraphics[width=3.6in, bb=0 0 540 240]{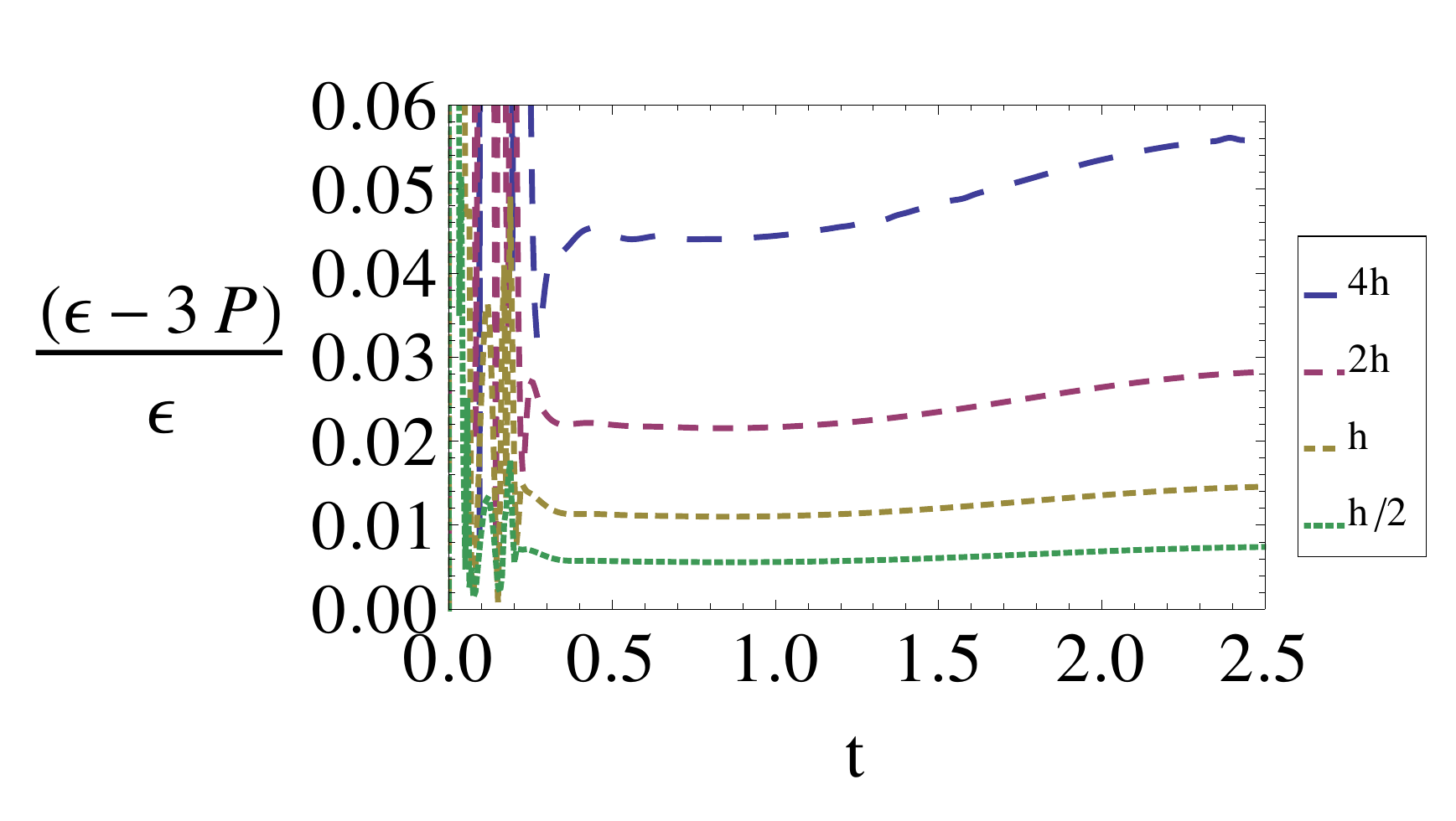}
        \includegraphics[width=3.6in, bb=0 0 540 240]{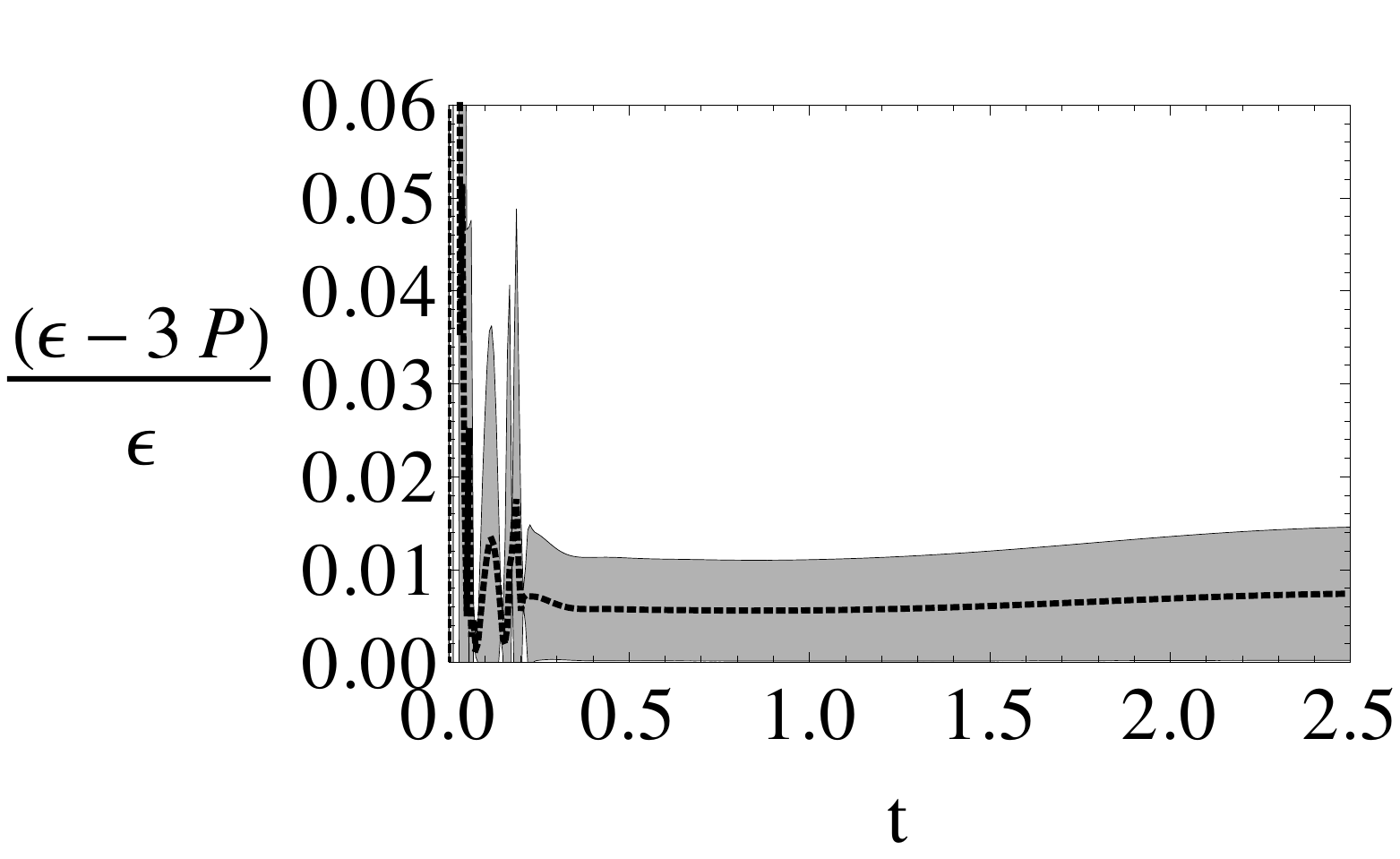}
\parbox{3.3in}{\caption{The quantity $(\epsilon-3P)/\epsilon$, which measures the extent 
to which the hydrodynamic description is that of a fluid with equation of state
$\epsilon=3P$. The first panel is constructed from simulations run at 4 different
resolutions, labeled by the mesh spacing relative to the highest resolution
run $h/2$. What is shown at each time is the $L^2$-norm of $(\epsilon-3P)/\epsilon$ 
taken over the entire grid. Except possibly for an early transient 
(see Fig.~\ref{fig:rh05_hyd_earlytime}), that this quantity is converging to zero 
shows that $\epsilon=3P$ to within truncation error.
In the second panel we interpret the finest resolution
data as the continuum solution to within an uncertainty (shaded region),
obtained from the magnitude of the difference between the highest and
next to highest resolution data (this assumes first order convergence
of the extracted quantities).
        }\label{fig:rh05_hyd}}
\end{figure}

\begin{figure}[!]
        \centering
        \includegraphics[width=3.6in, bb=0 0 540 240]{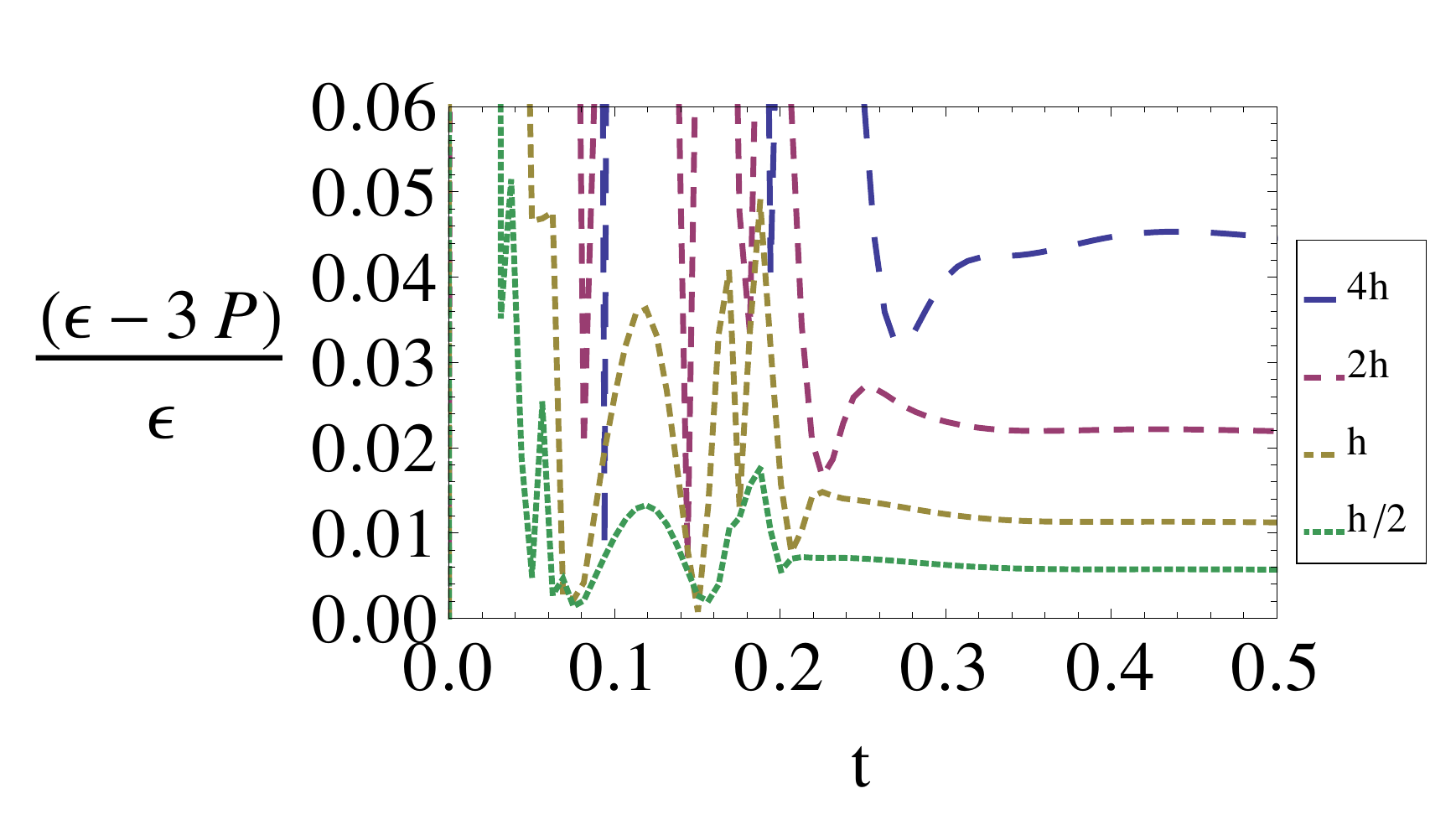}
        \includegraphics[width=3.6in, bb=0 0 540 240]{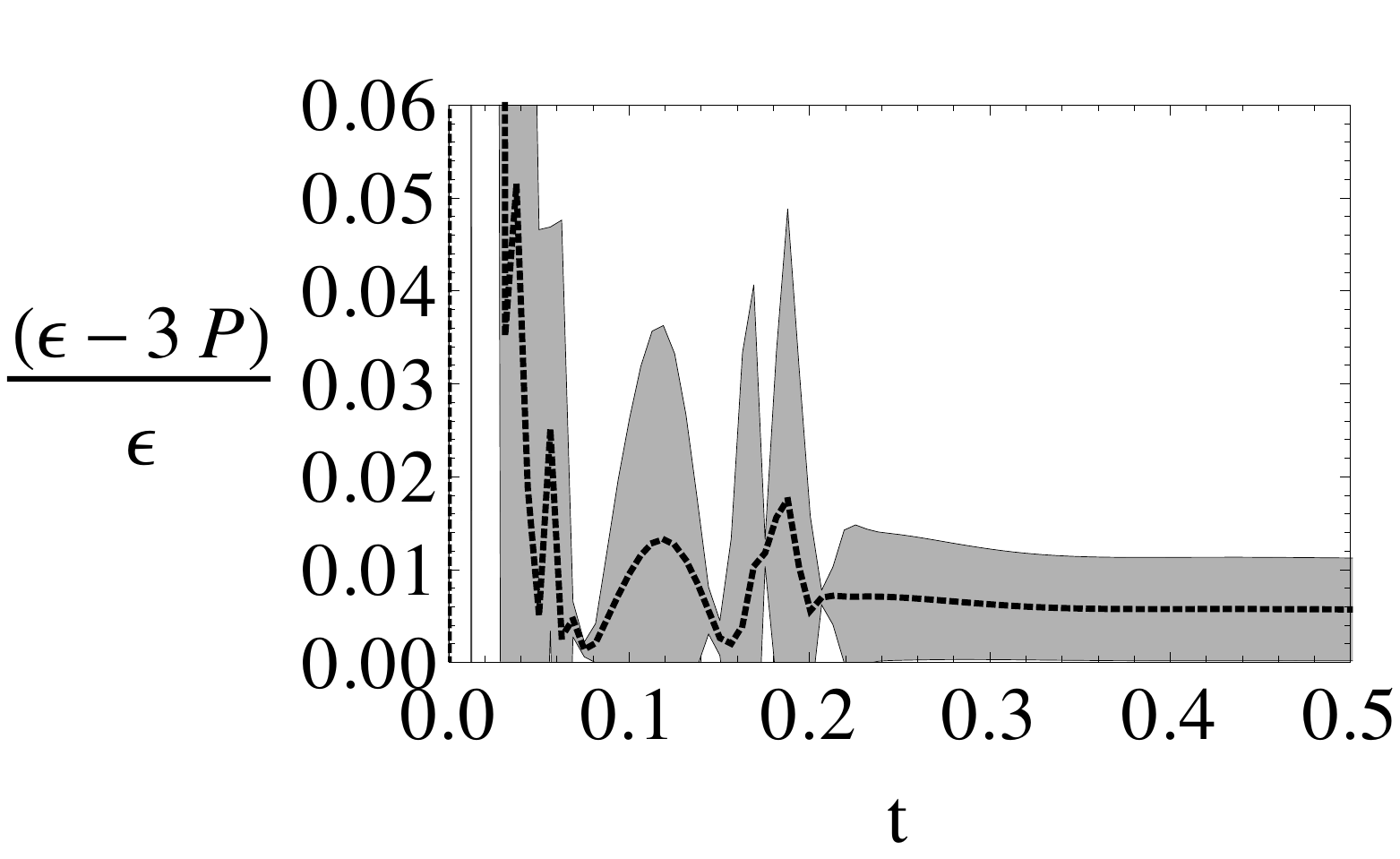}
\parbox{3.3in}{\caption{An expanded view of Fig.~\ref{fig:rh05_hyd} at early times. 
        }\label{fig:rh05_hyd_earlytime}}
\end{figure}

To support this claim, we first note that in all cases we have looked
at to date, we are able to perform the inversion (\ref{eqn:hydroextraction})
to obtain real-valued hydrodynamic variables that satisfy 
$\epsilon \ge 0$, $P \ge 0$ and $v \in (-1,1)$. Fig. \ref{fig:rh05_qsv} shows
an example of $v(t,\chi)$, from the $r_h=5$ quasi-normal ringdown case
discussed before. This suggests the stress tensor is consistent
with that of {\em some} fluid. To check that it is a conformal
fluid, in the top panel of Fig.~\ref{fig:rh05_hyd} we show the quantity $(\epsilon-3P)/\epsilon$ 
resulting from the evolution at several different resolutions, using the same 
initial data (to contrast with Fig.~\ref{fig:qstrace}, here we are
not merely testing that the stress tensor is traceless, but rather the more
restrictive property that it arises from fluid flow where in the rest frame
of the flow the pressure is isotropic). At any given resolution, this quantity is certainly nonzero 
due to truncation error; however, except for a transient at the initial time, 
the trends show that it is {\em converging to zero}, at or better than
first order\footnote{That the rate of convergence is closer
to first order is due to how we extrapolate the solution to the boundary
to extract the stress tensor components---the underlying solution for
the metric still shows second-order convergence as discussed in Sec.~\ref{sec:conv_tests}.}.
In other words, this says that after the initial transient, the boundary stress
tensor behaves like a fluid with equation of state $\epsilon=3P$ {\em to within
numerical truncation error}. Fig.~\ref{fig:rh05_hyd_earlytime} is a close-up
of the early transient behavior seen in Fig.~\ref{fig:rh05_hyd}, and as
before for the convergence of the solution discussed in Sec.~\ref{sec:conv_tests}, 
it appears to be converging away in the sense that it affects 
a smaller region in time as resolution increases.

An alternative way to present this information is displayed in 
the bottom panels of Fig.'s~\ref{fig:rh05_hyd} and ~\ref{fig:rh05_hyd_earlytime}.
Here, assuming first order convergence of the extracted quantities,
we take the finest resolution result as the continuum solution with
an uncertainty (from truncation error) given by the magnitude of the difference between
the finest and next-to-finest resolution results; this error is shown as the shaded
region about the finest resolution curve. 
For subsequent time-series plots we will display the
data in this format.

From this data we might also try to extract the lowest-order transport coefficient, namely
the shear viscosity $\eta$, but here we encounter the two primary limitations of this sub-section's 
extraction method. First, this method only allows us to extract the product $\eta\sigma$, so in order to 
find $\eta$ we must divide by an independent calculation of $\sigma$. Such a calculation may be 
achieved by substituting the extracted velocity field into (\ref{eqn:shear}), 
which yields $\sigma|_v$, so we can compute $\eta\sigma/\sigma|_v \approx \eta$.
Given our time-symmetric initial data, this $\sigma|_v$ has zeros at $t=0$, 
and periodically after that. Even though the ``true'' $\eta\sigma$ 
necessarily has zeros at exactly the same times, the extracted $\eta\sigma$ will not, 
because we have ignored the $\Pi_{\mu\nu}$ higher-order contributions in the map. 
Thus, such a calculation 
of $\eta\sigma / \sigma|_v \approx \eta$ has the unattractive feature of diverging whenever 
$\sigma|_v=0$. Secondly, and more crucially, this method does not readily extend to allow the 
extraction of the higher-order transport coefficients. To remedy these shortcomings, we will 
now make use of a more general method to analyze the hydrodynamic behavior of the CFT 
$T_{\mu \nu}$. 

\subsection{Higher-order Hydrodynamics}\label{section:higher_order_hydrodynamics}

In this section, we present an alternative method for comparing 
$T_{\mu \nu} \equiv \left< T_{\mu \nu} \right>_{\text{CFT}}$
with the stress tensor of an $\mathcal{N}=4$ SYM fluid. To accomplish this, let us first add as many 
higher-order terms $\Pi_{\mu \nu}$ to the hydrodynamic stress tensor (\ref{eqn:hydroset}) as is presently 
known, which includes viscous corrections up to second-order in the gradient expansion. This gives 
four additional higher-order transport coefficients to supplement the shear viscosity $\eta$: the stress 
relaxation time $\tau_\pi$, the shear vorticity coupling $\tau_\omega$, the shear-shear coupling 
$\xi_\sigma$, and the Weyl curvature coupling $\xi_C$. Our strategy here will be different from 
the one described in the previous sub-section, where we had first extracted all hydrodynamic 
variables from the CFT stress tensor components, then exhibited evidence that the conformal 
constitutive relations hold. Instead, we will now {\em assume} that the conformal constitutive 
relations hold at the outset, allowing us to reconstruct the hydrodynamic stress tensor order by 
order using only the energy density $\epsilon$ and the fluid four-velocity $u^\mu$. 

\begin{figure}[!]
        \centering
        \includegraphics[width=3.6in, bb=0 0 540 240]{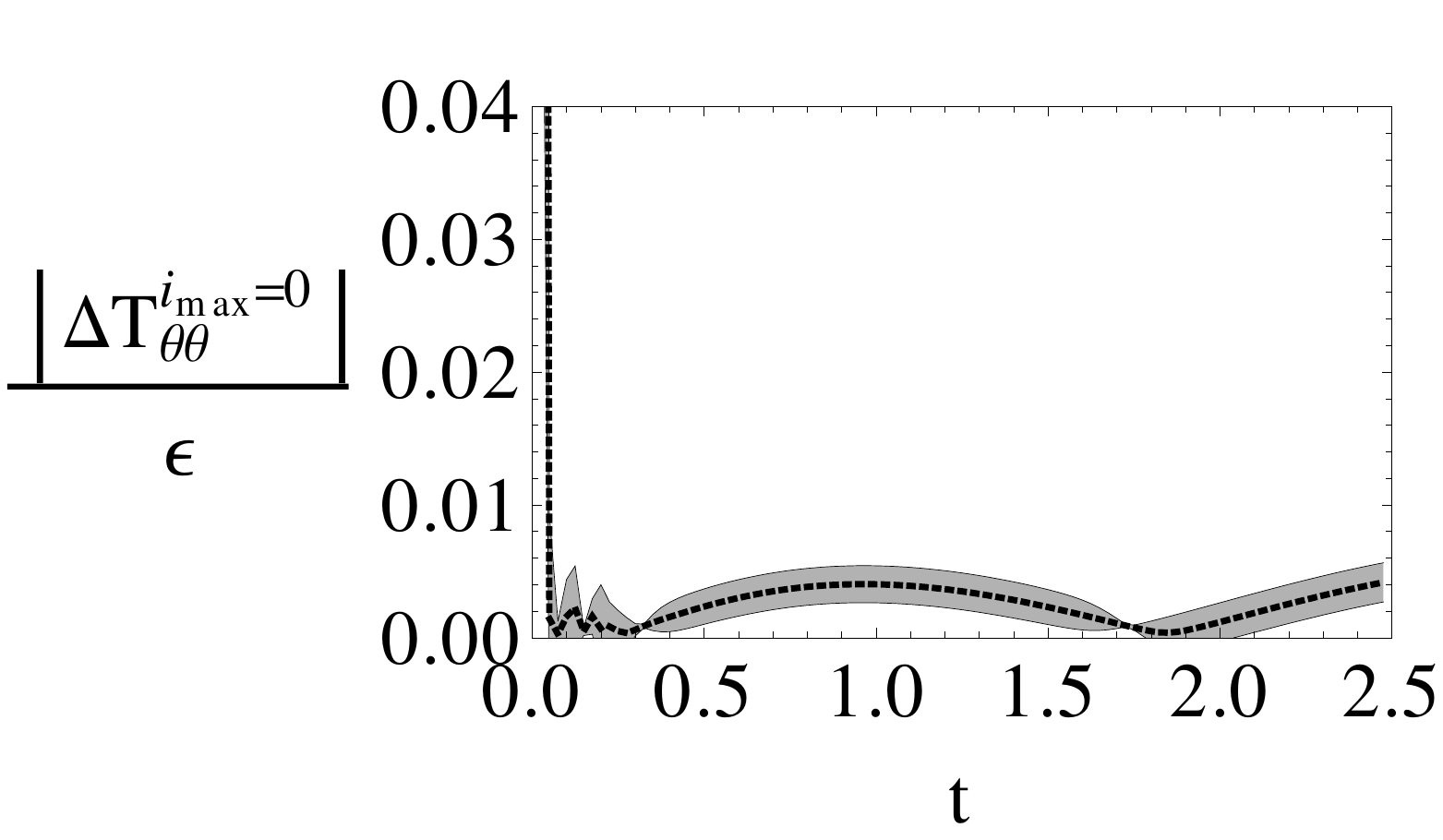}
        \includegraphics[width=3.6in, bb=0 0 540 240]{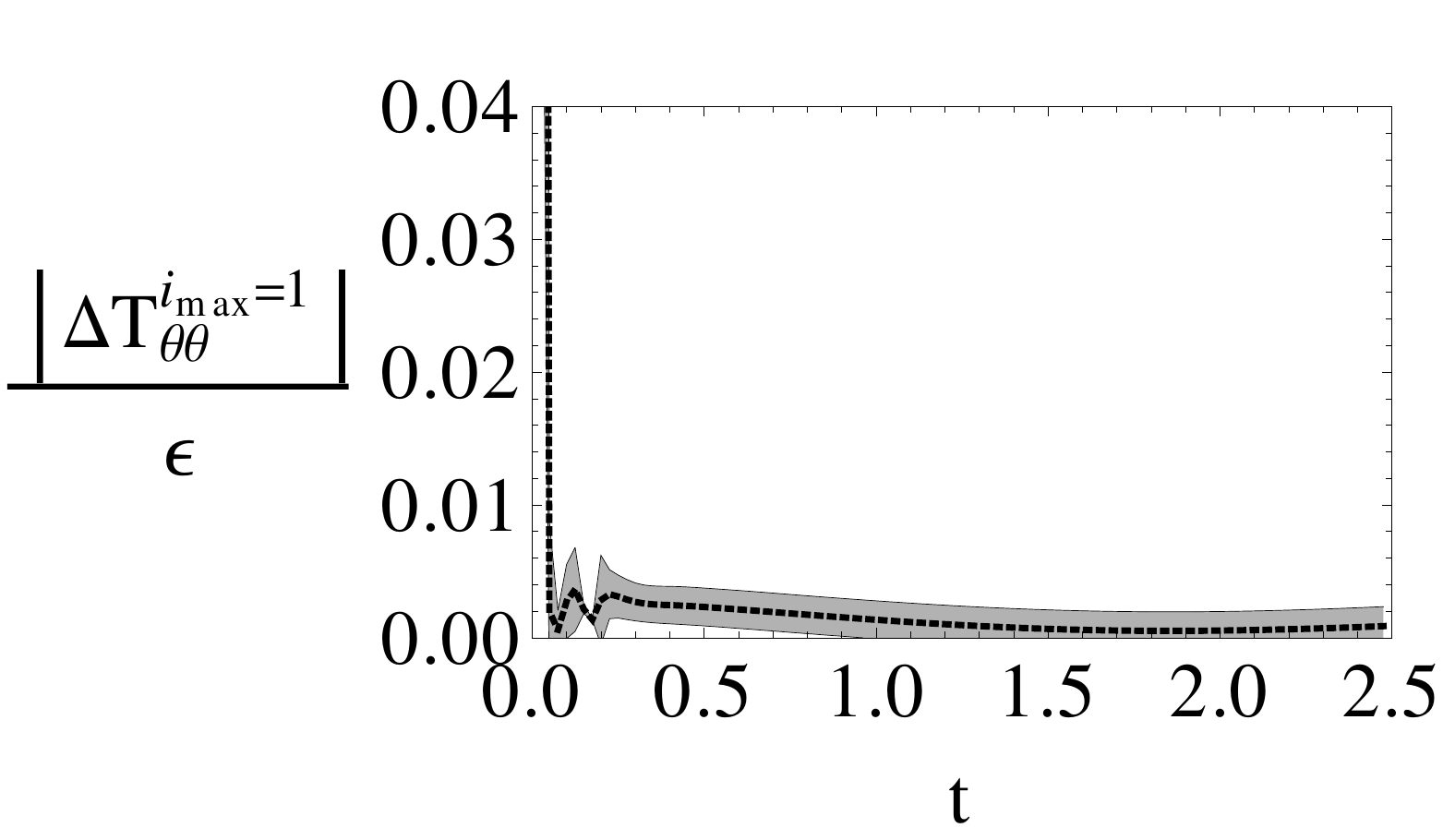}
        \includegraphics[width=3.6in, bb=0 0 540 240]{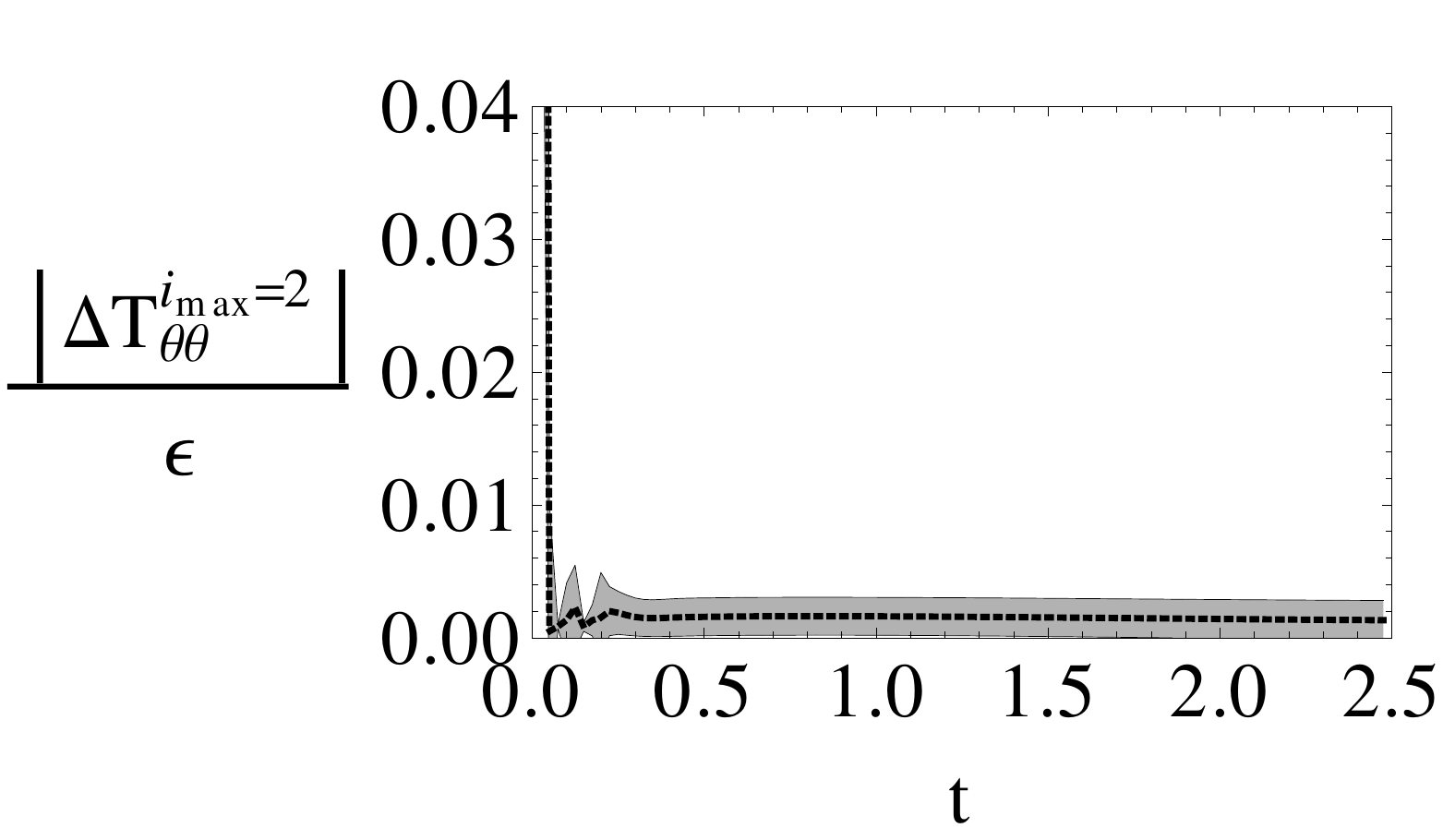}
\parbox{3.3in}{\caption{ The mismatch $\Delta T^{i_{max}}_{\theta\theta}$ between the CFT 
stress tensor and the reconstructed hydrodynamic stress tensor with viscous corrections up to order 
$i_{max}=0,1,2$, for a simulation run with $w_y/w_x=4$ initial data, whose final state 
black hole has horizon radius $r_h=5$. Here, 
$\Delta T^{i_{max}}_{\theta\theta} \equiv T_{\theta\theta} - \sum_{i=0}^{i_{max}} T_{\theta\theta}^{(i)}$, and the plots are normalized by the fluid energy density $\epsilon$. 
As discuss in the text and Fig.~\ref{fig:rh05_hyd}, what is shown
in each case is the data from the highest resolution run (dashed line) together with
an estimated uncertainty (shaded region).
Ideal hydrodynamics (top) shows a small residual mismatch; this is almost
gone including first order viscous corrections (middle), and zero to within truncation error
when second order corrections are added (bottom). 
        }\label{fig:viscous_medium}}
\end{figure}

\begin{figure}[!]
        \centering
        \includegraphics[width=3.6in, bb=0 0 540 240]{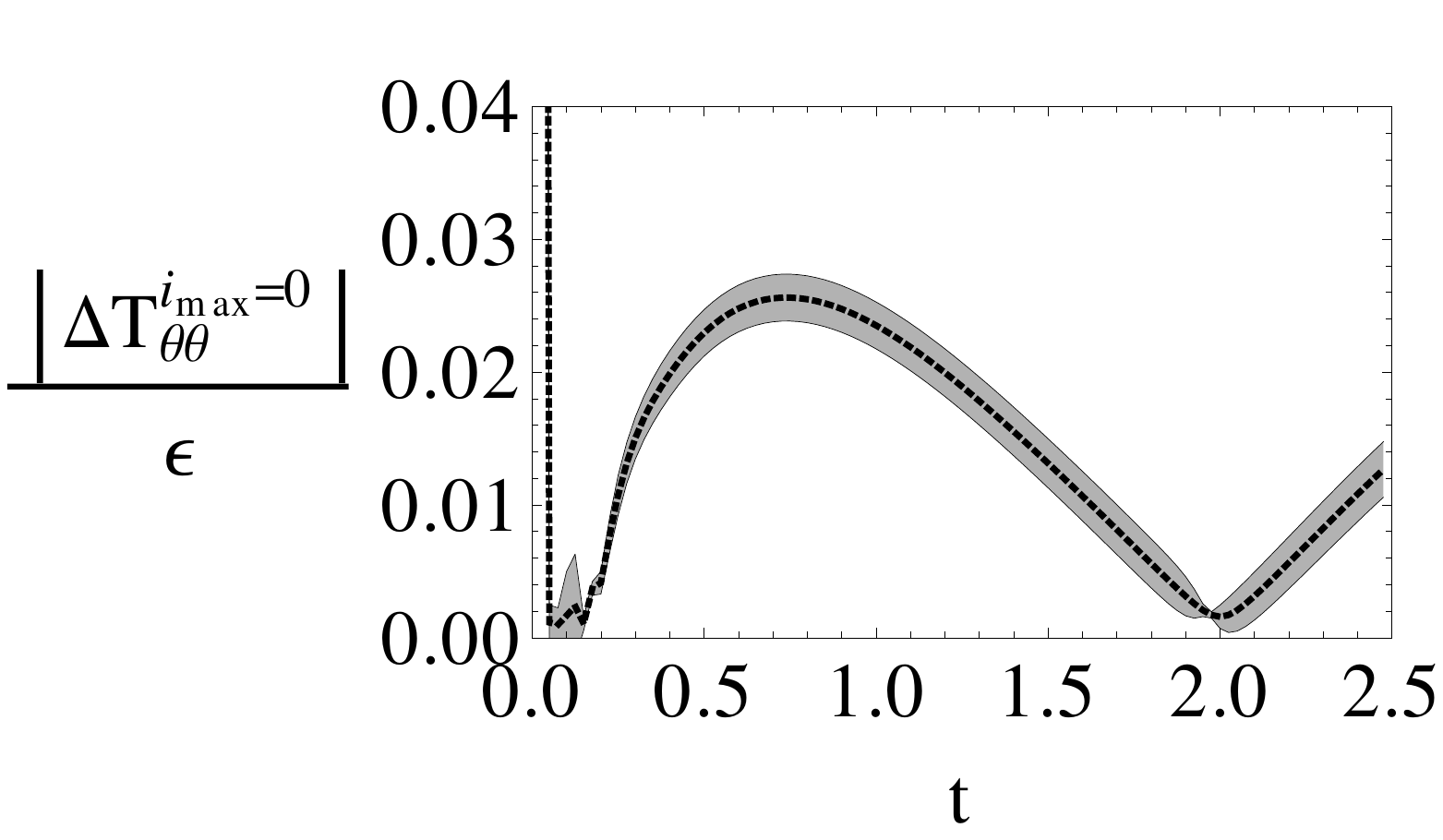}
        \includegraphics[width=3.6in, bb=0 0 540 240]{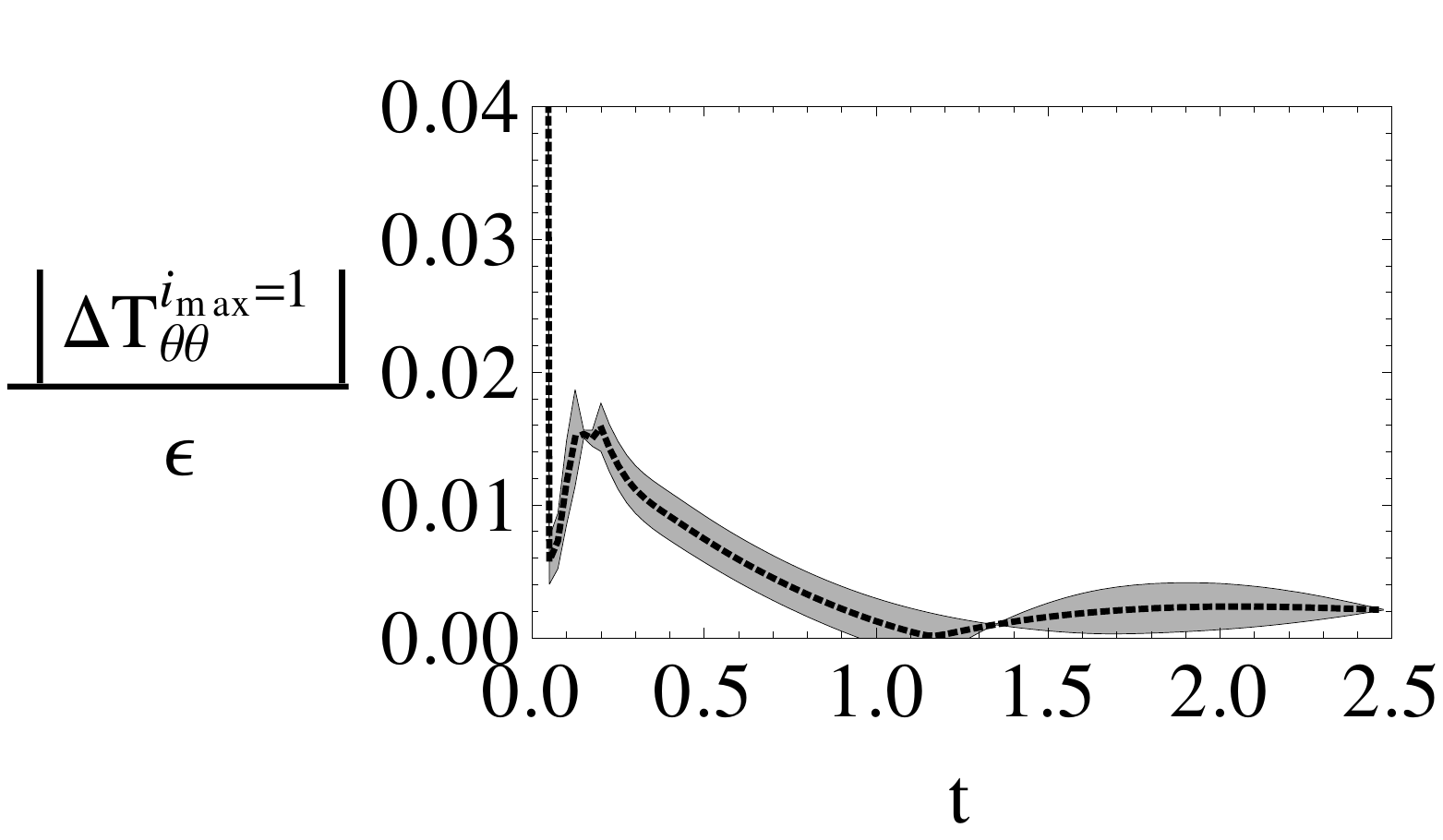}
        \includegraphics[width=3.6in, bb=0 0 540 240]{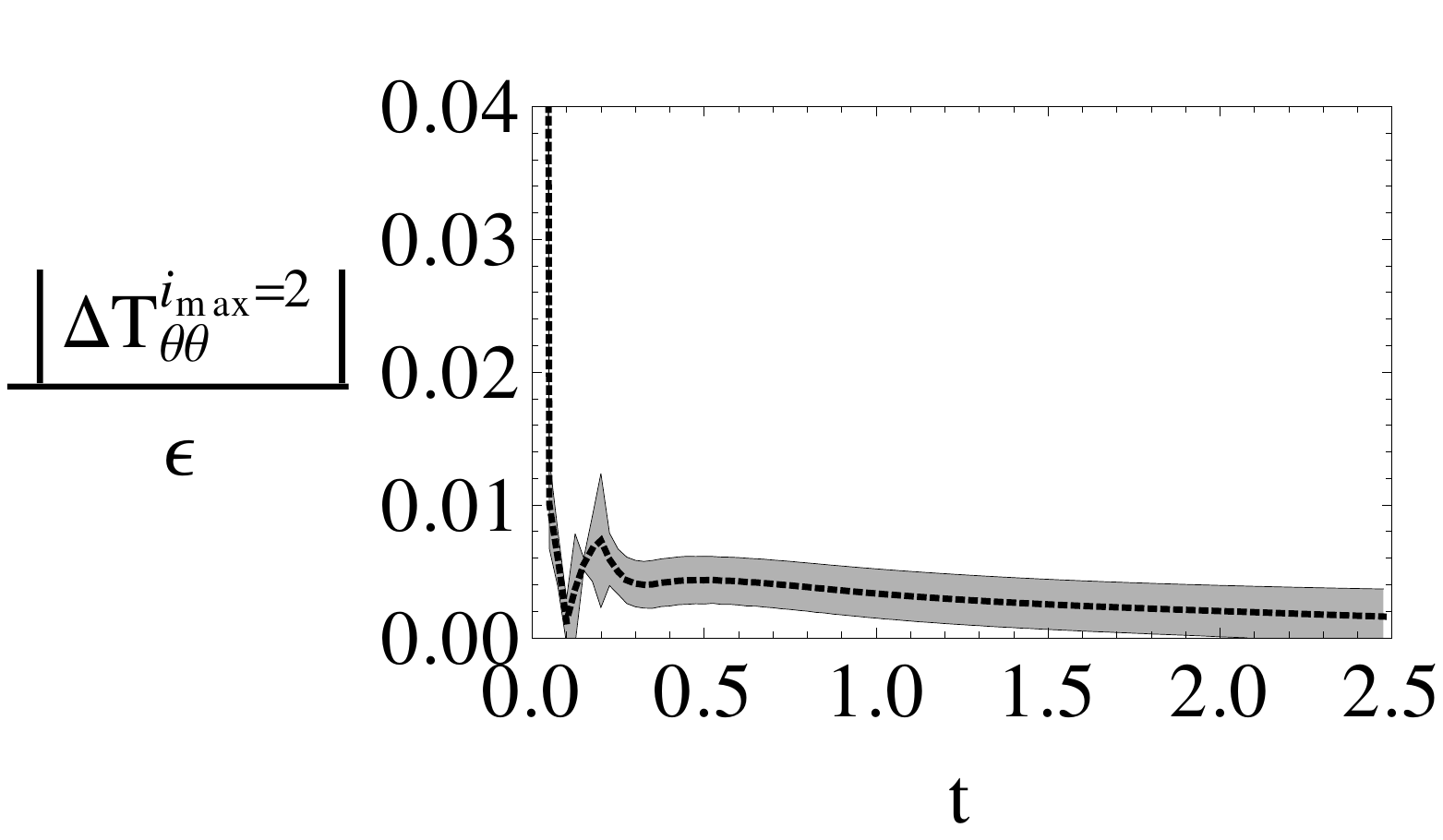}
\parbox{3.3in}{\caption{ The same set of plots as described in Fig.~\ref{fig:viscous_medium},
though here from the simulation of the $w_y/w_x=32$ initial data.
Notice the larger relative mismatches at a given order of the expansion 
compared to the $w_y/w_x=4$ case in Fig.~\ref{fig:viscous_medium},
and that there is still a small residual even after including all corrections through 
second order. The trend going from ideal (top) to second order viscous hydrodynamics (bottom) 
suggests that third and higher order corrections could further reduce the mismatch.
        }\label{fig:viscous_medium_wywx32}}
\end{figure}

This reconstruction takes the form
\begin{equation}\label{eqn:hydroset_full}
T_{\mu \nu} = \sum_{i=0}^{\infty} T_{\mu\nu}^{(i)}
\end{equation}
where $T_{\mu\nu}^{(0)}$ corresponds to the stress tensor of an ideal relativistic fluid, 
and $T_{\mu\nu}^{(i)}$ accounts for the $i^{th}$-order correction in a
velocity-gradient expansion. These corrections are explicitly given in~\cite{Loganayagam:2008is} 
for an $\mathcal{N}=4$ SYM fluid:
\begin{eqnarray}\label{eqn:hydroset_orderbyorder}
T_{\mu\nu}^{(0)} &=& \epsilon u_\mu u_\nu + P \perp_{\mu\nu} \nonumber \\
T_{\mu\nu}^{(1)} &=& -2 \eta \sigma_{\mu \nu} \nonumber \\
T_{\mu\nu}^{(2)} &=& -2 \eta \left[ -\tau_\pi u^\lambda \mathcal{D}_\lambda \sigma_{\mu\nu} 
                     + \tau_\omega \left( {\omega_\mu}^\lambda \sigma_{\lambda \nu} 
                     + {\omega_\nu}^\lambda \sigma_{\lambda \mu} \right) \right] \nonumber \\ 
                 & & + \xi_\sigma \left[ {\sigma_\mu}^\lambda \sigma_{\lambda \nu} 
                     - \frac{\perp_{\mu\nu}}{3} \sigma^{\alpha\beta} \sigma_{\alpha\beta} \right] 
                     + \xi_C C_{\mu\alpha\nu\beta} u^\alpha u^\beta \nonumber \\ 
\end{eqnarray}
where $C_{\mu \alpha \nu \beta}$ is the Weyl tensor and $\mathcal{D}$ is the Weyl covariant 
derivative defined in \cite{Loganayagam:2008is}. One further input are the constitutive relations 
for our $\mathcal{N}=4$ SYM fluid:
\begin{eqnarray}\label{eqn:constitutive_relations}
\epsilon    &=& \frac{3 {N_c}^2}{8 \pi^2} \left( \pi T \right)^4 = 3 P \nonumber \\
\eta        &=& \frac{{N_c}^2}{8 \pi^2} \left( \pi T \right)^3 \nonumber \\
\tau_\pi    &=& \frac{2-\ln 2}{2\pi T} \nonumber \\
\tau_\sigma &=& \frac{\ln 2}{2\pi T} \nonumber \\
\xi_\sigma  &=& \xi_C = \frac{4 \eta}{2\pi T}
\end{eqnarray}
where ${N_c}^2/(8\pi) = 1/(16\pi G)$, and $T$ is the temperature of the fluid which 
we measure from the energy density $\epsilon$. The raw materials from which each 
$T_{\mu\nu}^{(i)}$ is built can be written in terms of CFT stress tensor components as 
\begin{eqnarray}\label{eqn:hydroset_raw_materials}
\epsilon &=& \frac{1}{2} \left(T_{tt} -T_{\chi\chi} + \Xi \right) \nonumber \\
u^\chi   &=& \frac{-2 T_{t\chi}}{ \sqrt{-4 T_{t\chi}^2 + \left( T_{tt} + T_{\chi\chi} + \Xi \right)^2} }
\end{eqnarray}
where $\Xi$ is defined by (\ref{eqn:hydro_aux_var}). 

We are now in the position to ask whether the boundary flow is consistent
with an $\mathcal{N}=4$ SYM fluid, and if so, the extent to which higher-order corrections 
are important in describing it. We address these questions by 
comparing each of the reconstructed stress tensors $\sum_{i=0}^{i_{max}} T_{\mu\nu}^{(i)}$ 
for $i_{max}=0,1,2$ to the full boundary CFT stress tensor of the numerical solution.
The results of this comparison are summarized in Figs.~\ref{fig:viscous_medium} 
and~\ref{fig:viscous_medium_wywx32} for a representative stress tensor component
(normalized by the energy density $\epsilon$). 
The plots in Fig.~\ref{fig:viscous_medium} are from the $w_y/w_x=4$ solution, 
where the maximum velocity in the flow reaches $\approx 0.12$ (see Fig.~\ref{fig:rh05_qsv}),
and Fig.~\ref{fig:viscous_medium_wywx32} is from the most asymmetric initial
data case $w_y/w_x=32$ that exhibits a maximum velocity of $\approx 0.54$.
For the former case, the solution converges to behavior that is different 
from ideal hydrodynamics by at most $\approx0.5\%$ (modulo the transient
near $t=0$); the inclusion of first-order and second-order corrections 
successively decrease this difference to a level which is essentially
zero to within truncation error. 
The analogous plots for the larger asymmetry case in Fig.~\ref{fig:viscous_medium_wywx32}
show deviations from ideal hydrodynamics of up to $2.5\%$; adding first
order corrections reduces the maximum deviation to $\approx 1.5\%$, and
second-order corrections to just under $\approx 1\%$.
This is a rather remarkable level of consistency, in particular considering 
the trends in Fig.~\ref{fig:viscous_medium_wywx32} as succesively higher order
viscous corrections are added. 

\subsection{Passing to Minkowski Space}\label{section:passing_to_minkowski_space}

Thus far, in the field theory dual of deformed black holes, we have focused on fluid flows 
on the boundary of global AdS$_5$, namely $\mathbb{R} \times S^3$.  As is clear from 
section~\ref{subsec:bdy_stress_tensor}, and in particular from the second panel of 
Fig.~\ref{fig:rh05_qstt}, the fluid flow is essentially a compressive wave which starts with 
its peak at the equator of $S^3$ and then travels to the poles and back, oscillating and 
damping out toward a static configuration where the temperature on $S^3$ is everywhere 
constant.  In this section, we would like to make closer contact with real-world fluid 
flows by conformally mapping our extracted boundary solution on $\mathbb{R} \times S^3$ 
onto a corresponding flow in Minkowski space.  To do this, we first explain the main 
concepts behind the conformal mapping; then we provide numerical results based on our 
most anisotropic deformed black hole solutions.

Up to a conformal transformation, $\mathbb{R} \times S^3$ is the covering space of 
Minkowski space, $\mathbb{R}^{3,1}$.  Therefore, we can obtain fluid flows on $\mathbb{R}^{3,1}$ 
by restricting our results to an appropriate patch of $\mathbb{R} \times S^3$ and then 
conformally mapping to Minkowski space.  There are many ways of doing this, simply because 
there are many ways of positioning the Minkowski space patch within $\mathbb{R} \times S^3$.  
We will mostly focus on one particular choice of patch which leads to a flow reminiscent of 
a head-on heavy ion collision---though, as on $\mathbb{R} \times S^3$, the initial conditions 
are characterized by full stopping rather than approximate rapidity independence.  More 
specifically, the initial timeslice of our simulation, described as $t=0$ in $\mathbb{R} \times S^3$, 
will correspond to a Minkowski timeslice $t'=0$ in which the fluid is stationary, and compressed 
into a region which is axisymmetric and significantly oblate.  This is intended to be compared to 
the state of two heavy nuclei which have just achieved full overlap, though because the initial 
velocity profile is zero, a better analogy may be a quark-gluon plasma in a trap.  The subsequent 
evolution of the fluid comprises two distinct types of expansion: longitudinal expansion along 
the axis of symmetry, and radial expansion.  Overall, the flow preserves an $SO(3)$ subgroup of 
the conformal group $SO(4,2)$ (as it must since the black hole to which it is dual has an $SO(3)$ 
symmetry), but due to our choice of embedding of Minkowski space in $\mathbb{R} \times S^3$, this 
$SO(3)$ is not the obvious one composed of rotations around a point in a single timeslice.  
Rather, it is the conformal $SO(3)$ symmetry used in \cite{Gubser:2010ze,Gubser:2010ui,Staig:2011wj,Staig:2011as} 
to study generalizations of Bjorken flow.  Rotations around the axis of symmetry of the Minkowski space 
flow form an $SO(2)$ subgroup of the conformal $SO(3)$ symmetry.  The rest of this $SO(3)$ is composed 
of special conformal transformations, corresponding to conformal Killing vectors of Minkowski space.

In order to map $\mathbb{R} \times S^3$, covered by global coordinates $x^\mu=(\tilde{t},\tilde\chi,\tilde\theta,\tilde\phi)$, to $\mathbb{R}^{1,3}$, covered by coordinates $x^a=(t',x_1,x_2,x_3)$, we use the transformations
\begin{eqnarray}\label{eqn:bdyglobal_to_bdyminkowski}
t'/L  &=& \frac{\sin \tilde{t}/L}{\cos \tilde{t}/L + \cos\tilde\chi} \nonumber \\
x_1/L &=& \frac{\sin\tilde\chi}{\left( \cos \tilde{t}/L + \cos\tilde\chi \right)} \sin\tilde\theta \cos\tilde\phi \nonumber \\
x_2/L &=& \frac{\sin\tilde\chi}{\left( \cos \tilde{t}/L + \cos\tilde\chi \right)} \sin\tilde\theta \sin\tilde\phi \nonumber \\
x_3/L &=& \frac{\sin\tilde\chi}{\left( \cos \tilde{t}/L + \cos\tilde\chi \right)} \cos\tilde\theta
 \,.
\end{eqnarray}
The appearance of the AdS scale $L$ on the right hand side of \eqref{eqn:bdyglobal_to_bdyminkowski} is essential: only with this factor will \eqref{eqn:bdyglobal_to_bdyminkowski} lead to a conformal mapping of the metric
\begin{eqnarray}
ds_{\mathbb{R} \times S^3}^2 &=& g^{\mathbb{R} \times S^3}_{\mu\nu} dx^\mu dx^\nu  \nonumber \\
   &=& -d\tilde{t}^2 + L^2 (d\tilde\chi^2 + 
    \sin^2 \tilde\chi^2 d\tilde\theta^2  \nonumber \\
   &&\qquad{} + \sin^2 \tilde\chi^2 \sin^2 \tilde\theta^2 d\tilde\phi^2)
\end{eqnarray}
to the standard Minkowski metric
\begin{equation}
ds_{\mathbb{R}^{3,1}}^2 = g^{\mathbb{R}^{3,1}}_{ab} dx^a dx^b = 
  -(dt')^2 + (dx_1)^2 + (dx_2)^2 + (dx_3)^2 \,.
\end{equation}
The appearance of $L$ on the left hand side of \eqref{eqn:bdyglobal_to_bdyminkowski} is inessential: 
it could be replaced by any quantity with dimensions of length.  Doing so would amount to altering 
Minkowski space by an uniform dilation of both time and space.  The conformal 
mapping~\eqref{eqn:bdyglobal_to_bdyminkowski} is accompanied by the following rule for metric components:
\begin{equation}
g^{\mathbb{R}^{3,1}}_{ab} = W^2 {\partial x^\mu \over \partial x^a}
   {\partial x^\nu \over \partial x^b} g^{\mathbb{R} \times S^3}_{\mu\nu} \,,
\end{equation}
where
\begin{equation}\label{eqn:W}
W = {1 \over \cos \tilde{t}/L + \cos\tilde\chi} \,.
\end{equation}
The Minkowski space patch on $\mathbb{R} \times S^3$ is the connected region including 
$(\tilde{t},\tilde\chi) = (0,0)$ where $W>0$.  This region is easily seen to be the region 
$-\pi < \tilde{t}/L < \pi$ with $0 \leq \chi < \pi-|\tilde{t}/L|$.

We have used coordinates $(\tilde{t},\tilde\chi,\tilde\theta,\tilde\phi)$ on $\mathbb{R} \times S^3$, 
instead of our previous coordinates $(t,\chi,\theta,\phi)$, in order to preserve our freedom to 
position the patch in any way we wish on $\mathbb{R} \times S^3$.  For example, we could set 
$\tilde{t} = t-t_0$ in order to ``center'' the Minkowski space patch on a global time $t=t_0$.  
For our current purposes of describing a fluid flow reminiscent of a head-on heavy ion collision, 
the most useful choice is to set $\tilde{t} = t$, $\tilde\phi = \phi$, and
\begin{eqnarray}\label{eqn:S3_rotation}
\tilde\chi   &=& \frac{\pi}{2} - \arctan \left( \frac{\sin\chi \cos\theta}{\sqrt{\cos^2\chi + \sin^2\chi \sin^2\theta}} \right) \nonumber \\
\tilde\theta &=& \frac{\pi}{2} - \arctan \left( \cot\chi \csc\theta \right).
\end{eqnarray}
This mapping is an isometry of $S^3$, and composing it with the mapping \eqref{eqn:bdyglobal_to_bdyminkowski} 
gives
\begin{eqnarray}\label{eqn:bdyglobal_to_bdyminkowski_original}
t'/L  &=& \frac{\sin t/L}{\cos t/L + \sin\chi \cos\theta} \nonumber \\
x_1/L &=& \frac{\sin\chi}{\left( \cos t/L + \sin\chi \cos\theta \right)} \sin\theta \cos\phi \nonumber \\
x_2/L &=& \frac{\sin\chi}{\left( \cos t/L + \sin\chi \cos\theta \right)} \sin\theta \sin\phi \nonumber \\
x_3/L &=& \frac{\cos\chi}{\left( \cos t/L + \sin\chi \cos\theta \right)}.
\end{eqnarray}
Additionally, one may show by plugging \eqref{eqn:S3_rotation} into \eqref{eqn:W} that
\begin{equation}\label{eqn:minkowski_conformalfactor}
W = \frac{1}{\cos(t/L) + \cos\theta \sin\chi} \,.
\end{equation}
The reason that the final mapping~\eqref{eqn:bdyglobal_to_bdyminkowski_original} is a good idea for 
producing a flow reminiscent of a central heavy ion collision is that the origin of Minkowski space 
maps to an equatorial point $(\chi,\theta) = (\pi/2,0)$ on $S^3$, where the fluid density is at its 
peak on the initial timeslice.  (Note that the coordinate $\phi$ on $S^3$ becomes the usual angle 
$\phi$ around the symmetry axis in Minkowski space, but the coordinate $\theta$ on $S^3$ is more 
closely related to transverse radius from the symmetry axis in Minkowski space than it is to the 
angle of latitude with respect to the symmetry axis.)

\begin{figure*}[t!]
        \centering
        \vspace{-2.5cm}
        \includegraphics[width=7.0in, bb=0 0 1080 540]{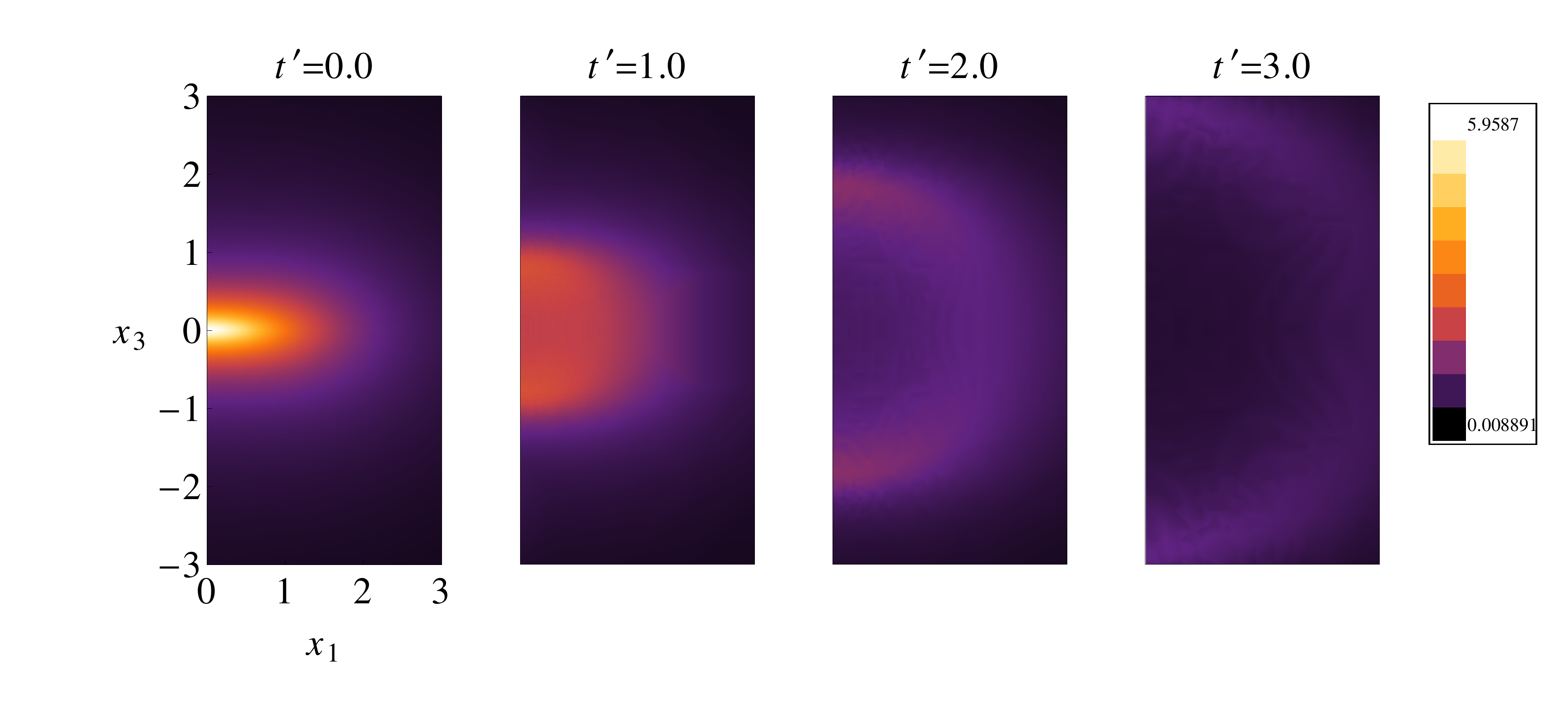}
\parbox{5.0in}{ \vspace{-0.5cm} \caption{ Temperature $T$ defined in (\ref{eqn:minkowski_temperature}),
for a simulation run with $w_y/w_x=32$ initial data, whose final state black hole has
horizon radius $r_h=5$. Each plot depicts the spatial profile of temperature in the 
$x_1 - x_3$ plane, taken at a constant $t^{'}$ slice and at $\phi=0$. One can recover 
the full spatial dependence by simply rotating each of these $x_1 - x_3$ profiles about
the $x_3$ axis. By interpreting the $x_3$ axis as the beam-line direction, and 
$x_1$ as the radius in the transverse plane, the initial data in the first panel can 
be thought of as approximating a head-on heavy ion collision at its moment of impact.
}\label{fig:minkowski_rotated}}
\end{figure*}

We now show how the boundary stress tensor 
$T^{\mathbb{R} \times S^3}_{\mu\nu} \equiv T_{\mu\nu} \equiv \left< T_{\mu \nu} \right>_{\text{CFT}}$ 
transforms under the above conformal mapping. First recall that we have effectively subtracted a term 
$t_{\mu\nu} = \underset{q \rightarrow 0}{\lim}{\frac{1}{q^2}} {}^{(q)} t_{\mu \nu}$ 
from $T^{\mathbb{R} \times S^3}_{\mu\nu}$, by removing the term ${}^{(q)} t_{\mu \nu}$ from 
the quasi-local stress tensor~(\ref{eqn:quasiset_subtracted}) prior to taking the boundary limit. 
Due to this subtraction, the correct rule for obtaining the stress tensor $T^{\mathbb{R}^{3,1}}_{ab}$ 
on Minkowski space obtained via the conformal mapping described above is:
\begin{equation}\label{eqn:conformal_transformation}
T^{\mathbb{R}^{3,1}}_{ab} = W^{-2} {\partial x^\mu \over \partial x^a}
   {\partial x^\nu \over \partial x^b} T^{\mathbb{R} \times S^3}_{\mu\nu} \,.
\end{equation}
Had we not subtracted $t_{\mu\nu}$ in our calculation of $T^{\mathbb{R} \times S^3}_{\mu\nu}$, 
we would have had to include an additional term in~(\ref{eqn:conformal_transformation}), 
identified in~\cite{Skenderis:2000in} as a consequence of the holographic Weyl anomaly. 
In other words, the full stress tensor is not a covariant object under the family of bulk diffeomorphisms 
that induce conformal transformations on the boundary. On the CFT, this anomalous term is not surprising: 
it is simply the usual Schwarzian derivative term that appears when one passes the stress tensor of 
an even-dimensional CFT through a conformal transformation. From the bulk perspective, the anomaly 
originates from picking out  a particular foliation before taking the boundary limit~(\ref{eqn:cftsetexpectation}).

To explicitly see that the rule~(\ref{eqn:conformal_transformation}) results in the correct
stress tensor on Minkowski space, it is instructive to consider the unsubtracted boundary stress tensor 
$T^0_{\mu \nu} = \underset{q \rightarrow 0}{\lim}{\frac{1}{q^2}} {}^{(q)} T^0_{\mu \nu}$. 
Since the anomalous term does not depend on the bulk dynamical variables $\bar{g}_{\mu\nu},\bar{H}_\mu,\bar{\phi}$, 
it suffices to evaluate this object in the case where $\bar{g}_{\mu\nu}=0,\bar{H}_\mu=0,\bar{\phi}=0$. 
Performing this calculation in global AdS$_5$, we use the foliator $q=1-\rho=1/(1+r)$ and find
that $T^{0,\mathbb{R} \times S^3}_{\mu \nu} = t_{\mu\nu}$. On the other hand, this same calculation 
on the Poincar\'e patch involves the foliator $q=1/(\sqrt(1+r^2/L^2) \cos(t/L) + r/L \sin\chi \cos\theta)$ 
(essentially the Poincar\'e radial coordinate) and one finds that $T^{0,\mathbb{R}^{3,1}}_{\mu \nu}=0$. 
This calculation concretely demonstrates that $T^0_{\mu \nu}$ is not a covariant object. It additionally shows that 
by subtracting $t_{\mu\nu}$ from the boundary stress tensor on $\mathbb{R} \times S^3$, one may pass the rest of the 
stress tensor $T^{\mathbb{R} \times S^3}_{\mu\nu} = T^{0,\mathbb{R} \times S^3}_{\mu \nu} - t_{\mu\nu}$ 
through the tensorial rule~(\ref{eqn:conformal_transformation}) and obtain the correct stress 
tensor on $\mathbb{R}^{3,1}$. 

If we now take the stress tensor $T^{\mathbb{R} \times S^3}_{\mu\nu}$ to have an inviscid hydrodynamical 
form $T^{\mathbb{R} \times S^3}_{\mu\nu} = \epsilon u_\mu u_\nu + \epsilon \perp_{\mu\nu} /3$ as was demonstrated in 
Sec.~\ref{section:higher_order_hydrodynamics}, then one can straightforwardly 
show that $u^{\mathbb{R}^{3,1}}_a = W {\partial x^\mu \over \partial x^a} u^{\mathbb{R} \times S^3}_\mu$ 
and $\epsilon^{\mathbb{R}^{3,1}} = W^{-4} \epsilon^{\mathbb{R} \times S^3}$.  Up to a constant factor 
specifying the number of degrees of freedom, the temperature of a conformal fluid is $T = \epsilon^{1/4}$.  
Setting $L=1$, the quantity we are going to plot is 
\begin{equation}\label{eqn:minkowski_temperature}
T \equiv W^{-1} \epsilon^{1/4} \,,
\end{equation}
where $T = T^{\mathbb{R}^{3,1}}$ is (up to the aforementioned constant factor) the temperature in 
Minkowski space and $\epsilon = \epsilon^{\mathbb{R} \times S^3}$ is the energy density on 
$\mathbb{R} \times S^3$. 

We are justified in plotting temperature starting at Minkowski time $t'=0$ because, by construction, 
our initial data perfectly conforms to the inviscid hydrodynamic ansatz.  Moreover, again by 
construction, the velocity field vanishes exactly on the initial timeslice.  Thus we are describing 
a version of the Landau-Khalatnikov full-stopping scenario \cite{Landau:1953gs,Khalatnikov:1954}, 
but with radial flow explicitly included.  The results of section~\ref{section:higher_order_hydrodynamics}, 
in particular Figs.~\ref{fig:viscous_medium} and~\ref{fig:viscous_medium_wywx32}, show that hydrodynamics 
remains approximately valid throughout the evolution, with modest contributions from shear viscosity and 
small contributions from second-order transport coefficients.

The four panels of Fig.~\ref{fig:minkowski_rotated} shows the temperature $T$ in Minkowski space 
reconstructed from a simulation with initial asymmetry $w_y/w_x=32$. Each panel in 
Fig.~\ref{fig:minkowski_rotated} is taken from a $t'=const$ slice, and at $\phi=0$ (i.e. $x_2=0$). 
Since the transformation to Minkowski space preserves the original $\phi$-symmetry in global AdS, 
one can recover the full spatial profile by simply rotating each of these $x_1 - x_3$ profiles about 
the $x_3$ axis. To quantify the initial anisotropy in the temperature profile, we use 
$\left< (x_1)^2 \right> / \left< (x_3)^2 \right>$, where 
$\left< A(x_1,0,x_3) \right> = {\int dx_1 dx_3  T(x_1,0,x_3)^4 A(x_1,0,x_3) / \int dx_1 dx_3  T(x_1,0,x_3)^4}$
is the energy-weighted average of some function $A(x_1,0,x_3)$ defined on the initial $t'=0$ slice and 
at $\phi=0$. This quantity evaluates to $6.94$ for the data displayed in 
Fig.~\ref{fig:minkowski_rotated}. We also calculate the widths $\delta_1$ and $\delta_3$ 
given by $T^4(\delta_1,0,0)=T^4(0,0,0)/2$ and $T^4(0,0,\delta_3)=T^4(0,0,0)/2$, then 
construct the ratio $(\delta_1)^2/(\delta_3)^2$; this evaluates to $18.7$ 
with the data\footnote{The extent to which $\left< (x_1)^2 \right> / \left< (x_3)^2 \right>$ and 
$(\delta_1)^2/(\delta_3)^2$ differ reflects the extent to which the temperature profile 
deviates from a perfect Gaussian. As a measure of anisotropy, the ratio of widths at half-max 
$(\delta_1)^2/(\delta_3)^2$ is the more faithful of the two.}.

\section{Conclusion}\label{section:conclusion}

We have described a generalized harmonic scheme for solving the Einstein 
field equations on asymptotically anti-de Sitter spacetimes in 4+1 
dimensions. Though restricting to $SO(3)$ symmetry in this initial code,
we expect that many of the methods developed here to achieve stable, consistent
evolution will carry over to scenarios with less symmetry. This will be needed
to tackle the main motivation for developing this code, namely to obtain the gravity
dual to a heavy ion collision, and its subsequent quark-gluon plasma formation.

As a first application, we studied the quasi-normal ringdown of highly distorted 
black holes, and the corresponding behavior of the stress energy tensor of the 
dual CFT on the boundary of the spacetime. We find quasi-normal mode
frequencies that are consistent with previously published linear modes,
as well as modes that can be modeled as arising from non-linear mode-coupling.
We further find purely decaying modes that we attribute to gauge. To be
certain this is the case would require transforming the solution to coordinates
fully compatible with perturbative calculations, which is a rather non-trivial
task numerically, and so we relegate this to a future study.

The boundary stress energy tensor exhibits correspondingly large initial fluctuations, 
yet we find that its behavior is consistent to better than $1\%$ with that of 
an $\mathcal{N}=4$ SYM fluid with equation of state $\epsilon=3P$, and with corresponding 
transport coefficients, essentially from $t=0$ onwards. This is in contrast to the numerical results 
reported in~\cite{Chesler:2010bi,Heller:2011ju,Wu:2011yd,Chesler:2011ds}, where only after a 
certain time did the boundary behavior approach that of a fluid. However, those studies 
looked at scenarios more akin to black hole formation, i.e. beginning with 
states that do not contain large black holes (or black branes), with the black 
holes forming at later times. Such processes are expected to be dual to 
thermalization, whereas our study is that of equilibration beginning from 
an inhomogeneous though thermal state. Nevertheless, it is curious that the link 
between Einstein and Navier-Stokes seems to be holding even in these 
far-from-equilibrium scenarios, and it will be interesting in future work to 
explore how far this relationship extends. Recently it was shown that the 
Rayleigh-Plateau instability in a fluid stream and the Gregory-Laflamme 
instability of a black string are at least qualitiatively 
similar~\cite{Lehner:2010pn}. These findings similarly suggest that, in 
some situations, the physics described by the Einstein and Navier-Stokes equations 
could exhibit similarities even in the most non-linear, near-singular regimes.

By passing to an appropriate Minkowski patch of the boundary of global AdS$_5$, 
we are able to extract a fluid flow which starts from a compressed disk and leads 
to expansion both in the radial and longitudinal directions.  The $SO(3)$ symmetry 
is still present, but as a conformal symmetry rather than an isometry of the flow.  
This flow exhibits the main qualitative features of the full-stopping scenario in 
heavy ion collisions, which, though generally considered implausible in light of 
asymptotic freedom, has some interesting phenomenological successes \cite{Steinberg:2004vy}.  
It would be interesting to pass our flow through a hadronization algorithm and extract the 
rapidity distribution of produced particles.  We leave this task for future work.

A further direction for future study, that could be accomplished within the 
simplification of the current $SO(3)$ symmetry, is to couple gravity to 
additional forms of matter. In the context of AdS/CFT, it is interesting to 
consider tachyonic scalar fields, since these amount to relevant operator 
insertions in the boundary CFT. As one might expect, the resulting deformation 
is dramatic: on the gravity side, these tachyonic scalars back-react in such a 
way as to significantly change the boundary conditions of the metric. The 
evolution of such systems in the full non-linear regime is unknown, though 
hopefully methods similar to those introduced here will be able to handle such 
spacetimes with ``deformed'' AAdS boundaries. These methods are also applicable 
to AAdS spacetimes with different boundary topology, which (with the appropriate 
matter content) are of interest to condensed matter applications of the duality.

\acknowledgments
We thank Alex Buchel, Paul Chesler and Andrei Starinets for valuable discussions, and the referee 
for insightful comments, particularly on the holographic Weyl anomaly and the extent of the 
Fefferman-Graham theorem's applicability. 
We gratefully acknowledge support from DOE Grant No.~DE-FG02-91ER40671 (SSG), NSF grant PHY-0745779 (FP), 
and the Alfred P. Sloan Foundation (FP). 
Simulations were run on the {\bf Woodhen} cluster at Princeton University.
\clearpage

\setcounter{secnumdepth}{1} 
\appendix

\section{Effect of Scalar Backreaction on Metric Fall-off}\label{app:effect_of_scalar_backreaction_on_metric_falloff}

Here we compute the effect that a static spherical distribution of massless 
scalar would have on the AAdS$_5$ metric. We follow the discussion in~\cite{Henneaux:2006hk}. 
The configuration only has radial dependence $\phi=\phi(r)$, so the metric must take the form
\begin{eqnarray}
ds^2= -\left( 1+r^2 + \mathcal{O}(r^{-1}) \right) dt^2 &+& \frac{1}{\left( 1 + r^2 
- \frac{\mu(r)}{r} \right)} dr^2 \nonumber \\ &+& r^2 d{\Omega_3}^2
\end{eqnarray}
where $\mu(r)$ grows slower than $r^3$ as $r \rightarrow \infty$ in order to 
preserve the value of the cosmological constant. We have set $L=1$, so that 
$\Lambda_5 = -6$. In a previous section, we had used the AdS$_5$ Klein-Gordon 
equation~(\ref{eqn:kgeqn_spherical}) in spherical symmetry to find the general 
leading order behavior~(\ref{eqn:scalarbcs}) of scalar fields in AdS$_5$. We now 
find the metric fall-off behavior when scalar back-reaction is taken into account, 
limiting ourselves to the $h_{\rho \rho}$ component of the metric deviation 
and using the compactified coordinate $\rho = r/(1+r)$. In order to find this 
fall-off, the strategy is to solve for $\mu(r)$ as a power series in $r$, from 
which we can reconstruct 
\begin{equation}
h_{\rho \rho} = \left( \frac{\partial r}{\partial \rho} \right)^2 h_{rr} = 
\frac{\mu(r)}{r}
\end{equation}
where we have used $\partial r/ \partial \rho = 1 / (1-\rho)^2 \sim r^2$ along 
with $h_{rr} = g_{rr} - \hat{g}_{rr}$ and
\begin{equation}
g_{rr} = \left( 1 + r^2 - \mu(r)/r \right)^{-1} = \frac{1}{r^2} \left[ 1 + 
\frac{\mu(r)}{r^3} + \mathcal{O}(r^{-5}) \right].
\end{equation}

We can deduce the asymptotic behavior of $\mu(r)$ from the
the Hamiltonian constraint\footnote{See Sec.~\ref{subsection:hamiltonian_constraint}.}
\begin{equation}\label{eqn:a1_constraint}
{}^{(4)}R - 2\Lambda_5 = 16 \pi \rho_E.
\end{equation}
where 
\begin{equation}
{}^{(4)}R = 2 \Lambda_5 + \frac{2 + r^2(5+3r^2)}{(r+r^3)^2} \mu'(r) + 
\frac{r(5+3r^2)}{(r+r^3)^2} \mu(r)
\end{equation}
is the Ricci scalar associated with the 4-metric on the slice, and 
\begin{equation}
\rho_E = \frac{1}{2} (1+r^2-\mu(r)/r)(\phi'(r))^2 + \frac{1}{2} m^2 \phi^2 
\end{equation}
is the scalar field energy density. Keeping only the terms 
in~(\ref{eqn:a1_constraint}) that dominate at large $r$, we obtain the form of 
the Hamiltonian constraint near the boundary 
\begin{equation}\label{eqn:a1_constraintnearbdy}
\frac{\mu'(r)}{r^2} + \frac{\mu(r)}{r} (\phi'(r))^2 \sim r^2 (\phi'(r))^2 + 
m^2 \phi^2.
\end{equation}
The scalar field goes as $\phi(r) \sim r^{-\Delta}$ near the boundary for some 
$\Delta$, so we see that the right-hand-side of~(\ref{eqn:a1_constraintnearbdy}) 
goes as $r^{-2\Delta}$. Matching the left and right-hand sides, we conclude 
that
\begin{equation}
\mu(r) \sim r^{3-2\Delta}
\end{equation}
and consequently,
\begin{equation}\label{eqn:metricbackreaction}
h_{\rho \rho} \sim r^{2-2\Delta}.
\end{equation}

To make use of what we have learnt, first suppose that we have a scalar with negative 
$m^2 < 0$. From~(\ref{eqn:scalarbcs}) and~(\ref{eqn:scalarpower}), 
a non-zero $A$ branch describes the 
leading behavior for large $r$, and is defined by a fall-off power of 
$\Delta = \Delta_+ = 2-\sqrt{4+m^2}$. Then equation~(\ref{eqn:metricbackreaction}) 
predicts that matter backreaction induces a metric deviation $h_{\rho \rho} 
\sim r^{-2+2\sqrt{4+m^2}}$ that is larger than the vacuum metric fall-off we 
assumed in~(\ref{eqn:metric_asymptotics}) and in~(\ref{eqn:metric_bcs}). 
This case corresponds to a tachyonic scalar configuration. These configurations 
are stable in AdS as long as its mass satisfies the Breitenlohner-Freedman 
bound~\cite{Breitenlohner:1982bm} given by $-(D-1)^2/(4 L^2) = -4 < m^2$ in 
$D=5$ dimensions: it is stable in the sense that its conserved 
energy functional remains positive in the function space of all fluctuations 
for which the energy functional remains a convergent integral. The qualitative 
reason for the positivity of this energy functional in AdS is that despite the 
negative unbounded potential of the tachyonic scalar, the positive kinetic 
terms in the energy functional dominate the negative potential terms so long 
as the scalar falls off sufficiently quickly near the boundary\footnote{A 
scalar with the critical fall-off behavior goes as $\phi \sim r^{\Delta_c}$ for 
${\Delta_c} = (D-1)/2 = 2$ in $D=5$ dimensions; the BF bound can be inferred 
from this observation.}. We shall elaborate on the behavior of these fields in 
a subsequent paper.

For the present paper, we consider the case of scalars with $m^2 \ge 0$. Again 
looking at (\ref{eqn:scalarbcs}), (\ref{eqn:scalarpower}), notice that we must 
now turn off the $A$ branch in order to have a scalar field that vanishes at 
infinity, which is required for any scalar field configuration with finite
energy. The result is a localized 
matter distribution defined by a fall-off power 
$\Delta = \Delta_+ = 2+\sqrt{4+m^2}$. Applying our result 
(\ref{eqn:metricbackreaction}) to this case, we find that backreaction induces 
a metric deviation $h_{\rho \rho} \sim r^{-2-2\sqrt{4+m^2}}$ that is subleading 
compared to the vacuum metric fall-off (\ref{eqn:metric_bcs}), and thus leaving 
it unchanged. 

\clearpage

\section{Boundary Conditions for Linearized Gravitational Perturbations of AdS}\label{app:boundary_operator}

\subsection{Fefferman-Graham Theorem}
A theorem due to Fefferman and Graham~\cite{fefferman} states that a 
distinguished coordinate system $(z, x^i)$ exists near the boundary of
any asymptotically AdS$_5$ spacetime $M$ in which the metric takes the form 
\begin{equation}\label{eqn:Pmetric}
G = \frac{L^2}{z^2} \left( dz^2 + g_{ij}dx^idx^j \right)
\end{equation}
and that there is a convergent power series solution for the $g_{ij}$ 
coefficients in~(\ref{eqn:Pmetric}) given by
\begin{equation}\label{eqn:Pcoefficients}
g_{ij} = g_{(0)ij} + g_{(2)ij} z^2 + g_{(4)ij} z^4 + 2 h_{(4)ij} z^4 \log z + \mathcal{O}(z^6)
\end{equation}
(the boundary $\partial M$ is located at $z=0$).

Let us first translate this statement in terms of global AdS$_5$ coordinates 
$(t,r,\chi,\theta,\phi)$. We follow the discussion in~\cite{Skenderis:2000in}, 
though we use a different Fefferman-Graham radial coordinate $z$ from the one 
use there, whose relation with the global radial coordinate $r$ is given by 
\begin{equation}
r^2 = \frac{1}{z^2} \left( 1-\frac{z^2}{4} \right)^2.
\end{equation}
To leading order, this relation between $z$ and $r$ near the boundary at $z=0$, 
or equivalently $r \rightarrow \infty$, simply reads
\begin{equation}
z = \frac{1}{r} + \mathcal{O}(1/r^2).
\end{equation}

In terms of the global coordinate $q=1/(1+r)$ introduced in Sec.~\ref{subsection:the_set_of_the_dual_cft}, 
to leading-order we have $z = 1/r + \mathcal{O}(1/r^2) = q + \mathcal{O}(q^2)$ near the $q=0$ 
boundary. So the Fefferman-Graham theorem's power series solution in these 
coordinates reads just like~(\ref{eqn:Pcoefficients}), replacing $z$ by $q$
\begin{equation}\label{eqn:Pcoefficients_global}
g_{ij} = g_{(0)ij} + g_{(2)ij} q^2 + g_{(4)ij} q^4 + 2 h_{(4)ij} q^4 \log q + \mathcal{O}(q^6)
\end{equation}
(the boundary $\partial M$ is located at $q=0$).

The $g_{(0)ij}$ corresponds to the non-radial components of the pure AdS$_5$ 
metric. The $g_{(2)ij}$ can be expressed, as was done explicitly 
in~\cite{deHaro:2000xn}, in terms of the Ricci tensor $R_{(0)ij}$ and the scalar 
curvature $R_{(0)}$ constructed from $g_{(0)ij}$, and is thus fixed. The leading-order 
dynamics then first appears\footnote {In $D=d+1$ dimensions, the leading-order dynamics 
appears in the $g_{(d)ij}$ term, since $g_{(n)ij}$ for $n<d$ are determined by 
$g_{(0)ij}$.} in the terms controlled by $g_{(4)ij}$. Inspecting~(\ref{eqn:Pcoefficients_global}), 
this implies that the leading-order asymptotics is $G_{ij} \sim q^2$, in agreement 
with the boundary conditions we wrote down in Sec.~\ref{sec:evo_vars}. 

\subsection{Ishibashi-Wald Boundary Conditions}

Here we show how our metric boundary conditions relate to those 
derived by Ishibashi and Wald in~\cite{Ishibashi:2004wx}. Working 
perturbatively, they derive conditions under which a metric perturbation
must have its boundary behavior enforced by explicit boundary 
conditions. The fields governing the perturbations, which we collectively denote by $f$, are decomposed in 
terms of spherical harmonics $\mathbb{S}_k(\Omega^i)$ according to 
\begin{equation}
f = \frac{1}{r^{3/2}} \sum_{k_S} \Phi_k (t,r) \mathbb{S}_k(\Omega^i).
\end{equation}
The $\Phi$ fields for the tensor and vectors modes of the gravitational 
perturbations are shown to require no boundary conditions at infinity 
(see Proposition 3.1 and its preceding discussion in \cite{Ishibashi:2004wx}). 
However, the scalar modes of the gravitational perturbations require boundary 
conditions, and their corresponding $\Phi$ are shown to behave asymptotically 
as 
\begin{equation}
\Phi = F_0(r) \left[ 2 a_0 \log(1/r) + b_0 + 2 L^2/r^2 \log(1/r) + ... \right]
\end{equation}
where the function $F_0(r)$ is asymptotically given by
\begin{equation}
F_0(r) \sim \frac{1}{r^{1/2}}.
\end{equation}

In terms of the global coordinate $q$, the implication is that for the 
scalar modes of the gravitational perturbations the near-boundary behavior is
\begin{equation}\label{eqn:selfadjointbcs}
G_{ij} \sim 2 a_0 q^2 \log(q) + b_0 q^2 + 2 c_0 q^4 \log(q) + ...
\end{equation}
For the scalar modes, imposing boundary conditions amounts to setting the ratio 
$b_0/a_0$. Comparing~(\ref{eqn:selfadjointbcs}) 
to~(\ref{eqn:Pmetric}),(\ref{eqn:Pcoefficients_global}), 
we see that the choice picked out by coordinates that are Fefferman-Graham-like near the 
boundary is $b_0/a_0 \rightarrow \infty$ i.e. $a_0=0$. This removes the logarithmic branch 
controlled by $a_0$, so the leading-order asymptotics is $G_{ij} \sim q^2$, as desired. 

\clearpage

\section{Tables of Scalar Quasinormal Mode Frequencies}\label{app:tables_of_scalar_qnm_frequencies}

\begin{widetext}

\begin{table}[b]
\begin{center}
\begin{tabular}{|ll|lll|}
\hline
&fund.		&k=0			&k=2			&k=4			\\
\hline
&$\frac{r_h}{L}$=12.2 	&$(2.95 \pm 0.13)\frac{r_h}{L}-i(2.67 \pm 0.02)\frac{r_h}{L}$	&$(3.96 \pm 0.03)\frac{r_h}{L}-i(2.33 \pm 0.07)\frac{r_h}{L}$	&$(4.93 \pm 0.007)\frac{r_h}{L}-i(1.83 \pm 0.05)\frac{r_h}{L}$ 	\\
&$\frac{r_h}{L}$=11.3 	&$(2.95 \pm 0.14)\frac{r_h}{L}-i(2.64 \pm 0.04)\frac{r_h}{L}$	&$(3.94 \pm 0.02)\frac{r_h}{L}-i(2.30 \pm 0.08)\frac{r_h}{L}$	&$(4.90 \pm 0.003)\frac{r_h}{L}-i(1.80 \pm 0.04)\frac{r_h}{L}$ 	\\
&$\frac{r_h}{L}$=10.5 	&$(2.96 \pm 0.14)\frac{r_h}{L}-i(2.64 \pm 0.05)\frac{r_h}{L}$	&$(3.95 \pm 0.01)\frac{r_h}{L}-i(2.29 \pm 0.08)\frac{r_h}{L}$	&$(4.92 \pm 0.002)\frac{r_h}{L}-i(1.78 \pm 0.04)\frac{r_h}{L}$ 	\\
&$\frac{r_h}{L}$=9.0	&$(2.99 \pm 0.15)\frac{r_h}{L}-i(2.63 \pm 0.07)\frac{r_h}{L}$	&$(3.95 \pm 0.002)\frac{r_h}{L}-i(2.26 \pm 0.09)\frac{r_h}{L}$	&$(4.90 \pm 0.02)\frac{r_h}{L}-i(1.72 \pm 0.02)\frac{r_h}{L}$ 	\\
&$\frac{r_h}{L}$=6.5 	&$(3.06 \pm 0.14)\frac{r_h}{L}-i(2.62 \pm 0.14)\frac{r_h}{L}$	&$(3.96 \pm 0.05)\frac{r_h}{L}-i(2.23 \pm 0.07)\frac{r_h}{L}$	&$(4.80 \pm 0.01)\frac{r_h}{L}-i(1.65 \pm 0.003)\frac{r_h}{L}$ 	\\
&$\frac{r_h}{L}$=5.0 	&$(3.11 \pm 0.09)\frac{r_h}{L}-i(2.62 \pm 0.20)\frac{r_h}{L}$	&$(3.94 \pm 0.05)\frac{r_h}{L}-i(2.21 \pm 0.04)\frac{r_h}{L}$	&$(4.79 \pm 0.08)\frac{r_h}{L}-i(1.60 \pm 0.02)\frac{r_h}{L}$ 	\\
&$\frac{r_h}{L}$=4.5 	&$(3.18 \pm 0.06)\frac{r_h}{L}-i(2.67 \pm 0.22)\frac{r_h}{L}$	&$(4.00 \pm 0.06)\frac{r_h}{L}-i(2.24 \pm 0.04)\frac{r_h}{L}$	&$(4.87 \pm 0.11)\frac{r_h}{L}-i(1.69 \pm 0.04)\frac{r_h}{L}$ 	\\
&$\frac{r_h}{L}$=4.0 	&$(3.21 \pm 0.001)\frac{r_h}{L}-i(2.67 \pm 0.23)\frac{r_h}{L}$	&$(3.99 \pm 0.06)\frac{r_h}{L}-i(2.24 \pm 0.03)\frac{r_h}{L}$	&$(4.83 \pm 0.12)\frac{r_h}{L}-i(1.76 \pm 0.04)\frac{r_h}{L}$ 	\\
&$\frac{r_h}{L}$=3.3 	&$(3.33 \pm 0.08)\frac{r_h}{L}-i(2.71 \pm 0.20)\frac{r_h}{L}$	&$(4.05 \pm 0.08)\frac{r_h}{L}-i(2.27 \pm 0.03)\frac{r_h}{L}$	&$(4.9 \pm 0.1)\frac{r_h}{L}-i(1.89 \pm 0.05)\frac{r_h}{L}$ 	\\
\hline
\end{tabular}
\end{center}
\centering
\parbox{5.0in}{\caption{The fundamental ($n=0$) quasi-normal mode frequencies $\omega_r - i\omega_i$ 
extracted from the scalar field variable $\bar{\phi}$, from evolution of initial
data as described in Sec.~\ref{sec:qnr}. These are shown for 
various horizon radii $r_h$ of the final state AdS-Schwarzschild black hole 
with $SO(4)$ quantum numbers $k\ge 0,l=0,m=0$. Uncertainties are estimated from convergence
studies.
	}\label{tab:scalarqnms_fastmodes_n0}}
\end{table}

\begin{table}[!]
\begin{center}
\begin{tabular}{|ll|lll|}
\hline
&$1^{st}$ overt.&k=0			        &k=2			        &k=4			\\
\hline
&$\frac{r_h}{L}$=12.2 	&$(5.66 \pm 0.24)\frac{r_h}{L}-i(2.90 \pm 0.28)\frac{r_h}{L}$	&$(6.36 \pm 0.26)\frac{r_h}{L}-i(3.28 \pm 0.15)\frac{r_h}{L}$	&$(7.60 \pm 0.17)\frac{r_h}{L}-i(4.11 \pm 0.17)\frac{r_h}{L}$ 	\\
&$\frac{r_h}{L}$=11.3 	&$(5.65 \pm 0.27)\frac{r_h}{L}-i(2.91 \pm 0.30)\frac{r_h}{L}$	&$(6.36 \pm 0.31)\frac{r_h}{L}-i(3.32 \pm 0.16)\frac{r_h}{L}$	&$(7.59 \pm 0.23)\frac{r_h}{L}-i(4.31 \pm 0.14)\frac{r_h}{L}$ 	\\
&$\frac{r_h}{L}$=10.5 	&$(5.69 \pm 0.30)\frac{r_h}{L}-i(2.94 \pm 0.32)\frac{r_h}{L}$	&$(6.41 \pm 0.35)\frac{r_h}{L}-i(3.37 \pm 0.14)\frac{r_h}{L}$	&$(7.63 \pm 0.26)\frac{r_h}{L}-i(4.56 \pm 0.14)\frac{r_h}{L}$ 	\\
&$\frac{r_h}{L}$=9.0	&$(5.76 \pm 0.38)\frac{r_h}{L}-i(3.00 \pm 0.36)\frac{r_h}{L}$	&$(6.50 \pm 0.45)\frac{r_h}{L}-i(3.52 \pm 0.16)\frac{r_h}{L}$	&$(7.74 \pm 0.39)\frac{r_h}{L}-i(5.22 \pm 0.10)\frac{r_h}{L}$ 	\\
&$\frac{r_h}{L}$=6.5 	&$(5.98 \pm 0.62)\frac{r_h}{L}-i(3.19 \pm 0.42)\frac{r_h}{L}$	&$(6.78 \pm 0.71)\frac{r_h}{L}-i(4.02 \pm 0.08)\frac{r_h}{L}$	&$(8.04 \pm 0.79)\frac{r_h}{L}-i(7.92 \pm 0.41)\frac{r_h}{L}$ 	\\
&$\frac{r_h}{L}$=5.0 	&$(6.20 \pm 0.96)\frac{r_h}{L}-i(3.54 \pm 0.51)\frac{r_h}{L}$	&$(7.15 \pm 1.02)\frac{r_h}{L}-i(4.96 \pm 0.17)\frac{r_h}{L}$	&$(7.40 \pm 1.43)\frac{r_h}{L}-i(8.69 \pm 1.87)\frac{r_h}{L}$ 	\\
&$\frac{r_h}{L}$=4.5 	&$(6.41 \pm 1.13)\frac{r_h}{L}-i(3.82 \pm 0.57)\frac{r_h}{L}$	&$(7.45 \pm 1.19)\frac{r_h}{L}-i(5.61 \pm 0.32)\frac{r_h}{L}$	&$(7.63 \pm 2.19)\frac{r_h}{L}-i(8.78 \pm 1.74)\frac{r_h}{L}$ 	\\
&$\frac{r_h}{L}$=4.0 	&$(6.58 \pm 1.36)\frac{r_h}{L}-i(4.24 \pm 0.68)\frac{r_h}{L}$	&$(7.71 \pm 1.44)\frac{r_h}{L}-i(6.67 \pm 0.76)\frac{r_h}{L}$	&$(7.77 \pm 1.82)\frac{r_h}{L}-i(9.94 \pm 3.02)\frac{r_h}{L}$ 	\\
&$\frac{r_h}{L}$=3.3 	&$(7.02 \pm 1.78)\frac{r_h}{L}-i(5.34 \pm 1.04)\frac{r_h}{L}$	&$(8.04 \pm 1.74)\frac{r_h}{L}-i(9.35 \pm 2.38)\frac{r_h}{L}$	&$(7.68 \pm 1.32)\frac{r_h}{L}-i(12.7 \pm 4.7)\frac{r_h}{L}$ 	\\
\hline
\end{tabular}
\end{center}
\centering
\parbox{5.0in}{\caption{The first overtone ($n=1$) quasi-normal mode frequencies $\omega_r - i\omega_i$ 
extracted from the scalar field variable $\bar{\phi}$, from evolution of initial
data as described in Sec.~\ref{sec:qnr}. These are shown for 
various horizon radii $r_h$ of the final state AdS-Schwarzschild black hole 
and for harmonics with $SO(4)$ quantum numbers 
$k\ge 0,l=0,m=0$. Uncertainties are estimated from convergence
studies.
	}\label{tab:scalarqnms_fastmodes_n1}}
\bigskip
\end{table}

\end{widetext}

\clearpage

\bibliography{paper_jan_10_2012}

\begin{thebibliography}{72}
\expandafter\ifx\csname natexlab\endcsname\relax\def\natexlab#1{#1}\fi
\expandafter\ifx\csname bibnamefont\endcsname\relax
  \def\bibnamefont#1{#1}\fi
\expandafter\ifx\csname bibfnamefont\endcsname\relax
  \def\bibfnamefont#1{#1}\fi
\expandafter\ifx\csname citenamefont\endcsname\relax
  \def\citenamefont#1{#1}\fi
\expandafter\ifx\csname url\endcsname\relax
  \def\url#1{\texttt{#1}}\fi
\expandafter\ifx\csname urlprefix\endcsname\relax\def\urlprefix{URL }\fi
\providecommand{\bibinfo}[2]{#2}
\providecommand{\eprint}[2][]{\url{#2}}

\bibitem[{\citenamefont{Maldacena}(1998)}]{Maldacena:1997re}
\bibinfo{author}{\bibfnamefont{J.~M.} \bibnamefont{Maldacena}},
  \bibinfo{journal}{Adv. Theor. Math. Phys.} \textbf{\bibinfo{volume}{2}},
  \bibinfo{pages}{231} (\bibinfo{year}{1998}), \eprint{hep-th/9711200}.

\bibitem[{\citenamefont{Gubser et~al.}(1998)\citenamefont{Gubser, Klebanov, and
  Polyakov}}]{Gubser:1998bc}
\bibinfo{author}{\bibfnamefont{S.}~\bibnamefont{Gubser}},
  \bibinfo{author}{\bibfnamefont{I.~R.} \bibnamefont{Klebanov}},
  \bibnamefont{and} \bibinfo{author}{\bibfnamefont{A.~M.}
  \bibnamefont{Polyakov}}, \bibinfo{journal}{Phys.Lett.}
  \textbf{\bibinfo{volume}{B428}}, \bibinfo{pages}{105} (\bibinfo{year}{1998}),
  \eprint{hep-th/9802109}.

\bibitem[{\citenamefont{Witten}(1998)}]{Witten:1998qj}
\bibinfo{author}{\bibfnamefont{E.}~\bibnamefont{Witten}},
  \bibinfo{journal}{Adv. Theor. Math. Phys.} \textbf{\bibinfo{volume}{2}},
  \bibinfo{pages}{253} (\bibinfo{year}{1998}), \eprint{hep-th/9802150}.

\bibitem[{\citenamefont{Gubser and Karch}(2009)}]{Gubser:2009md}
\bibinfo{author}{\bibfnamefont{S.~S.} \bibnamefont{Gubser}} \bibnamefont{and}
  \bibinfo{author}{\bibfnamefont{A.}~\bibnamefont{Karch}},
  \bibinfo{journal}{Ann. Rev. Nucl. Part. Sci.} \textbf{\bibinfo{volume}{59}},
  \bibinfo{pages}{145} (\bibinfo{year}{2009}), \eprint{0901.0935}.

\bibitem[{\citenamefont{Herzog}(2009)}]{Herzog:2009xv}
\bibinfo{author}{\bibfnamefont{C.~P.} \bibnamefont{Herzog}},
  \bibinfo{journal}{J.Phys.A} \textbf{\bibinfo{volume}{A42}},
  \bibinfo{pages}{343001} (\bibinfo{year}{2009}), \eprint{0904.1975}.

\bibitem[{\citenamefont{Hartnoll}(2009)}]{Hartnoll:2009sz}
\bibinfo{author}{\bibfnamefont{S.~A.} \bibnamefont{Hartnoll}},
  \bibinfo{journal}{Class.Quant.Grav.} \textbf{\bibinfo{volume}{26}},
  \bibinfo{pages}{224002} (\bibinfo{year}{2009}), \eprint{0903.3246}.

\bibitem[{\citenamefont{McGreevy}(2010)}]{McGreevy:2009xe}
\bibinfo{author}{\bibfnamefont{J.}~\bibnamefont{McGreevy}},
  \bibinfo{journal}{Adv. High Energy Phys.} \textbf{\bibinfo{volume}{2010}},
  \bibinfo{pages}{723105} (\bibinfo{year}{2010}), \eprint{0909.0518}.

\bibitem[{\citenamefont{Sachdev}(2010)}]{Sachdev:2010ch}
\bibinfo{author}{\bibfnamefont{S.}~\bibnamefont{Sachdev}}
  (\bibinfo{year}{2010}), \eprint{1002.2947}.

\bibitem[{\citenamefont{Gubser}(2010{\natexlab{a}})}]{Gubser:2010nc}
\bibinfo{author}{\bibfnamefont{S.~S.} \bibnamefont{Gubser}}
  (\bibinfo{year}{2010}{\natexlab{a}}), \eprint{1012.5312}.

\bibitem[{\citenamefont{Horowitz}(2010)}]{Horowitz:2010gk}
\bibinfo{author}{\bibfnamefont{G.~T.} \bibnamefont{Horowitz}}
  (\bibinfo{year}{2010}), \eprint{1002.1722}.

\bibitem[{\citenamefont{Gubser}(2011)}]{Gubser:2011qv}
\bibinfo{author}{\bibfnamefont{S.~S.} \bibnamefont{Gubser}}
  (\bibinfo{year}{2011}), \eprint{1103.3636}.

\bibitem[{\citenamefont{Pretorius}(2007)}]{Pretorius:2007nq}
\bibinfo{author}{\bibfnamefont{F.}~\bibnamefont{Pretorius}}
  (\bibinfo{year}{2007}), \eprint{0710.1338}.

\bibitem[{\citenamefont{Duez}(2010)}]{Duez:2009yz}
\bibinfo{author}{\bibfnamefont{M.~D.} \bibnamefont{Duez}},
  \bibinfo{journal}{Class.Quant.Grav.} \textbf{\bibinfo{volume}{27}},
  \bibinfo{pages}{114002} (\bibinfo{year}{2010}), \eprint{0912.3529}.

\bibitem[{\citenamefont{Campanelli et~al.}(2010)\citenamefont{Campanelli,
  Lousto, Mundim, Nakano, Zlochower et~al.}}]{Campanelli:2010ac}
\bibinfo{author}{\bibfnamefont{M.}~\bibnamefont{Campanelli}},
  \bibinfo{author}{\bibfnamefont{C.~O.} \bibnamefont{Lousto}},
  \bibinfo{author}{\bibfnamefont{B.~C.} \bibnamefont{Mundim}},
  \bibinfo{author}{\bibfnamefont{H.}~\bibnamefont{Nakano}},
  \bibinfo{author}{\bibfnamefont{Y.}~\bibnamefont{Zlochower}},
  \bibnamefont{et~al.}, \bibinfo{journal}{Class.Quant.Grav.}
  \textbf{\bibinfo{volume}{27}}, \bibinfo{pages}{084034}
  (\bibinfo{year}{2010}), \eprint{1001.3834}.

\bibitem[{\citenamefont{Centrella et~al.}(2010)\citenamefont{Centrella, Baker,
  Kelly, and van Meter}}]{Centrella:2010mx}
\bibinfo{author}{\bibfnamefont{J.}~\bibnamefont{Centrella}},
  \bibinfo{author}{\bibfnamefont{J.~G.} \bibnamefont{Baker}},
  \bibinfo{author}{\bibfnamefont{B.~J.} \bibnamefont{Kelly}}, \bibnamefont{and}
  \bibinfo{author}{\bibfnamefont{J.~R.} \bibnamefont{van Meter}},
  \bibinfo{journal}{Rev. Mod. Phys.} \textbf{\bibinfo{volume}{82}},
  \bibinfo{pages}{3069} (\bibinfo{year}{2010}), \eprint{1010.5260}.

\bibitem[{\citenamefont{Pretorius and Choptuik}(2000)}]{Pretorius:2000yu}
\bibinfo{author}{\bibfnamefont{F.}~\bibnamefont{Pretorius}} \bibnamefont{and}
  \bibinfo{author}{\bibfnamefont{M.~W.} \bibnamefont{Choptuik}},
  \bibinfo{journal}{Phys.Rev.} \textbf{\bibinfo{volume}{D62}},
  \bibinfo{pages}{124012} (\bibinfo{year}{2000}), \eprint{gr-qc/0007008}.

\bibitem[{\citenamefont{Husain and Olivier}(2001)}]{Husain:2000vm}
\bibinfo{author}{\bibfnamefont{V.}~\bibnamefont{Husain}} \bibnamefont{and}
  \bibinfo{author}{\bibfnamefont{M.}~\bibnamefont{Olivier}},
  \bibinfo{journal}{Class.Quant.Grav.} \textbf{\bibinfo{volume}{18}},
  \bibinfo{pages}{L1} (\bibinfo{year}{2001}), \eprint{gr-qc/0008060}.

\bibitem[{\citenamefont{Hwang et~al.}(2011)\citenamefont{Hwang, Kim, and
  Yeom}}]{Hwang:2011mn}
\bibinfo{author}{\bibfnamefont{D.-i.} \bibnamefont{Hwang}},
  \bibinfo{author}{\bibfnamefont{H.}~\bibnamefont{Kim}}, \bibnamefont{and}
  \bibinfo{author}{\bibfnamefont{D.-h.} \bibnamefont{Yeom}}
  (\bibinfo{year}{2011}), \eprint{1105.1371}.

\bibitem[{\citenamefont{Bizon and Rostworowski}(2011)}]{Bizon:2011gg}
\bibinfo{author}{\bibfnamefont{P.}~\bibnamefont{Bizon}} \bibnamefont{and}
  \bibinfo{author}{\bibfnamefont{A.}~\bibnamefont{Rostworowski}}
  (\bibinfo{year}{2011}), \eprint{1104.3702}.

\bibitem[{\citenamefont{Garfinkle and Pando~Zayas}(2011)}]{Garfinkle:2011hm}
\bibinfo{author}{\bibfnamefont{D.}~\bibnamefont{Garfinkle}} \bibnamefont{and}
  \bibinfo{author}{\bibfnamefont{L.~A.} \bibnamefont{Pando~Zayas}}
  (\bibinfo{year}{2011}), \eprint{1106.2339}.

\bibitem[{\citenamefont{Husain et~al.}(2003)\citenamefont{Husain, Kunstatter,
  Preston, and Birukou}}]{Husain:2002nk}
\bibinfo{author}{\bibfnamefont{V.}~\bibnamefont{Husain}},
  \bibinfo{author}{\bibfnamefont{G.}~\bibnamefont{Kunstatter}},
  \bibinfo{author}{\bibfnamefont{B.}~\bibnamefont{Preston}}, \bibnamefont{and}
  \bibinfo{author}{\bibfnamefont{M.}~\bibnamefont{Birukou}},
  \bibinfo{journal}{Class. Quant. Grav.} \textbf{\bibinfo{volume}{20}},
  \bibinfo{pages}{L23} (\bibinfo{year}{2003}), \eprint{gr-qc/0210011}.

\bibitem[{\citenamefont{Birukou et~al.}(2002)\citenamefont{Birukou, Husain,
  Kunstatter, Vaz, and Olivier}}]{Birukou:2002ge}
\bibinfo{author}{\bibfnamefont{M.}~\bibnamefont{Birukou}},
  \bibinfo{author}{\bibfnamefont{V.}~\bibnamefont{Husain}},
  \bibinfo{author}{\bibfnamefont{G.}~\bibnamefont{Kunstatter}},
  \bibinfo{author}{\bibfnamefont{E.}~\bibnamefont{Vaz}}, \bibnamefont{and}
  \bibinfo{author}{\bibfnamefont{M.}~\bibnamefont{Olivier}},
  \bibinfo{journal}{Phys.Rev.} \textbf{\bibinfo{volume}{D65}},
  \bibinfo{pages}{104036} (\bibinfo{year}{2002}).

\bibitem[{\citenamefont{Chesler and Yaffe}(2011)}]{Chesler:2010bi}
\bibinfo{author}{\bibfnamefont{P.~M.} \bibnamefont{Chesler}} \bibnamefont{and}
  \bibinfo{author}{\bibfnamefont{L.~G.} \bibnamefont{Yaffe}},
  \bibinfo{journal}{Phys. Rev. Lett.} \textbf{\bibinfo{volume}{106}},
  \bibinfo{pages}{021601} (\bibinfo{year}{2011}), \eprint{1011.3562}.

\bibitem[{\citenamefont{Wu and Romatschke}(2011)}]{Wu:2011yd}
\bibinfo{author}{\bibfnamefont{B.}~\bibnamefont{Wu}} \bibnamefont{and}
  \bibinfo{author}{\bibfnamefont{P.}~\bibnamefont{Romatschke}}
  (\bibinfo{year}{2011}), \eprint{1108.3715}.

\bibitem[{\citenamefont{Chesler and Teaney}(2011)}]{Chesler:2011ds}
\bibinfo{author}{\bibfnamefont{P.~M.} \bibnamefont{Chesler}} \bibnamefont{and}
  \bibinfo{author}{\bibfnamefont{D.}~\bibnamefont{Teaney}}
  (\bibinfo{year}{2011}), \eprint{1112.6196}.

\bibitem[{\citenamefont{Heller et~al.}(2011)\citenamefont{Heller, Janik, and
  Witaszczyk}}]{Heller:2011ju}
\bibinfo{author}{\bibfnamefont{M.~P.} \bibnamefont{Heller}},
  \bibinfo{author}{\bibfnamefont{R.~A.} \bibnamefont{Janik}}, \bibnamefont{and}
  \bibinfo{author}{\bibfnamefont{P.}~\bibnamefont{Witaszczyk}}
  (\bibinfo{year}{2011}), \eprint{1103.3452}.

\bibitem[{\citenamefont{Pretorius}(2005{\natexlab{a}})}]{Pretorius:2004jg}
\bibinfo{author}{\bibfnamefont{F.}~\bibnamefont{Pretorius}},
  \bibinfo{journal}{Class.Quant.Grav.} \textbf{\bibinfo{volume}{22}},
  \bibinfo{pages}{425} (\bibinfo{year}{2005}{\natexlab{a}}),
  \eprint{gr-qc/0407110}.

\bibitem[{\citenamefont{Pretorius}(2005{\natexlab{b}})}]{Pretorius:2005gq}
\bibinfo{author}{\bibfnamefont{F.}~\bibnamefont{Pretorius}},
  \bibinfo{journal}{Phys.Rev.Lett.} \textbf{\bibinfo{volume}{95}},
  \bibinfo{pages}{121101} (\bibinfo{year}{2005}{\natexlab{b}}),
  \eprint{gr-qc/0507014}.

\bibitem[{\citenamefont{Gubser et~al.}(2008)\citenamefont{Gubser, Pufu, and
  Yarom}}]{Gubser:2008pc}
\bibinfo{author}{\bibfnamefont{S.~S.} \bibnamefont{Gubser}},
  \bibinfo{author}{\bibfnamefont{S.~S.} \bibnamefont{Pufu}}, \bibnamefont{and}
  \bibinfo{author}{\bibfnamefont{A.}~\bibnamefont{Yarom}},
  \bibinfo{journal}{Phys. Rev.} \textbf{\bibinfo{volume}{D78}},
  \bibinfo{pages}{066014} (\bibinfo{year}{2008}), \eprint{0805.1551}.

\bibitem[{\citenamefont{Gubser et~al.}(2009)\citenamefont{Gubser, Pufu, and
  Yarom}}]{Gubser:2009sx}
\bibinfo{author}{\bibfnamefont{S.~S.} \bibnamefont{Gubser}},
  \bibinfo{author}{\bibfnamefont{S.~S.} \bibnamefont{Pufu}}, \bibnamefont{and}
  \bibinfo{author}{\bibfnamefont{A.}~\bibnamefont{Yarom}},
  \bibinfo{journal}{JHEP} \textbf{\bibinfo{volume}{11}}, \bibinfo{pages}{050}
  (\bibinfo{year}{2009}), \eprint{0902.4062}.

\bibitem[{\citenamefont{Kovtun and Starinets}(2006)}]{Kovtun:2006pf}
\bibinfo{author}{\bibfnamefont{P.}~\bibnamefont{Kovtun}} \bibnamefont{and}
  \bibinfo{author}{\bibfnamefont{A.}~\bibnamefont{Starinets}},
  \bibinfo{journal}{Phys.Rev.Lett.} \textbf{\bibinfo{volume}{96}},
  \bibinfo{pages}{131601} (\bibinfo{year}{2006}), \eprint{hep-th/0602059}.

\bibitem[{\citenamefont{Janik and Peschanski}(2006)}]{Janik:2006gp}
\bibinfo{author}{\bibfnamefont{R.~A.} \bibnamefont{Janik}} \bibnamefont{and}
  \bibinfo{author}{\bibfnamefont{R.~B.} \bibnamefont{Peschanski}},
  \bibinfo{journal}{Phys. Rev.} \textbf{\bibinfo{volume}{D74}},
  \bibinfo{pages}{046007} (\bibinfo{year}{2006}), \eprint{hep-th/0606149}.

\bibitem[{\citenamefont{Friess et~al.}(2007)\citenamefont{Friess, Gubser,
  Michalogiorgakis, and Pufu}}]{Friess:2006kw}
\bibinfo{author}{\bibfnamefont{J.~J.} \bibnamefont{Friess}},
  \bibinfo{author}{\bibfnamefont{S.~S.} \bibnamefont{Gubser}},
  \bibinfo{author}{\bibfnamefont{G.}~\bibnamefont{Michalogiorgakis}},
  \bibnamefont{and} \bibinfo{author}{\bibfnamefont{S.~S.} \bibnamefont{Pufu}},
  \bibinfo{journal}{JHEP} \textbf{\bibinfo{volume}{04}}, \bibinfo{pages}{080}
  (\bibinfo{year}{2007}), \eprint{hep-th/0611005}.

\bibitem[{\citenamefont{Ishibashi and Wald}(2004)}]{Ishibashi:2004wx}
\bibinfo{author}{\bibfnamefont{A.}~\bibnamefont{Ishibashi}} \bibnamefont{and}
  \bibinfo{author}{\bibfnamefont{R.~M.} \bibnamefont{Wald}},
  \bibinfo{journal}{Class.Quant.Grav.} \textbf{\bibinfo{volume}{21}},
  \bibinfo{pages}{2981} (\bibinfo{year}{2004}), \eprint{hep-th/0402184}.

\bibitem[{\citenamefont{Henneaux and Teitelboim}(1985)}]{Henneaux:1985tv}
\bibinfo{author}{\bibfnamefont{M.}~\bibnamefont{Henneaux}} \bibnamefont{and}
  \bibinfo{author}{\bibfnamefont{C.}~\bibnamefont{Teitelboim}},
  \bibinfo{journal}{Commun. Math. Phys.} \textbf{\bibinfo{volume}{98}},
  \bibinfo{pages}{391} (\bibinfo{year}{1985}).

\bibitem[{\citenamefont{Fefferman and Graham}(1985)}]{fefferman}
\bibinfo{author}{\bibfnamefont{C.}~\bibnamefont{Fefferman}} \bibnamefont{and}
  \bibinfo{author}{\bibfnamefont{C.~R.} \bibnamefont{Graham}},
  \bibinfo{journal}{Ast\'erisque} pp. \bibinfo{pages}{95--116}
  (\bibinfo{year}{1985}), ISSN \bibinfo{issn}{0303-1179}, \bibinfo{note}{the
  mathematical heritage of {\'E}lie Cartan (Lyon, 1984)}.

\bibitem[{\citenamefont{de~Haro et~al.}(2001)\citenamefont{de~Haro, Solodukhin,
  and Skenderis}}]{deHaro:2000xn}
\bibinfo{author}{\bibfnamefont{S.}~\bibnamefont{de~Haro}},
  \bibinfo{author}{\bibfnamefont{S.~N.} \bibnamefont{Solodukhin}},
  \bibnamefont{and}
  \bibinfo{author}{\bibfnamefont{K.}~\bibnamefont{Skenderis}},
  \bibinfo{journal}{Commun.Math.Phys.} \textbf{\bibinfo{volume}{217}},
  \bibinfo{pages}{595} (\bibinfo{year}{2001}), \eprint{hep-th/0002230}.

\bibitem[{\citenamefont{Henneaux et~al.}(2007)\citenamefont{Henneaux, Martinez,
  Troncoso, and Zanelli}}]{Henneaux:2006hk}
\bibinfo{author}{\bibfnamefont{M.}~\bibnamefont{Henneaux}},
  \bibinfo{author}{\bibfnamefont{C.}~\bibnamefont{Martinez}},
  \bibinfo{author}{\bibfnamefont{R.}~\bibnamefont{Troncoso}}, \bibnamefont{and}
  \bibinfo{author}{\bibfnamefont{J.}~\bibnamefont{Zanelli}},
  \bibinfo{journal}{Annals Phys.} \textbf{\bibinfo{volume}{322}},
  \bibinfo{pages}{824} (\bibinfo{year}{2007}), \eprint{hep-th/0603185}.

\bibitem[{\citenamefont{Papadimitriou and
  Skenderis}(2005)}]{Papadimitriou:2005ii}
\bibinfo{author}{\bibfnamefont{I.}~\bibnamefont{Papadimitriou}}
  \bibnamefont{and}
  \bibinfo{author}{\bibfnamefont{K.}~\bibnamefont{Skenderis}},
  \bibinfo{journal}{JHEP} \textbf{\bibinfo{volume}{0508}}, \bibinfo{pages}{004}
  (\bibinfo{year}{2005}), \eprint{hep-th/0505190}.

\bibitem[{\citenamefont{Brown and York}(1993)}]{Brown:1992br}
\bibinfo{author}{\bibfnamefont{J.~D.} \bibnamefont{Brown}} \bibnamefont{and}
  \bibinfo{author}{\bibfnamefont{J.~W.} \bibnamefont{York},
  \bibfnamefont{Jr.}}, \bibinfo{journal}{Phys. Rev.}
  \textbf{\bibinfo{volume}{D47}}, \bibinfo{pages}{1407} (\bibinfo{year}{1993}),
  \eprint{gr-qc/9209012}.

\bibitem[{\citenamefont{Balasubramanian and
  Kraus}(1999)}]{Balasubramanian:1999re}
\bibinfo{author}{\bibfnamefont{V.}~\bibnamefont{Balasubramanian}}
  \bibnamefont{and} \bibinfo{author}{\bibfnamefont{P.}~\bibnamefont{Kraus}},
  \bibinfo{journal}{Commun. Math. Phys.} \textbf{\bibinfo{volume}{208}},
  \bibinfo{pages}{413} (\bibinfo{year}{1999}), \eprint{hep-th/9902121}.

\bibitem[{\citenamefont{Arnowitt et~al.}(1962)\citenamefont{Arnowitt, Deser,
  and Misner}}]{Arnowitt:1962hi}
\bibinfo{author}{\bibfnamefont{R.~L.} \bibnamefont{Arnowitt}},
  \bibinfo{author}{\bibfnamefont{S.}~\bibnamefont{Deser}}, \bibnamefont{and}
  \bibinfo{author}{\bibfnamefont{C.~W.} \bibnamefont{Misner}}
  (\bibinfo{year}{1962}), \bibinfo{note}{gravitation: an introduction to
  current research, Louis Witten ed. (Wilew 1962), chapter 7, pp 227-265},
  \eprint{gr-qc/0405109}.

\bibitem[{\citenamefont{Gourgoulhon}(2007)}]{Gourgoulhon:2007tn}
\bibinfo{author}{\bibfnamefont{E.}~\bibnamefont{Gourgoulhon}},
  \bibinfo{journal}{J. Phys. Conf. Ser.} \textbf{\bibinfo{volume}{91}},
  \bibinfo{pages}{012001} (\bibinfo{year}{2007}), \eprint{0704.0149}.

\bibitem[{\citenamefont{{Garfinkle}}(2002)}]{2002APS..APRC12004G}
\bibinfo{author}{\bibfnamefont{D.}~\bibnamefont{{Garfinkle}}},
  \bibinfo{journal}{APS Meeting Abstracts} pp. \bibinfo{pages}{12004--+}
  (\bibinfo{year}{2002}).

\bibitem[{\citenamefont{Lindblom et~al.}(2006)\citenamefont{Lindblom, Scheel,
  Kidder, Owen, and Rinne}}]{Lindblom:2005qh}
\bibinfo{author}{\bibfnamefont{L.}~\bibnamefont{Lindblom}},
  \bibinfo{author}{\bibfnamefont{M.~A.} \bibnamefont{Scheel}},
  \bibinfo{author}{\bibfnamefont{L.~E.} \bibnamefont{Kidder}},
  \bibinfo{author}{\bibfnamefont{R.}~\bibnamefont{Owen}}, \bibnamefont{and}
  \bibinfo{author}{\bibfnamefont{O.}~\bibnamefont{Rinne}},
  \bibinfo{journal}{Class.Quant.Grav.} \textbf{\bibinfo{volume}{23}},
  \bibinfo{pages}{S447} (\bibinfo{year}{2006}), \eprint{gr-qc/0512093}.

\bibitem[{\citenamefont{{Friedrich}}(1985)}]{1985CMaPh.100..525F}
\bibinfo{author}{\bibfnamefont{H.}~\bibnamefont{{Friedrich}}},
  \bibinfo{journal}{Communications in Mathematical Physics}
  \textbf{\bibinfo{volume}{100}}, \bibinfo{pages}{525} (\bibinfo{year}{1985}).

\bibitem[{\citenamefont{Gundlach et~al.}(2005)\citenamefont{Gundlach,
  Martin-Garcia, Calabrese, and Hinder}}]{Gundlach:2005eh}
\bibinfo{author}{\bibfnamefont{C.}~\bibnamefont{Gundlach}},
  \bibinfo{author}{\bibfnamefont{J.~M.} \bibnamefont{Martin-Garcia}},
  \bibinfo{author}{\bibfnamefont{G.}~\bibnamefont{Calabrese}},
  \bibnamefont{and} \bibinfo{author}{\bibfnamefont{I.}~\bibnamefont{Hinder}},
  \bibinfo{journal}{Class. Quant. Grav.} \textbf{\bibinfo{volume}{22}},
  \bibinfo{pages}{3767} (\bibinfo{year}{2005}), \eprint{gr-qc/0504114}.

\bibitem[{\citenamefont{Garfinkle and Duncan}(2001)}]{Garfinkle:2000hd}
\bibinfo{author}{\bibfnamefont{D.}~\bibnamefont{Garfinkle}} \bibnamefont{and}
  \bibinfo{author}{\bibfnamefont{G.~C.} \bibnamefont{Duncan}},
  \bibinfo{journal}{Phys. Rev.} \textbf{\bibinfo{volume}{D63}},
  \bibinfo{pages}{044011} (\bibinfo{year}{2001}), \eprint{gr-qc/0006073}.

\bibitem[{\citenamefont{Hubeny et~al.}(2011)\citenamefont{Hubeny, Minwalla, and
  Rangamani}}]{Hubeny:2011hd}
\bibinfo{author}{\bibfnamefont{V.~E.} \bibnamefont{Hubeny}},
  \bibinfo{author}{\bibfnamefont{S.}~\bibnamefont{Minwalla}}, \bibnamefont{and}
  \bibinfo{author}{\bibfnamefont{M.}~\bibnamefont{Rangamani}}
  (\bibinfo{year}{2011}), \eprint{1107.5780}.

\bibitem[{\citenamefont{Stephani et~al.}(2003)\citenamefont{Stephani, Kramer,
  Maccallum, Hoenselaers, and Herlt}}]{Stephani:2003tx}
\bibinfo{author}{\bibfnamefont{H.}~\bibnamefont{Stephani}},
  \bibinfo{author}{\bibfnamefont{D.}~\bibnamefont{Kramer}},
  \bibinfo{author}{\bibfnamefont{M.}~\bibnamefont{Maccallum}},
  \bibinfo{author}{\bibfnamefont{C.}~\bibnamefont{Hoenselaers}},
  \bibnamefont{and} \bibinfo{author}{\bibfnamefont{E.}~\bibnamefont{Herlt}},
  \emph{\bibinfo{title}{Exact Solutions of Einstein's Field Equations}}
  (\bibinfo{publisher}{Cambridge University Press},
  \bibinfo{address}{Cambridge, UK}, \bibinfo{year}{2003}).

\bibitem[{\citenamefont{Kreiss and Oliger}(1973)}]{KO}
\bibinfo{author}{\bibfnamefont{H.-O.} \bibnamefont{Kreiss}} \bibnamefont{and}
  \bibinfo{author}{\bibfnamefont{J.}~\bibnamefont{Oliger}},
  \bibinfo{journal}{Global Atmospheric Research Program Publication}
  \textbf{\bibinfo{volume}{10}} (\bibinfo{year}{1973}).

\bibitem[{\citenamefont{Calabrese et~al.}(2004)\citenamefont{Calabrese, Lehner,
  Reula, Sarbach, and Tiglio}}]{Calabrese:2003vx}
\bibinfo{author}{\bibfnamefont{G.}~\bibnamefont{Calabrese}},
  \bibinfo{author}{\bibfnamefont{L.}~\bibnamefont{Lehner}},
  \bibinfo{author}{\bibfnamefont{O.}~\bibnamefont{Reula}},
  \bibinfo{author}{\bibfnamefont{O.}~\bibnamefont{Sarbach}}, \bibnamefont{and}
  \bibinfo{author}{\bibfnamefont{M.}~\bibnamefont{Tiglio}},
  \bibinfo{journal}{Class.Quant.Grav.} \textbf{\bibinfo{volume}{21}},
  \bibinfo{pages}{5735} (\bibinfo{year}{2004}), \eprint{gr-qc/0308007}.

\bibitem[{\citenamefont{Pfeiffer and York}(2005)}]{Pfeiffer:2005jf}
\bibinfo{author}{\bibfnamefont{H.~P.} \bibnamefont{Pfeiffer}} \bibnamefont{and}
  \bibinfo{author}{\bibfnamefont{J.~W.} \bibnamefont{York},
  \bibfnamefont{Jr.}}, \bibinfo{journal}{Phys. Rev. Lett.}
  \textbf{\bibinfo{volume}{95}}, \bibinfo{pages}{091101}
  (\bibinfo{year}{2005}), \eprint{gr-qc/0504142}.

\bibitem[{\citenamefont{Horowitz and Hubeny}(2000)}]{Horowitz:1999jd}
\bibinfo{author}{\bibfnamefont{G.~T.} \bibnamefont{Horowitz}} \bibnamefont{and}
  \bibinfo{author}{\bibfnamefont{V.~E.} \bibnamefont{Hubeny}},
  \bibinfo{journal}{Phys. Rev.} \textbf{\bibinfo{volume}{D62}},
  \bibinfo{pages}{024027} (\bibinfo{year}{2000}), \eprint{hep-th/9909056}.

\bibitem[{\citenamefont{Cardoso et~al.}(2003)\citenamefont{Cardoso, Konoplya,
  and Lemos}}]{Cardoso:2003cj}
\bibinfo{author}{\bibfnamefont{V.}~\bibnamefont{Cardoso}},
  \bibinfo{author}{\bibfnamefont{R.}~\bibnamefont{Konoplya}}, \bibnamefont{and}
  \bibinfo{author}{\bibfnamefont{J.~P.~S.} \bibnamefont{Lemos}},
  \bibinfo{journal}{Phys. Rev.} \textbf{\bibinfo{volume}{D68}},
  \bibinfo{pages}{044024} (\bibinfo{year}{2003}), \eprint{gr-qc/0305037}.

\bibitem[{\citenamefont{Kovtun and Starinets}(2005)}]{Kovtun:2005ev}
\bibinfo{author}{\bibfnamefont{P.~K.} \bibnamefont{Kovtun}} \bibnamefont{and}
  \bibinfo{author}{\bibfnamefont{A.~O.} \bibnamefont{Starinets}},
  \bibinfo{journal}{Phys. Rev.} \textbf{\bibinfo{volume}{D72}},
  \bibinfo{pages}{086009} (\bibinfo{year}{2005}), \eprint{hep-th/0506184}.

\bibitem[{\citenamefont{Teaney}(2006)}]{Teaney:2006nc}
\bibinfo{author}{\bibfnamefont{D.}~\bibnamefont{Teaney}},
  \bibinfo{journal}{Phys. Rev.} \textbf{\bibinfo{volume}{D74}},
  \bibinfo{pages}{045025} (\bibinfo{year}{2006}), \eprint{hep-ph/0602044}.

\bibitem[{\citenamefont{Berti et~al.}(2009)\citenamefont{Berti, Cardoso, and
  Starinets}}]{Berti:2009kk}
\bibinfo{author}{\bibfnamefont{E.}~\bibnamefont{Berti}},
  \bibinfo{author}{\bibfnamefont{V.}~\bibnamefont{Cardoso}}, \bibnamefont{and}
  \bibinfo{author}{\bibfnamefont{A.~O.} \bibnamefont{Starinets}},
  \bibinfo{journal}{Class. Quant. Grav.} \textbf{\bibinfo{volume}{26}},
  \bibinfo{pages}{163001} (\bibinfo{year}{2009}), \eprint{0905.2975}.

\bibitem[{\citenamefont{Avery}(2000)}]{Avery:2000tx}
\bibinfo{author}{\bibfnamefont{J.}~\bibnamefont{Avery}},
  \emph{\bibinfo{title}{Hyperspherical Harmonics and Generalized Sturmians}}
  (\bibinfo{publisher}{Kluwer Academic Publishers},
  \bibinfo{address}{Dordrecht, Netherlands}, \bibinfo{year}{2000}).

\bibitem[{\citenamefont{Kodama and Ishibashi}(2003)}]{Kodama:2003jz}
\bibinfo{author}{\bibfnamefont{H.}~\bibnamefont{Kodama}} \bibnamefont{and}
  \bibinfo{author}{\bibfnamefont{A.}~\bibnamefont{Ishibashi}},
  \bibinfo{journal}{Prog. Theor. Phys.} \textbf{\bibinfo{volume}{110}},
  \bibinfo{pages}{701} (\bibinfo{year}{2003}), \eprint{hep-th/0305147}.

\bibitem[{\citenamefont{Emparan}(2011)}]{Emparan:2011br}
\bibinfo{author}{\bibfnamefont{R.}~\bibnamefont{Emparan}}
  (\bibinfo{year}{2011}), \eprint{1106.2021}.

\bibitem[{\citenamefont{Loganayagam}(2008)}]{Loganayagam:2008is}
\bibinfo{author}{\bibfnamefont{R.}~\bibnamefont{Loganayagam}},
  \bibinfo{journal}{JHEP} \textbf{\bibinfo{volume}{05}}, \bibinfo{pages}{087}
  (\bibinfo{year}{2008}), \eprint{0801.3701}.

\bibitem[{\citenamefont{Gubser}(2010{\natexlab{b}})}]{Gubser:2010ze}
\bibinfo{author}{\bibfnamefont{S.~S.} \bibnamefont{Gubser}},
  \bibinfo{journal}{Phys. Rev.} \textbf{\bibinfo{volume}{D82}},
  \bibinfo{pages}{085027} (\bibinfo{year}{2010}{\natexlab{b}}),
  \eprint{1006.0006}.

\bibitem[{\citenamefont{Gubser and Yarom}(2011)}]{Gubser:2010ui}
\bibinfo{author}{\bibfnamefont{S.~S.} \bibnamefont{Gubser}} \bibnamefont{and}
  \bibinfo{author}{\bibfnamefont{A.}~\bibnamefont{Yarom}},
  \bibinfo{journal}{Nucl. Phys.} \textbf{\bibinfo{volume}{B846}},
  \bibinfo{pages}{469} (\bibinfo{year}{2011}), \eprint{1012.1314}.

\bibitem[{\citenamefont{Staig and Shuryak}(2011{\natexlab{a}})}]{Staig:2011wj}
\bibinfo{author}{\bibfnamefont{P.}~\bibnamefont{Staig}} \bibnamefont{and}
  \bibinfo{author}{\bibfnamefont{E.}~\bibnamefont{Shuryak}},
  \bibinfo{journal}{Phys. Rev.} \textbf{\bibinfo{volume}{C84}},
  \bibinfo{pages}{044912} (\bibinfo{year}{2011}{\natexlab{a}}),
  \eprint{1105.0676}.

\bibitem[{\citenamefont{Staig and Shuryak}(2011{\natexlab{b}})}]{Staig:2011as}
\bibinfo{author}{\bibfnamefont{P.}~\bibnamefont{Staig}} \bibnamefont{and}
  \bibinfo{author}{\bibfnamefont{E.}~\bibnamefont{Shuryak}},
  \bibinfo{journal}{J. Phys.} \textbf{\bibinfo{volume}{G38}},
  \bibinfo{pages}{124039} (\bibinfo{year}{2011}{\natexlab{b}}),
  \eprint{1106.3243}.

\bibitem[{\citenamefont{Skenderis}(2001)}]{Skenderis:2000in}
\bibinfo{author}{\bibfnamefont{K.}~\bibnamefont{Skenderis}},
  \bibinfo{journal}{Int. J. Mod. Phys.} \textbf{\bibinfo{volume}{A16}},
  \bibinfo{pages}{740} (\bibinfo{year}{2001}), \eprint{hep-th/0010138}.

\bibitem[{\citenamefont{Landau}(1953)}]{Landau:1953gs}
\bibinfo{author}{\bibfnamefont{L.~D.} \bibnamefont{Landau}},
  \bibinfo{journal}{Izv. Akad. Nauk SSSR Ser. Fiz.}
  \textbf{\bibinfo{volume}{17}}, \bibinfo{pages}{51} (\bibinfo{year}{1953}).

\bibitem[{\citenamefont{Khalatnikov}(1954)}]{Khalatnikov:1954}
\bibinfo{author}{\bibfnamefont{I.~M.} \bibnamefont{Khalatnikov}},
  \bibinfo{journal}{ZhETF} \textbf{\bibinfo{volume}{27}}, \bibinfo{pages}{529}
  (\bibinfo{year}{1954}).

\bibitem[{\citenamefont{Lehner and Pretorius}(2010)}]{Lehner:2010pn}
\bibinfo{author}{\bibfnamefont{L.}~\bibnamefont{Lehner}} \bibnamefont{and}
  \bibinfo{author}{\bibfnamefont{F.}~\bibnamefont{Pretorius}},
  \bibinfo{journal}{Phys.Rev.Lett.} \textbf{\bibinfo{volume}{105}},
  \bibinfo{pages}{101102} (\bibinfo{year}{2010}), \eprint{1006.5960}.

\bibitem[{\citenamefont{Steinberg}(2005)}]{Steinberg:2004vy}
\bibinfo{author}{\bibfnamefont{P.}~\bibnamefont{Steinberg}},
  \bibinfo{journal}{Acta Phys. Hung.} \textbf{\bibinfo{volume}{A24}},
  \bibinfo{pages}{51} (\bibinfo{year}{2005}), \eprint{nucl-ex/0405022}.

\bibitem[{\citenamefont{Breitenlohner and
  Freedman}(1982)}]{Breitenlohner:1982bm}
\bibinfo{author}{\bibfnamefont{P.}~\bibnamefont{Breitenlohner}}
  \bibnamefont{and} \bibinfo{author}{\bibfnamefont{D.~Z.}
  \bibnamefont{Freedman}}, \bibinfo{journal}{Phys. Lett.}
  \textbf{\bibinfo{volume}{B115}}, \bibinfo{pages}{197} (\bibinfo{year}{1982}).

\end{thebibliography}

\end{document}